\documentclass{aastex61}
\received{\today}
\revised{}
\accepted{}
\submitjournal{ApJ}

\shorttitle{Chemistry of COMs}
\shortauthors{Suzuki et al.}

\begin{document}

\title{The Difference in Abundances between N-bearing and O-bearing Species in High-Mass Star-Forming Regions}

\author[0000-0003-3278-2513]{Taiki Suzuki}
\affil{Department of Astronomy, the Graduate University for Advanced Studies (SOKENDAI), Osawa 2-21-1, Mitaka, Tokyo 181-8588, Japan}
\affil{Astrobiology Center, Osawa 2-21-1, Mitaka, Tokyo 181-8588, Japan}
\email{taiki.suzuki@nao.ac.jp}

\author{Masatoshi Ohishi}
\affil{Department of Astronomy, the Graduate University for Advanced Studies (SOKENDAI), Osawa 2-21-1, Mitaka, Tokyo 181-8588, Japan}
\affil{National Astronomical Observatory of Japan, Osawa 2-21-1, Mitaka, Tokyo 181-8588, Japan}

\author{Masao Saito}
\affil{Department of Astronomy, the Graduate University for Advanced Studies (SOKENDAI), Osawa 2-21-1, Mitaka, Tokyo 181-8588, Japan}
\affil{National Astronomical Observatory of Japan, Osawa 2-21-1, Mitaka, Tokyo 181-8588, Japan}

\author{Tomoya Hirota}
\affil{Department of Astronomy, the Graduate University for Advanced Studies (SOKENDAI), Osawa 2-21-1, Mitaka, Tokyo 181-8588, Japan}
\affil{National Astronomical Observatory of Japan, Osawa 2-21-1, Mitaka, Tokyo 181-8588, Japan}

\author{Liton Majumdar}
\affiliation{Jet Propulsion Laboratory, California Institute of Technology, 4800 Oak Grove Drive, Pasadena, CA 91109, USA}
\email{liton.majumdar@jpl.nasa.gov}

\author{Valentine Wakelam}
\affil{Laboratoire d'astrophysique de Bordeaux, Univ. Bordeaux, CNRS, B18N, all\'{e}e Geoffroy Saint-Hilaire, 33615 Pessac, France}



\begin{abstract}
The different spatial distributions of N-bearing and O-bearing species, as is well known towards Orion~KL, is one of the long-lasting mysteries. We conducted a survey observation and chemical modeling study to investigate if the different distributions of O- and N-bearing species are widely recognized in general star-forming regions.
First, we report our observational results of complex organic molecules (COMs) with the 45~m radio telescope at the Nobeyama Radio Observatory towards eight star-forming regions. Through our spectral survey ranging from 80 to 108~GHz, we detected CH$_3$OH, HCOOCH$_3$, CH$_3$OCH$_3$, (CH$_3$)$_2$CO, CH$_3$CHO, CH$_3$CH$_2$CN, CH$_2$CHCN, and NH$_2$CHO. Their molecular abundances were derived via the rotation diagram and the least squares methods. 
We found that N-bearing molecules, tend to show stronger correlations with other N-bearing molecules rather than O-bearing molecules.
While G10.47+0.03 showed high fractional abundances of N-bearing species, those in NGC6334F were not so rich, being less than 0.01 compared to CH$_3$OH. 
Then, the molecular abundances towards these sources were evaluated by chemical modeling with NAUTILUS three-phase gas-grain chemical code. Through the simulations of time evolutions for the abundances of COMs, we suggest that observed correlations of fractional abundances between COMs can be explained by the combination of the different temperature structures inside the hot cores and the different evolutionary phase. Since our modeling could not fully explain the observed excitation temperatures, it is important to investigate the efficiency of grain surface reactions and their activation barriers, and the binding energy of COMs to further promote our understanding.
\end{abstract}




\section{Introduction}
In the interstellar medium (ISM), almost 200 molecules (as listed by CDMS\footnote{\url{https://www.astro.uni-koeln.de/cdms/molecules}}) ranging from simple linear molecules to complex organic molecules (COMs) were detected, mainly towards dark clouds, low-mass and high-mass star-forming regions.
While relatively simple species have been prime target for the previous astronomical studies, the study of interstellar chemistry of COMs are still limited and has a lot of room to debate.

Among COMs, methanol (CH$_3$OH), dimethyl either (CH$_3$OCH$_3$) and methyl formate (HCOOCH$_3$) have been detected towards various sources.
Since the first detection by \cite{Ball70}, CH$_3$OH was detected toward various high-mass and low-mass star-forming regions, and circumstellar shell.
HCOOCH$_3$ was firstly detected by \cite{Brown75}, and then HCOOCH$_3$ has been reported towards several star forming regions \citep[e.g.,][]{Bisschop07}.
Since the first detection by \cite{Snyder74}, CH$_3$OCH$_3$ is known as one of the abundant species in hot cores \citep{Ikeda01} and hot corinos such as IRAS~16293-2422 \citep{Cazaux03,Bottinelli04}.
N-bearing COMs have been less studied compared to these O-bearing species.
Well known N-bearing species are vinyl cyanide (CH$_2$CHCN) and ethyl cyanide (CH$_3$CH$_2$CN), which have been detected towards Sgr~B2, Orion~KL, and other star-forming regions \citep[e.g.,][]{Gardner75,Johnson77}.
Recently, our group has successfully detected CH$_2$NH and CH$_3$NH$_2$ towards several star-forming regions \citep{Suzuki16,Ohishi17}.

However, the overall picture of COM evolution along with star-formation is still under debate.
The correlations of molecular abundances among certain species would be one of the good clues to connect the evolution of COMs and the physical condition of hot cores.
Such molecular correlation has been discussed in Orion~KL.
\cite{Blake87} claimed that O- and N-bearing species had different radial velocities.
The recent interferometric observations enabled us to see the different spatial distributions of COMs directly \citep[e.g.,][]{Friedel08,Peng13}, giving us the direct evidence that O-bearing species are enhanced in the Orion Compact Ridge and N-bearing species are rich in Orion Hot Core.
\cite{Caselli93} discussed the origin of chemical difference in Hot Core and Compact Ridge by chemical modeling study.
They claimed that the different chemistry would be due to the efficiency of thermal evaporation before the warm-up of the cores, assuming that the initial temperature of Hot Core (40~K) would be hotter than Compact Ridge (20~K).
They reported that H$_2$CO was efficiently converted to CH$_3$OH in Compact ridge than Hot Core, resulting in increase of not only CH$_3$OH, but also CH$_3$OCH$_3$ and HCOOCH$_3$, through the formation of CH$_3$OCH$_3$ and HCOOCH$_3$ via gas phase reactions using CH$_3$OH.
Their result also suggest that while the difference in molecular abundances are caused by their chemical connections, but strongly enhanced by the source properties.

Since the chemical difference in hot cores would be a key to improve our knowledge about chemical evolution of COMs, it would be quite important to know if such the correlation of molecular abundances are ubiquitous phenomena or limited to only special sources.
The first survey of O- and N-bearing COMs, CH$_2$CHCN, CH$_3$CH$_2$CN, and CH$_3$OCH$_3$, was conducted by \cite{Fontani07}, reporting the correlation of not only ``CH$_2$CHCN and CH$_3$CH$_2$CN'', but also ``CH$_2$CHCN and CH$_3$OCH$_3$'' and ``CH$_3$CH$_2$CN and CH$_3$OCH$_3$''.
They reported that the correlations of molecular abundances of O- and N-bearing species.
Therefore, the clear evidence of different spatial distributions between O- and N- bearing species were not obtained.
In this paper, we will present the results of our single-dish survey observations of COMs towards high-mass star-forming regions, with large number of COMs than \cite{Fontani07}.
Since we could not obtain the spatial distributions of COMs, we use the radial velocities to discuss if molecules co-exist.
Then, we will investigate if the difference in molecular abundances of O- and N-bearing species are widely recognized.
Furthermore, we will discuss the connection of molecular correlation and the key physical condition of sources using our state-of-art chemical model.
As a result, we will report that, not only the connection of chemical reactions, the degree of absorption onto grains determined by gas temperature may strongly affect the molecular abundances.
The observational method will be described in section~2 (Observations); our results will be described in section~3 (Results);
We will compare the modeling results and the observed abundances in section~4 (Discussion).
We will summarize our work in section 5 (Conclusion).

\section{Observations}
\begin{figure}
 \begin{tabular}{l}
\includegraphics[scale=.6]{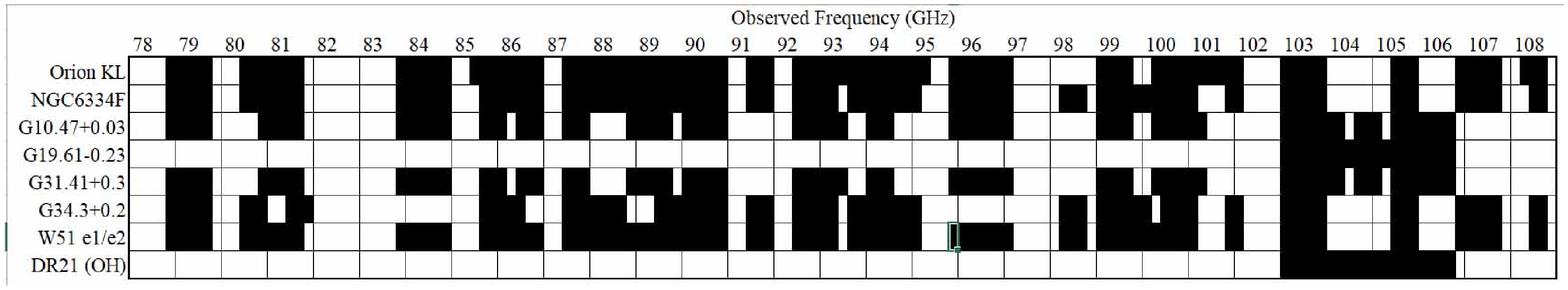}\\
  \end{tabular}
\caption{
Observed frequency ranges for our sources are shown with black color.
\label{fig:observed_frequency}
}
\end{figure}

In this paper, we report the observational results of well known COMs in hot cores: CH$_3$OH, CH$_3$OCH$_3$, (CH$_3$)$_2$CO, CH$_3$CHO, NH$_2$CHO, CH$_2$CHCN, and CH$_3$CH$_2$CN.
We conducted survey observations of COMs with the 45~m radio telescope at the Nobeyama Radio Observatory (NRO), National Astronomical Observatory of Japan.
Our data was simultaneously obtained with our previous observations aimed to detect CH$_2$NH and CH$_3$NH$_2$ \citep{Suzuki16,Ohishi17}, from March 2013 to May 2014.
We used the TZ receiver \citep{Nakajima13} in the dual polarization mode to obtain spectra between 78.8 to 108.4~GHz.
Since the molecular line data in this paper was side products of our previous surveys, our observed frequency ranges did not entirely cover this frequency range and {\bf differ} from source to source.
The observed frequency ranges for each source are shown with the black color in Figure~\ref{fig:observed_frequency}, and its actual values are summarized in Table~\ref{table:observed_frequency}.
We note that the observed frequency ranges are limited for DR21(OH) and G19.61-0.23, where CH$_3$NH$_2$ survey was not conducted.

The SAM45 spectrometer was used for the backend with a frequency resolution of 122~kHz, corresponding to a velocity resolution of 0.35~km~s$^{-1}$ at 105~GHz, but 0.46~km~s$^{-1}$ at 79~GHz.
The system temperatures ranged from 150 to 300~K.
The main beam efficiency ($\eta_{\rm mb}$) was 0.37 for more than 90~GHz and 0.41 for less than 90~GHz, and the beam size (FWHM) was linearly decreased from 20.6" at 80~ to 15.5" at 106~GHz.
Observations were performed in the position switching mode.
The pointing accuracy was checked by observing nearby SiO masers, and the pointing error was typically 5", wheres 10", under windy condition.

Our target were eight hot core sources, G10.47+0.03, G31.41+0.3, Orion~KL, NGC6334F, G34.3+0.2, W51~e1/e2, G19.61-0.23, and DR21~(OH), where we detected CH$_2$NH \citep{Suzuki16}.
\cite{Suzuki16} conducted a survey observation of CH$_2$NH towards hot core sources, where CH$_3$OH were known to be rich.
Since CH$_3$OH is known to be formed via hydrogenation processes on grains, CH$_3$OH would be a good indicator for other grain surface reactions, which would form various COMs.
The most CH$_2$NH abundant source in this survey was G10.47+0.03, with its fractional abundance compared to H$_2$ molecules of 3.1 $\times$10$^{-8}$.
The abundances of CH$_2$NH were almost 10 times less than G10.47+0.03, in W51~e1/e2, NGC6334F, and G19.61-0.23, where CH$_2$NH abundances were, respectively, 2.8 $\times$10$^{-9}$, 2.4 $\times$10$^{-9}$, and 1.4 $\times$10$^{-9}$.
Their different fractional abundances of CH$_2$NH indicate that other N-bearing species may be enhanced in these sources, as is known in Orion~KL Hot Core and Compact Ridge.
The observed sources and their properties are summarized in Table~\ref{table:sources}.

\section{Results}
\subsection{Observed Spectra}
In our survey, we detected more than 1000 molecular lines of CH$_3$OH, CH$_3$OCH$_3$, HCOOCH$_3$, CH$_3$CHO, (CH$_3$)$_2$CO, NH$_2$CHO, CH$_2$CHCN, and CH$_3$CH$_2$CN.

The rms noise levels per channel were ranging from 5 to $\sim$20~mK, depending on the source and the frequency.
This difference is due to the system temperature and the condition of TZ receiver.
The rms noise levels sometimes different from source to source due to our original purpose to detect CH$_2$NH and CH$_3$NH$_2$ and the strategy to achieve it.

An example of the observed spectra for all sources from 105.6 to 105.9~GHz is shown in Figure~\ref{fig:observed_spectra}.
Although the observed frequency ranges are different from source to source (Figure~\ref{fig:observed_frequency}), this frequency range enable us to compare the spectra for all observed sources.
These species are well known in hot cores and most of the detections were also supported by many authors through radio observations \cite[see for example,][]{Ikeda01,Zernickel12,Widicus17}
The detailed line parameters, such as radial velocities and line widths, were obtained by the task of ``Gaussian fitting'' with ``NEWSTAR''\footnote{\url{http://www.nro.nao.ac.jp/~nro45mrt/html/obs/newstar/}}, which was developed by NRO.
As the results of this procedure, our molecular lines were relatively well fitted unless they are blended with other transitions.
These line parameters will be described in the subsequent subsection for each source.
All the observed lines are available on-line as the supplementary file.
%

\subsection{Derivation of Abundance}
Here, we will describe the methodologies in deriving fractional abundances of COMs.
In sources where enough number of transitions were available, we calculated column densities (N$_{\rm rot}$) of COMs using the rotation-diagram method described in \cite{Turner91}, and the following equation was employed:
\begin{equation}
\log \frac{3kW}{8\pi^3 \nu \mu^{2}S g_{\rm I}g_{\rm K}} = \log \frac{N_{\rm rot}}{Q(T_{\rm rot})} - \frac{E_{\rm u}}{k} \frac{\log e}{T_{\rm rot}}
\end{equation}
where $W$ is the integrated intensity, $S$ is the intrinsic line strength, $d$ is the permanent electric dipole moment, $g_{\rm I}$ and $g_{\rm K}$ are the nuclear spin degeneracy and the $K$-level degeneracy, respectively, $N$ is the column density, $Q_{\rm rot}$ is the rotational partition function, $E_{\rm u}$ is the upper level energy, and $T_{\rm rot}$ is the rotation temperature.
The column density is derived from the interception of a diagram, and its slope will give us the excitation temperature. 
In this derivation, the source size of 10" was assumed.
As an example, the rotation diagrams of G10.47+0.03 and NGC6334F are shown in Figures~\ref{fig:G10_rotation_diagrams} and ~\ref{fig:NGC6334F_rotation_diagrams}.
While rotation diagram method assumes that observed transitions are optically thin, this assumption is not always correct.
Also, the source size may have large uncertainty.
A more accurate way to derive abundances is to use the least squares fitting method with the equations of radiative transfer model described in \cite{Suzuki92}.
This method does not require the assumption of optically thin property of observed transitions, but assumes LTE condition.
As we will show the result below, the observed transitions are well explained by this method, suggesting this assumption is not so bad.
In our calculation, we also assumed the effect of the source size, which lead to the beam dilution.
Then, if the column density and excitation temperature are given, the brightness temperature can be calculated with the following formula:
\begin{equation}
T_{\rm B} = (J(T_{\rm ex})-J(T_{\rm bb}))(1-\exp(-\tau))\eta_{\rm mb}(\frac{\Theta_{\rm Source}}{\Theta_{\rm beam}})^2,
\end{equation}
where
\begin{equation}
J(T) = \frac{{\rm h}\nu}{k}\frac{1}{\exp(\frac{{\rm h}\nu}{kT})-1}
\end{equation}
\begin{equation}
\tau = 1.248 \times 10^{-13} (\frac{\mu}{D})^2 S (\frac{N_{\rm rot}}{\rm{cm^{-2}}}) (\frac{\Delta v}{\rm{km~s^{-1}}})^{-1} \frac{f(T,E_{\rm u})}{Q(T_{\rm ex})}
\end{equation}
\begin{equation}
f(T,E_{\rm u}) = \exp(\frac{{\rm h}\nu}{kT}-1)\times\exp(-E_{\rm u}/kT)
\end{equation}

T$_{\rm bb}$ is the brightness temperature of the cosmic background radiation (2.7~K), $\eta_{\rm mb}$ is the main beam efficiency, $\Theta_{\rm Source}$ is the source size, and $\Theta_{\rm beam}$ is the beam size.

Thus, if more than three transitions are available, the best-estimated source size, the column density, and the excitation temperatures, which make the $\chi^2$ minimum, can be derived via utilizing the Marquard method:

\begin{equation}
\chi^2(N,T,\Theta) = \sum_{i}(Ta^*(i)-T_{\rm calc}(N,T,\Theta,i))^2
\end{equation}

``$i$'' represents the number of observed transitions.
In this method, $T_{\rm calc}(N,T,\Theta,i)$ is theoretically predicted brightness temperature of a certain transition number ``$i$'' via the equation (2), under a given column density of $N$, the excitation temperature of $T_{\rm ex}$, and the source size of $\Theta$.
In the equation (6), squares of subtractions of observed $T$a*($i$) values from corresponding predicted $T_{\rm calc}(N,T,\Theta,i)$ values are calculated for all ``$i$'', and the $\chi^2(N,T,\Theta)$ is given as their sum.
The initial column density and the excitation temperature are assumed from the rotation diagram method.
The initial column density for the Marquard method was given by multiplying the factor of $(\frac{10"}{\Theta_{continuum}})^2$ to the $N_{\rm rot}$, where $\Theta_{continuum}$ is the size of continuum emission:

\begin{equation}
N_{\rm initial}=N_{\rm rot} \times (\frac{10"}{\Theta_{continuum}})^2
\end{equation}

\subsection{Observational Results and the Obtained Abundances for Each Source}
Our results can be found in Tables~\ref{abundances_rotation_diagram} and \ref{abundances_compact} for (1) rotational diagram method assuming a source size of 10" and (2) using the $\chi^2$ method with a variable source size respectively. 
Although the value of source size of 10" is quite uncertain, to estimate the fractional abundance assuming 10" source size and H$_2$ column density from CO would be one of a good way to estimate the fractional abundance \citep{Suzuki16}.
The column densities and the excitation temperatures were used as the initial values in the analysis of the least squares method.
We note that, if available, the source sizes derived in the least squares method would be better one.
With the error propagation from rms noise level for observed transitions, we calculated the error of column densities with one sigma level.
In addition, we added the additional error of 15$\%$ compared to the main value as the calibration error when the transitions were obtained from the different setup. 
With the least squares method, the source sizes were always smaller than 10".
We will refer to it as the compact source case.
If the numbers of detected lines were small, we fixed the source size and/or the excitation temperatures. 
We summarize these exceptions, line properties for individual sources, and comparison with previous studies in the subsequent subsections.

The typical excitation temperatures for CH$_3$CHO was $\sim$17~K, suggesting that CH$_3$CHO would exist in the envelopes.
These low excitation temperatures of CH$_3$CHO agree with the result by \cite{Ikeda01}.
Therefore, we will exclude CH$_3$CHO in the below discussion for hot cores.
The average of fractional abundances for the compact source sizes were 9.8$\times$10$^{-7}$, 3.6$\times$10$^{-7}$, 4.3$\times$10$^{-8}$, and 2.2$\times$10$^{-9}$ for CH$_3$OH, HCOOCH$_3$, CH$_3$OCH$_3$, and CH$_3$CHO.
Their dispersions were 6.3$\times$10$^{-7}$, 3.3$\times$10$^{-7}$, 2.7$\times$10$^{-8}$, and 8.8$\times$10$^{-10}$ respectively, which were lower than their averaged fractional abundances.
On the other hand, the average of fractional abundances for the compact source sizes were 1.1$\times$10$^{-8}$, 4.0$\times$10$^{-8}$, and 1.6$\times$10$^{-8}$ for NH$_2$CHO, CH$_2$CHCN, and CH$_3$CH$_2$CN.
Their dispersions of fractional abundances, 2.1$\times$10$^{-8}$, 5.6$\times$10$^{-8}$, and 2.5$\times$10$^{-8}$, respectively,were larger than their main values of fractional abundances.
Hence, N-bearing molecules tend to show wide range of fractional abundances than O-bearing molecules, suggesting that abundances of N-bearing molecules would be more affected by the physical conditions of the sources.

In Table~\ref{correlation}, we summarized the correlation coefficients between the observed abundances of COMs based on Table~\ref{abundances_compact}.
We added CH$_2$NH abundances from \cite{Suzuki16} in this table.
The correlation coefficients of ``CH$_3$OH and CH$_3$OCH$_3$", ``NH$_2$CHO and CH$_3$CH$_2$CN", and ``CH$_2$CHCN and CH$_2$NH" were higher than 0.7, suggesting strong positive correlation.
In addition, observed N-bearing molecules tended to have larger correlation coefficients with other N-bearing molecules, rather than O-bearing molecules.
The correlation coefficients between N-bearing species and O-bearing species were usually less than 0.2 and sometimes even negative, although the correlation coefficient of ``CH$_3$OH and CH$_2$CHCN" has exceptionally higher value of 0.51. 
(CH$_3$)$_2$CO was a unique molecule which showed higher correlation coefficients with CH$_3$OH, HCOOCH$_3$, CH$_3$CH$_2$N, CH$_2$CHCN, and CH$_2$NH.

Since our source number is small, it would be essential to see the probability that our correlation coefficients are identical to those of parent population, i.e., general star forming region.
The t-test, described in detailed in section 2.2 of \cite{Wall96}, is a powerful tool to evaluate if the derived correlation coefficients represent that of parent population.
We performed the t-test to see the significance of these correlation coefficients. With a correlation coefficient, r, it is known that a parameter of $t=\frac{r\sqrt{n-2}}{\sqrt{1-r^2}}$ obeys with t distribution. We summarize this value in Table~\ref{t-test}. In t-test, this value of t presents the probability that the correlation coefficients of our sample are the same as those of the parent population. This probability also depends on the degree of freedom of t-distribution, n-2, with n being the number of sources where molecules were detected. In this work, n is six for CH$_3$OH, NH$_2$OH, CH$_3$CH$_2$CN, and CH$_2$NH, while seven for the other sources. Then, the significance level of the correlation coefficients is more than 95$\%$ if t is larger than 2.78 for n=6 and 2.57 for n=7, respectively. 

As a result, we can confirm enough significance level only for the correlations with NH$_2$CHO and CH$_2$CHCN, CH$_3$CH$_2$CN and CH$_2$NH, and (CH$_3$)$_2$CO and CH$_2$NH.
While NH$_2$CHO and (CH$_3$)$_2$CO contain O atom, these species also showed the correlations with other N-bearing species. This result suggests that the correlation among different species are not determined by only atoms they contain.
For the correlation coefficients between other species, the significance level is not enough for most cases with the current source number.
The expansion of survey observations would be required to confirm their correlations.
Although we should have more data to confirm the general trend, there are some sources where N-bearing species, NH$_2$CHO, CH$_2$CHCN, CH$_3$CH$_2$CN, and CH$_2$NH are depleted compared to CH$_3$OH, a representative species of O-bearing species.
With the fractional abundances under the compact source size, the all abundances of above four N-bearing species compared to CH$_3$OH are less than 0.01 for NGC6334F and W51~e1/e2.
Therefore, at least NGC6334F and W51~e1/e2 are special sources where N-bearing species are poor compared to O-bearing species.
The physical conditions to lead he different abundances of N-bearing species will be discussed in the proceeding section, using the chemical abundances of representative sources, G10.47+0.03 and NGC6334F, where N-bearing species are rich and not so rich.

\subsubsection{G10.47+0.03}
For extended source size, the abundances of O-bearing species, CH$_3$OCH$_3$, CH$_3$OH, and CH$_3$CHO were consistent with \cite{Ikeda01} within a factor of 3, considering that \cite{Ikeda01} used a source size of 20".
The line widths were typically 9~km~s$^{-1}$ for observed COMs except for CH$_3$OCH$_3$, whose transitions of EE, EA, AE, and AA state were heavily blended with each other. 
While the radial velocity of CH$_2$CHCN and CH$_3$CH$_2$CN were 67.7$\pm0.4$ and 66.9$\pm0.6$~km~s$^{-1}$, CH$_3$OH and NH$_2$CHO showed the slightly lower averaged radial velocity of 66.2$\pm0.3$ and 66.4$\pm0.3$~km~s$^{-1}$.
However, although the radial velocity of CH$_2$CHCN is the suggestive of the different peak position, the difference of radial velocity for other species is small compared to the standard deviation of the radial velocities. 

In G10.47+0.03, we determined the source sizes of observed species with previous mapping observations \citep{Rolffs11}.
Then, the column densities and the excitation temperatures were given by the least squares method.
The number of observed transitions were not enough to analyze the abundances of (CH$_3$)$_2$CO via LTE analysis.
Assuming that the distribution of (CH$_3$)$_2$CO was the same as CH$_3$CH$_2$CN similar to Orion~KL \citep{Peng13}, we employed the source size of 2" for the case of compact source, and the excitation temperature of 71~K.
For other species, the source sizes for the case of compact source were determined by the least squares method. 
The column densities of HCOOCH$_3$, CH$_2$CHCN and CH$_3$CH$_2$CN for the compact source size were consistent with the results by \cite{Fontani07}, where the source size was estimated from the line widths.
\cite{Rolffs11} also reported the abundances of CH$_3$OH, HCOOCH$_3$, CH$_3$OCH$_3$, (CH$_3$)$_2$CO, NH$_2$CHO, CH$_3$CH$_2$CN, and CH$_2$CHCN, with the interferometric observations.
We note that our column densities of CH$_3$OCH$_3$, CH$_3$CH$_2$CN, and CH$_2$CHCN was almost one order of magnitude lower than the report by \cite{Rolffs11}, where they used smaller source size of 1.5", which is corresponding to the densest part of the core.
We got more extended source sizes and lower column densities through our least squares method probably because our single dish observation could not resolve such structure.

\subsubsection{Orion~KL}
Our column densities under the extended source size were consistent with \cite{Ikeda01} for CH$_3$OH, HCOOCH$_3$, CH$_3$CH$_2$CN, and CH$_2$CHCN.
In addition, the comparable abundance of NH$_2$CHO was reported by \cite{Turner91}.
For the abundances under the compact source size, the column density of CH$_3$CH$_2$CN and CH$_2$CHCN were consistent with \cite{Rizzo17} for the Hot Core, but our column densities of (CH$_3$)$_2$CO was three times higher than \cite{Peng13}.
This discrepancy would have come from the assumption of excitation temperature of 100~K in our survey, due to lack of observed transitions with wide range of the energy level.

As it has been already discussed by many authors, the characteristics of molecular lines are clearly different for O-bearing molecules and N-bearing molecules, which originate Compact Ridge and Hot Core, respectively.
These different distributions were also indicated from the different radial velocities obtained in our survey.
HCOOCH$_3$, CH$_3$OCH$_3$ and CH$_3$OH had the averaged radial velocities of 7.6$\pm0.3$, 7.6$\pm0.6$, and 7.7$\pm0.4$~km~s$^{-1}$, respectively, while CH$_2$CHCN and CH$_3$CH$_2$CN showed the radial velocities of 4.9$\pm0.8$ and 4.6$\pm0.9$~km~s$^{-1}$.

We calculated the abundances for compact source sizes based on previously reported spatial distributions of the observed species.
We used 5.2" for CH$_3$OH \citep{Peng12}, 2.8" for CH$_3$CH$_2$CN and (CH$_3$)$_2$CO \citep{Peng13}, CH$_3$OCH$_3$ \citep{Brouillet13}, 3.5" for HCOOCH$_3$ \citep{Favre11}.
We employed a source size of 2.8" for CH$_2$CHCN considering the distribution of CH$_3$CH$_2$CN, assuming that the distributions of them would similar to that of (CH$_3$)$_2$CO \citep{Peng13}.

\subsubsection{NGC6334F}
The previous observations of COMs to compare with our results were limited for NGC6334F, however, our column densities of CH$_3$OH and CH$_3$OCH$_3$ agreed with \cite{Ikeda01}.

The typical line width was $\sim$4.0~km~s$^{-1}$ in this source.
The averaged radial velocities of O-bearing species, CH$_3$OH, HCOOCH$_3$, and CH$_3$OCH$_3$ were -7.8$\pm0.6$, -7.5$\pm0.4$, and -7.7$\pm0.8$~km~s$^{-1}$, while that of CH$_3$CH$_2$CN is -6.7$\pm0.6$~km~s$^{-1}$.
Hence, in this source, it is difficult to claim the different radial velocities with their standard deviations and this result would suggest that COMs would exist in the same position.
%

Due to the lack of available data, the excitation temperatures of (CH$_3$)$_2$CO and CH$_2$CHCN were assumed to be 80~K referring to that of CH$_3$CH$_2$CN.
The least squares fitting method was employed to estimate abundances under the compact source sizes of CH$_3$OH, HCOOCH$_3$, and CH$_3$OCH$_3$.
The initial source size was assumed to be 3.5" from the source size of continuum emission \citep{Hernandez14}.
For CH$_3$CH$_2$CN, CH$_2$CHCN and (CH$_3$)$_2$CO, and NH$_2$CHO, we fixed the source size to be 3.5" due to the lack of enough number of emission lines.

\subsubsection{W51~e1/e2}
While our column densities of CH$_3$OH, HCOOCH$_3$, and CH$_3$OCH$_3$ agreed with the previous studies \citep{Ikeda01,Fuchs05}, those of CH$_3$CH$_2$CN and CH$_2$CHCN were lower than \cite{Ikeda01} by a factor of four.

The averaged radial velocities of CH$_2$CHCN and CH$_3$CH$_2$CN were 58.1$\pm1.1$ and 57.8$\pm1.3$~km~s$^{-1}$, and their line widths were about 10.0~km~s$^{-1}$.
On the other hand, the radial velocities of CH$_3$OH and HCOOCH$_3$ were, respectively, 56.3$\pm0.4$ and 56.6$\pm0.9$~km~s$^{-1}$ and their typical line widths were 8.0~km~s$^{-1}$.
It is known that W51~e1/e2 contains two UCHII regions named e2 and  e8 \citep{Zhang98}.
Since the systematic velocity for e2 and e8 are 55 and 59~km~s$^{-1}$, O-bearing and N-bearing molecules would be dominant in e2 and e8, respectively.
While the difference in radial velocities of CH$_2$CHCN, CH$_3$CH$_2$CN, CH$_3$OH and HCOOCH$_3$ may due to the origin from the different UCHII regions, the difference is not clear due to relatively high standard deviation.
%

%
%

%
%
NH$_2$CHO showed the low excitation temperature of 27~K, however, we note that the observed energy range is so limited for this species to determine an accurate excitation temperature.
For HCOOCH$_3$, we could not obtain the reliable abundance by the usual rotation diagram method due to large scattering. 
This would be due to the optical thickness of some transitions.
Considering that the strongest HCOOCH$_3$ line show Ta* of $\sim$400~mK, we selected only weak (Ta* $<$ 0.1~K, corresponding to the quarter of 400~mK) line to select potentially optical thin lines in the rotation diagram method.
As the results, we obtained its excitation temperature and the column density of 113~K and 3.5$\times$10$^{16}$~cm$^{-2}$ assuming 10" source size.
We used the least squares method using all observed lines, which can calculate the effect of optical thickness, with this excitation temperature and the column density of 1.3$\times$10$^{18}$~cm$^{-2}$, which was obtained by multiplying a factor of (10"/2")$^2$ to the column density obtained in 10" source size.
We note that there is an uncertainty to use the column density and the excitation temperatures obtained through this method as the initial value for the least squares method due to the effect of optically thickness.
Therefore, as the initial values for the least squares method, we changed the excitation temperature from 50 to 150~K and the initial column density from 5$\times$10$^{17}$ to 5$\times$10$^{17}$~cm$^{-2}$, to obtain the range of these parameters
Finally we obtained the excitation temperature and the column density of 98$\pm$20~K and 2.5$\pm$1$\times$10$^{18}$~cm$^{-2}$.

The abundances under compact source size for (CH$_3$)$_2$CO was derived by assuming that the source size is the same as the continuum emission size of 2" \citep{Hernandez14}, and assuming the excitation temperatures of 60~K, referring to that of CH$_2$CHCN.
The excitation temperature of CH$_2$CHCN and NH$_2$CHO were fixed to be 60~K and 27~K, respectively in deriving the abundances for compact source size due to the limited number of observed transitions and the limited range of observed energy level.

\subsubsection{G31.41+0.03}
The observed column densities of CH$_3$OH, HCOOCH$_3$, CH$_3$CH$_2$CN, and CH$_2$CHCN agreed with \cite{Ikeda01} and \cite{Fontani07}.
\cite{Ikeda01} reported higher column density of CH$_3$OCH$_3$ by a factor of four compared to our abundance.

The averaged radial velocities of CH$_3$OH, CH$_2$CHCN, CH$_3$CH$_2$CN, and NH$_2$CHO were fallen into 97.5$\pm$0.4 km~s$^{-1}$.
It would be difficult to discuss their distributions from these results with our velocity resolution of 0.4~km~s$^{-1}$.
While the radial velocity for HCOOCH$_3$ of 98.5$\pm$1.0~km~s$^{-1}$ is relatively high, its standard deviation makes it difficult to be distinguished from other species. 

The source sizes of CH$_3$OCH$_3$ and HCOOCH$_3$ were determined based on previous mapping observations \citep{Rivilla17}.
The abundance of CH$_3$CH$_2$CN was not determined from the rotation diagram method, presumably due to optically thickness of observed transitions with low energy levels.
Therefore we determined its abundance by $\chi^2$ method, with the initial column density of 1.0$\times$10$^{17}$~cm$^{-2}$, the initial excitation temperature of 118~K, and the initial source size of 1.1" from \cite{Beltran05}.
Since the numbers of observed transition were not enough for CH$_2$CHCN, and (CH$_3$)$_2$CO, their source sizes were fixed to be the continuum size of 1.7".
Although the excitation temperature of CH$_3$CH$_2$CN would be a good indicator for those of CH$_2$CHCN, and (CH$_3$)$_2$CO, the excitation temperature of CH$_3$CH$_2$CN has large error.
An excitation temperature of 100~K was assumed for CH$_2$CHCN and (CH$_3$)$_2$CO, considering the higher excitation temperature of CH$_3$CH$_2$CN reported in \cite{Beltran05}.
An excitation temperature of 10~K was assumed for CH$_3$CHO.

\subsubsection{G34.3+0.2}
The observed column densities of CH$_3$OH, CH$_3$CH$_2$CN, and CH$_2$CHCN agreed with \cite{Ikeda01}.
Although the column density of HCOOCH$_3$ agreed with \cite{Fontani07}, our column density of HCOOCH$_3$ was higher than \cite{Ikeda01} by an order of magnitude.

In this source, the averaged radial velocities for CH$_3$OH, CH$_2$CHCN, CH$_3$CH$_2$CN, and NH$_2$CHO were fallen into 97.5$\pm$0.4~km~s$^{-1}$.
This difference would not be enough to claim the different spatial distributions with our velocity resolution of 0.35~km~s$^{-1}$.
{\bf HCOOCH$_3$ showed slightly higher radial velocity of 98.5$\pm$1.2~km~s$^{-1}$, however, its standard deviation is not enough to be distinguished from other species.}

The excitation temperature of HCOOCH$_3$ was not determined from rotation diagram.
This would be because some transitions are optically thick and led to scattering of plots in rotation diagram.
%
%
%
(CH$_3$)$_2$CO and NH$_2$CHO abundances were derived under the excitation temperatures of 60~K, considering that the excitation temperatures of CH$_3$CH$_2$CN and HCOOCH$_3$ derived with the least squares method are not so high, as described in the latter part of this subsection.
The least squares method was applied for CH$_3$CH$_2$CN, HCOOCH$_3$, and CH$_3$CH$_2$CN, while this method was not applied for CH$_2$CHCN and CH$_3$OCH$_3$, due to lack of enough number of observed transitions.
In Table~\ref{abundances_compact}, abundances under compact source sizes for (CH$_3$)$_2$CO, CH$_3$OCH$_3$, NH$_2$CHO, and CH$_2$CHCHN were simply derived by compensating an effect of beam dilution.

\subsubsection{G19.61-0.23}
Our column densities of CH$_3$OH, NH$_2$CHO, and CH$_3$CH$_2$CN were consistent with the interferometric observation by \cite{Qin10}.
However, we note that the column density of CH$_3$CH$_2$CN was by an order of magnitude higher than \cite{Fontani07}.

The obtained spectra of this source was limited from 103 to 105~GHz.
Despite this limitation, some CH$_2$CHCN, CH$_3$CH$_2$CN, and CH$_3$OH lines were detected.
{\bf CH$_3$OH and NH$_2$CHO showed the averaged radial velocities of $\sim$37.5$\pm0.3$ and $\sim$37.5$\pm0.7$~km~s$^{-1}$, while CH$_2$CHCN showed that of $\sim$38.5$\pm$1.1~km~s$^{-1}$.
The line width for CH$_3$OH of 5.9~km~s$^{-1}$ was narrower than CH$_2$CHCN and CH$_3$CH$_2$CN, which showed line widths of 7.8 and 7.3~km~s$^{-1}$, respectively.}

For the assumption of compact source sized, molecular abundances for this source were calculated assuming a continuum source size of 2.5" due to the lack of enough observed transitions.
The excitation temperatures of CH$_2$CHCN and CH$_3$CH$_2$CN were fixed at 60~K from the relatively lower excitation temperature of NH$_2$CHO for this source.

\subsubsection{DR21(OH)}
Although the obtained  spectra of this source was limited from 103 to 105~GHz in this source, CH$_3$OH and HCOOCH$_3$ were detected.
We avoided to discuss the spatial structure of this source from line parameters, since only three lines were available in total and their S/N ratios were not low.
Their abundances were obtained assuming the excitation temperature of 150~K, based on \cite{Ikeda01}.

\section{Evaluation by Chemical Modeling}
\subsection{The Chemical Model}
To simulate the abundance of various COMs in hot cores, the NAUTILUS gas-grain chemical model \citep{Ruaud16} was used with the fast warm-up model presented in \cite{Garrod13}.
This chemical model computes the time evolution of species in three-phases, gas-phase, grain surface, and grain mantle.
All the physico-chemical processes included in the model can be found with their corresponding equations in \cite{Ruaud16}.
The gas-phase chemistry is described by the public network kida.uva.2014 \citep{Wakelam15}. 
We incorporated reactions regarding CH$_2$NH and CH$_3$NH$_2$ from \cite{Suzuki16} and reactions for COMs from \cite{Garrod13}.
The important chemical reactions, cited from kida.uva.2014, \cite{Suzuki16}, and \cite{Garrod13} for the interested species are summarized in Table~\ref{gas_reactions} and Table~\ref{dust_reactions}.

For the initial composition, we used the initial abundances from \cite{Ruaud16} assuming that the chemical evolution started from atomic form except for H$_2$.
Elements with an ionization potential below 13.6~eV, corresponding to that of hydrogen, C, S, Si, Fe, Na, Mg, Cl, and P were initially singly ionized.

\subsection{The Physical Model}
The physical evolution starts from diffuse cloud phase, where gas density increases by gravitational collapse.
This phase is called collapsing phase.
The initial gas density, $n$, was increased up to the peak density as a function of time, $t$, along with the modified free-fall differential equation shown below \citep{Nejad90}:

\begin{equation}
\frac{{\rm d}n}{{\rm d}t}=B\left(\frac{n^4}{n_{\rm i}}\right)^{1/3}\left\{24\pi G m_H n_i \left[\left(\frac{n}{n_i}\right)^{1/3}-1\right]\right\}^{1/2}
\end{equation}

where n$_{\rm i}$ is the initial density, $m_H$ is the mass of the hydrogen atom, and G is the gravitational constant.
However, the actual gravitational collapse would be somewhat slowed down than free-fall process due to the resistance of turbulent, thermal motion, or magnetic field.
Thus, the collapsing speed is adjusted by a parameter of B.
The factor B=1 corresponds to the usual free-fall case.
The smaller values of B than unity is used to assume the situation of non-free-fall cases, as \cite{Garrod13} employed B=0.7 assuming that internal physical processes would be against free-fall process.
While the gas kinetic temperature was fixed at 10~K in this collapsing phase, the dust temperature, $T_{\rm dust}$, decreased from 19 to 8~K as was presented in \cite{Garrod11}.

Once the peak density was achieved, we fixed the density, and the gas and dust temperatures were raised to their peak temperatures.
After that, chemical evolution continued with fixing their temperature and density to their peak values.

In section 4.3., we will discuss potentially important parameter in our modeling, by changing the initial density, n$_{\rm i}$, the parameter of B, the peak density, the timescale of warm-up phase and the peak temperature.
If one of these factors largely contribute to difference of the chemical compositions of hot cores, they may be keys to explain the observed chemical difference.

\subsection{Parameter Dependence of the Chemical Network Simulations}
\subsubsection{Methodology}
To discuss the difference in modeling results under different conditions, we computed ``the degree of proximity'' (hereafter DoP) at each time step using the following formula:
\begin{equation}
DoP(t) = \frac{1}{N_{total}} \sum_{i} |\log_{10}\frac{n_A(X)_{i}(t)}{n_B(X)_{i}(t)}|,
\end{equation}

where n$_A$(X)$_{i}$(t) and n$_B$(X)$_{i}$(t) are the calculated abundance of species $i$ at a certain time step $t$, with different models A and B.
N$_{total}$ is the total number of species involved in the comparison.
This formula will also be used to evaluate the distance from observational results to our modeling, by replacing $n_B(X)_{i}$ by observed fractional abundances.
DoP is similar to the distance of disagreement: the smaller value of DoP means a better agreement between predicted abundances by modelings and/or observed abundances.
Since the simulated fractional abundances range up to 10$^{-5}$ to less than 10$^{-10}$,  the comparison using the squares of the difference tends to underestimate the importance of less abundant species.
The comparison using logarithm in the above formula have an advantage in giving the same weight for both overestimated and underestimated by a factor of k ($|\log_{10}{k}|$ = $|\log_{10}k^{-1}|$).
This method was employed by many authors since suggested by \cite{Wakelam06} for the first time.
The observed species except for CH$_3$CHO were used to calculate DoPs since CH$_3$CHO would be in low temperature envelope considering its low excitation temperature ($\sim$20~K).
We will use the derived abundance of observed COMs from Table~\ref{abundances_compact} and CH$_2$NH from \cite{Suzuki16}.

The parameters to test their importance are the following.
(1) initial densities of 0.1, 1, and 10~cm$^{-3}$, (2) warm-up timescales of 7.1$\times$10$^2$, 7.1$\times$10$^3$, and 7.1$\times$10$^4$ years, (3) peak densities of 1$\times$10$^6$, 1$\times$10$^7$, and 1$\times$10$^8$~cm$^{-3}$, (4) the degree for resistance to gravitational collapsing, B, of 1, 0.7, 0.2, and 0.1, and (5) different peak temperatures of 30, 60, 90, 120, 150, and 200~K.
We calculated DoPs compared to a standard simulations, which employed the initial density of 1~cm$^{-3}$, B=0.7, the peak density of 1$\times$10$^7$~cm$^{-3}$, the warm-up timescale of 7.1$\times$10$^4$~years, and the peak temperature of 200~K.
The calculated DoPs are shown in Figure~\ref{fig:DoPs}.
Longer timescale than 1$\times$10$^6$ years is doubtful for the lifetime of hot cores \citep{Wilner01}.
Within this timescale, DoPs are small at any time in Figure~\ref{fig:DoPs} (1) to (4).
By contrast, DoPs were quite large under the different temperatures in Figure~\ref{fig:DoPs}~(5), suggesting that the peak temperature would be a key parameter to change chemical compositions of hot cores.

In Figure~\ref{fig:different_peak_temperature}, the simulated abundances of COMs were compared under the different temperatures ranging from 30~K to 200~K.
The different degree of depletions of each species can be explained by the following reasons. 

\begin{itemize}
\item [1.] Each species has different binding energies, resulting in different evaporation rates. 

\item [2.] A part of COMs were non-thermally liberated from grains with the formation heat \citep{Garrod07}.
For instance, the radicals such as OH, CH$_3$, and HCO were not evaporate if the temperature is low ($\sim$60~K), leading to the formation of some kinds of COMs like CH$_3$OCH$_3$ and HCOOCH$_3$ through thermal hopping and a part of the products were released at that time due to the reaction heat.

\item [3.] If species were susceptible to the hydrogenation processes, they were converted on grain surface to be more complex molecules when the temperature is low.
Such species, CH$_2$CHCN, NH$_2$CHO, and CH$_2$NH, tend to decrease through the below hydrogenation processes (\cite{Theule11} and \cite{Suzuki16} for CH$_2$NH, \cite{Caselli93} for CH$_2$CHCN, and \cite{Noble15} for NH$_2$CHO):

CH$_2$NH + H $\longrightarrow$ CH$_2$NH$_2$

CH$_2$CHCN + H $\longrightarrow$ CH$_2$CH$_2$CN

NH$_2$CHO + H $\longrightarrow$ HNCHO + H$_2$
\end{itemize}

\subsection{Comparison with Observed Data}
In this subsection, we will discuss if observed abundances can be explained with the different structure of the temperature inside of the hot cores.
For comparison, we selected two representative hot cores, G10.47+0.03 and NGC6334F.
G10.47+0.03 is rich in N-bearing species compared to NGC6334F, providing us with preferable samples to discuss chemical differences.
In addition to the observed abundances of CH$_3$OH, HCOOCH$_3$, CH$_3$OCH$_3$, (CH$_3$)$_2$CO, CH$_2$CHCN, and NH$_2$CHO in our survey, we will use the abundances of CH$_2$NH for these sources from \cite{Suzuki16}, where CH$_2$NH abundances of G10.47+0.03 and NGC6334F were reported to be 3.1$\times$10$^{-8}$ and 2.4$\times$10$^{-9}$.
Although CH$_3$CH$_2$CN is observed, this species is not included in the calculation of DoP since further discussion would be needed for CH$_3$CH$_2$CN chemistry.

In actual hot cores, there would be a temperature gradient from inner hot region to outer warm region.
We approximately represented this temperature gradient in hot cores with a two-layer model, where the temperature of inner core was fixed to be 200~K and the warm region with lower temperature was surrounding the inner core.
Although the this hot core would be further surrounded by the cold envelope, we did not consider the contribution of the envelope in this model.
The temperature and the volume of the warm region were free parameters in this model.
The temperatures of the warm region, T$_{\rm warm}$, were set to be 30, 60, 90, 120, and 150~K.
The different volume ratios of the 200~K and the warm regions, V$_{\rm warm}$/V$_{\rm 200K}$ was assumed to be 1, 10, 100, and 1000, to represent the different temperature structure of hot cores.
Then, the chemical compositions are calculated as $\frac{X_{\rm 200K}V_{\rm 200K}+X_{\rm warm}V_{\rm warm}}{V_{\rm 200K}+V_{\rm warm}}$, where $X_{\rm warm}$ and $X_{\rm 200K}$, respectively, represent the simulated fractional abundances under temperatures of T$_{\rm warm}$ and 200~K.
{\bf In this calculation, the ratio of V$_{\rm warm}$/V$_{\rm 200K}$ does not depend on the distance to the sources from the Earth, since both of the volumes of V$_{\rm warm}$ and V$_{\rm 200K}$ will proportionally decrease to the distance.
}

With this simulated abundances, DoPs for G10.47+0.03 and NGC6334F under the different T$_{\rm warm}$ and V$_{\rm warm}$/V$_{\rm 200K}$ were calculated along with the physical evolution of the core.
In Figure~\ref{fig:DoP_different_peak_temperature}, the time evolution of DoPs under T$_{\rm warm}$=120~K were compared by assuming the different volume ratios of V$_{\rm warm}$/V$_{\rm 200K}$.
While DoPs of G10.47+0.03 showed the minimum values at 8.0$\times$10$^{5}$~years and V$_{\rm warm}$/V$_{\rm 200K}$ of 100, the minimum of DoP for NGC6334F was achieved at 6.5$\times$10$^{5}$~years and V$_{\rm warm}$/V$_{\rm 200K}$ of 1000.
The values of DoP and its best age is associated with the error due to the uncertainties of observed abundance.
While chemical reactions in our models also contain uncertainties with their reaction coefficients, we neglect such error of chemistry to determine the best model since we have employed the identical chemical dataset for all models and the uncertainties of DoP among models are determined only by the error of observed abundances.
Then the propagation of the error in Figure~\ref{abundances_compact} associated with the abundances of species give us the error of DoP of 0.04 for G10.47+0.03 and 0.05 for NC6334F.
With these errors, we summarized the minimum values of DoPs, achieved under other combinations of T$_{\rm warm}$ and V$_{\rm warm}$/V$_{\rm 200K}$ in Table~\ref{DoP_G10} and \ref{DoP_NGC6334F}.
Finally we found that (T$_{\rm warm}$,V$_{\rm warm}$/V$_{\rm 200K}$) = (120~K, 1000) for NGC6334F, was the best combinations of the parameters to explain the observed abundances. 
For G10.47+0.03, both (T$_{\rm warm}$,V$_{\rm warm}$/V$_{\rm 200K}$) = (120~K, 100) and (120~K, 10) were appropriate under the error of 0.05.
The temperature of 120~K decreased the abundances of CH$_2$NH, CH$_2$CHCN, and NH$_2$CHO than CH$_3$OH and CH$_3$OCH$_3$ through the absorption of these species and grain surface reactions. 
For the age of the core, our results indicated that G10.47+0.03 is more evolved than NGC6334F, whereas hot region would be larger than NGC6334F.
The abundances of CH$_3$OH, HCOOCH$_3$, and CH$_3$OCH$_3$ decrease via the destruction reactions in the gas phase, whereas those of CH$_2$NH, CH$_2$CHCN, and NH$_2$CHO increase through their formation processes in the gas phase.
Therefore, observed N-bearing species tend to be rich in the evolved hot cores.

The best simulated abundances were compared with our observed abundances in Table~\ref{compare_model_with_obs}.
The symbols, s and o, respectively, represent simulated abundances and observed abundance.
Observed abundances are presented in bold face if the difference from the simulated one is within a factor of 10.
We found that the observed abundances for most of species can be explained from our chemical model, except for underestimation of HCOOCH$_3$ and overestimation of NH$_2$CHO in NGC6334F.
These discrepancy would come from the uncertainties which are associated with the desorption energies, and very simplified physical structure.

Are the observed correlations in Table~\ref{correlation} explained by our model by assuming the different temperature structures inside hot cores and/or their different age?
In this paper, we reported the interesting correlation between CH$_3$OH, HCOOCH$_3$, and CH$_3$OCH$_3$, and between CH$_2$NH, CH$_2$CHCN, and NH$_2$CHO, respectively.
With the two-layer model, if we change the ratio of V$_{\rm warm}$/V$_{\rm 200K}$ by fixing the age of the core to be 8.0$\times$10$^{5}$~years and T$_{\rm warm}$ to be 120~K, the fractional abundance of CH$_2$NH, CH$_2$CHCN and NH$_2$CHO are almost linearly decreased, leading the correlation coefficients among these species to be unity.
While high correlation coefficients between N-bearing species are consistent with our observation, the different V$_{\rm warm}$/V$_{\rm 200K}$ ratios could not explain the negative correlation coefficients between O- and N- bearing species.
Another important factor for the correlations of molecular abundances would be the age of the core.
In Table~\ref{correlation-model}, we showed the correlation coefficient between CH$_3$OH, HCOOCH$_3$, CH$_3$OCH$_3$, CH$_2$NH, CH$_2$CHCN, and NH$_2$CHO using the simulated abundances by the two-layer model in different ages, fixing V$_{\rm warm}$/V$_{\rm 200K}$ of 100 and T$_{\rm warm}$ of 120~K.
In the calculation of correlation coefficients, we used the chemical compositions of different ages ranging from 1.0$\times$10$^{5}$ to 1.0$\times$10$^{6}$~years since the beginning of the warm-up phase by the interval of 5.0$\times$10$^{4}$~years.
While CH$_3$OH, HCOOCH$_3$, and CH$_3$OCH$_3$ showed the high correlation coefficients, these O-bearing species have the negative correlation coefficients with CH$_2$NH and CH$_2$CHCN.
These results are consistent with our observational results.
However, the correlation coefficients between CH$_3$OH and CH$_2$NH and CH$_2$CHCN are too low compared to our observation.
In addition, NH$_2$CHO shows the correlation with CH$_3$OH and HCOOCH$_3$ rather than CH$_2$NH and CH$_2$CHCN, contradicting with the observation.
Therefore, the age of the core could only partly explain the observed correlations.
We suggest that both of the different age of the cores and the V$_{\rm warm}$/V$_{\rm 200K}$ ratios would contribute to the observed correlations between CH$_3$OH, HCOOCH$_3$, CH$_3$OCH$_3$, CH$_2$NH, CH$_2$CHCN, and NH$_2$CHO.

Although our model successfully explained the observed abundances, the temperature of 120~K seems higher considering the observed excitation temperatures.
Since our physical structure and its time evolution would be oversimplified, revision of physical structure would also be a key to improve our knowledge regarding the evolution of COMs. 
In addition, giving older age for G10.47+0.03 than NGC6334F do not agree with \cite{Suzuki16}, where we suggested the possibility that G10.47+0.03 is evolved than NGC6334F based on the very weak hydrogen recombination line in G10.47+0.03.
One possibility to explain this discrepancy would be that the distributions of molecules and hydrogen recombination are not overlapped with each other and we have observed the different component simultaneously.
Towards G10.47+0.03, \cite{Cesaroni10} detected two HCHII region indicative of the very early evolutionary phase of star-formation.
NGC6334F is known to have an UCHII region and hot cores associated with molecular emission (SMA~3 and SMA~1 and 2 respectively, in \cite{Hunter06}).
Considering the non detection of the free-free emission in \cite{Hunter06}, the age of the hot cores, SMA~1 and 2, are thought to be younger than SMA~3.
However, the detailed comparison of the ages of hot cores in G10.47+0.03 and NGC6334F is difficult with currently available data.
The mapping observations of molecular emission and hydrogen recombination lines with ALMA will be helpful to resolve the detailed molecular distributions and their evolutionary phase for further discussion.

\subsection{Comparison with \cite{Caselli93}}
It would be worth to discuss the difference of this work from the previous study by \cite{Caselli93}.
They investigated the chemical difference of Orion Hot Core and Compact Ridge, where N- and O-bearing species are abundant, respectively.
They prepared a Hot Core and a Compact Ridge models, where chemical evolution starts from 40~K and 20~K, followed by the sudden increase of temperatures to  200~K and 100~K, respectively.
For the initial gas densities, the densities of 1$\times$10$^{5}$ and 2$\times$10$^{4}$~cm$^{-3}$ were employed for the Hot Core and Compact Ridge models.
They found that N-bearing species such as CH$_2$CHCN and CH$_3$CH$_2$CN are abundant in the Hot Core model than the Compact Ridge model, since the higher temperature enables radicals to efficiently move to form C$_3$N, followed by successive hydrogenation to form CH$_2$CHCN and CH$_3$CH$_2$CN.
On the other hand, the abundance of CH$_3$OH was much higher in the Compact Ridge model, due to efficient hydrogenation processes.
Other O-bearing species like CH$_3$OCH$_3$ and HCOOCH$_3$ were produced using the liberated CH$_3$OH after warm-up phase.

While our results that the chemical difference between Hot Core and Compact Ridge might be reconciled by temperature is consistent with \cite{Caselli93}, we emphasize the importance of temperature in the different way from \cite{Caselli93}.
We assumed that the evolution of the cloud starts from the diffuse cloud phase, where the temperature and the density are as low as 1~cm$^{-3}$.
This assumption would be more reasonable than the one in \cite{Caselli93} as the initial condition for the evolution of the cores.
In this case, CH$_3$OH is equally produced during the collapsing phase independent of the peak temperature in the following warm-up phase, inconsistent with \cite{Caselli93}.
Furthermore, in contrary to \cite{Caselli93}, the origins of CH$_3$OCH$_3$ and HCOOCH$_3$ in our model were not gas phase, due to the update of the rate coefficients of gas phase reactions based on current understanding that gas phase recombination processes of positive ions were inefficient to produce COMs \citep{Geppert06}.

In our model, the higher abundances of O-bearing species in the hot regions were simply due to their binding energies.
For N-bearing species, our model suggested that the differences in the abundances were not due to the formation processes before the warm-up phase as suggested by \cite{Caselli93}, but due to the evaporation rates after the warm-up.
In this work, the difference of the chemical compositions is explained by the different temperatures after the warm-up phase.
N-bearing species, such as CH$_2$CHCN, NH$_2$CHO, and CH$_2$NH, are formed in the gas phase with updated gas phase chemical reaction dataset than \cite{Caselli93}.
However, if the temperature is not high enough, a part of desorbed molecules were converted to the further complex molecules on the grains.
Therefore, the depletion of N-bearing species in Compact Ridge can be explained due to adsorption and the following destruction process on grains under lower temperature than Hot Core.

\subsection{Chemistry in Hot Core Envelopes}
We found the very low excitation temperature ($\sim$20~K) of CH$_3$CHO.
In addition, NH$_2$CHO in W51~e1/e2 and CH$_2$CHCN in G34.3+0.2 clearly showed the excitation temperature of lower than 40~K.
These species would mainly exist in the envelope.

We performed chemical modeling assuming the condition of envelope to see if we can explain the observed abundances of these species.
We assumed the peak temperature of 30~K and the peak density of 1$\times$10$^6$cm$^{-3}$ and the other physical conditions and chemical reactions were fixed with our standard model.
In Figure~\ref{fig:envelope}~(a), we showed the simulated fractional abundances of CH$_3$CHO, CH$_2$CHCN, and NH$_2$CHO with the envelope model.
We lowered the binding energies of these species by 500 and 1000~K, respectively, in Figure~\ref{fig:envelope}~(b) and (c).
Since thermal evaporation processes are negligible in 30~K, chemical desorption processes and photo-evaporation by UV photons contribute to the liberation of frozen species on grains.
We used the equations of non-thermal desorption processes from \cite{Garrod07}.

In Figure~\ref{fig:envelope}~(a), the peak abundances of CH$_3$CHO, and NH$_2$CHO were, respectively, 2.6$\times$10$^{-10}$ and 1.9$\times$10$^{-11}$, which were more than 10 times less than our observed abundances.
Therefore, for these species, their observed fractional abundances were not explained at all.
Their fractional abundances has changed only slightly even if we changed the binding energies in Figure~\ref{fig:envelope}~(b) and (c), suggesting that their discrepancies would not be due to the uncertainties of the desorption energies.

We found that the peak of simulated fractional abundance of CH$_2$CHCN was 6.4$\times$10$^{-10}$.
Although the fractional abundance of CH$_2$CHCN was enhanced by chemical desorption process in this case, we have to be careful of its efficiency since the latest modeling by \cite{Wakelam17} claimed that chemical desorption processes would be less effective than previously thought.
It is obvious that further detailed studies are required for the chemistry in the envelope of hot cores.

\section{Conclusion}
The main results of this paper can be summarized as follows:

\begin{enumerate}
\item We conducted the survey observations of COMs towards high-mass star-forming regions. 
We reported more than 1000 transitions of CH$_3$OH, HCOOCH$_3$, CH$_3$OCH$_3$, CH$_3$CHO, (CH$_3$)$_2$CO, NH$_2$CHO, CH$_2$CHCN, and CH$_3$CH$_2$CN.
Their abundances and excitation temperatures were analyzed using the rotation diagram method and the least squares method.

\item In this work, we aimed at testing whether different spatial distributions of N- versus O-bearing COMs could be probed by spatially unresolved data, as previous work \citep{Blake87} indicated a shift in radial velocities between N- and O-bearing species. Our results indicate that any existing velocity shifts are generally smaller than the uncertainties on the measured velocities and hence do not allow us to draw any conclusions on different spatial distribution of O- and N-bearing COMs from single-dish data. Future spatially resolved observations are therefore crucial to investigate the origin of the emission from N- and O-bearing COMs in hot cores.

\item Among our sources NGC6334F and W51~e1/e2 are rich in N-bearing species than O-bearing species.
In addition, we found the possible correlations among N-bearing species based on our data.
However, their significance levels were not enough except for ``CH$_2$CHCN and NH$_2$CHO'' and ``CH$_3$CH$_2$CN and CH$_2$NH'' due to the limited number of sources.
It would be essentially important to increase sources to confirm if such correlations are general in star-forming regions.

\item With our chemical modeling study, we searched for essential parameter of star-forming regions that may dramatically change the chemical evolution.
Through simulations under different physical conditions, i.e., initial densities, warm-up speed, peak density, timescale of collapsing phase, and peak temperatures, we found that different temperature had a significant impact on the chemistry, while the effects from other parameters were small.
With observed chemical compositions of G10.47+0.03, and NGC6334F, where N-bearing species are rich and not so rich, respectively, we discussed if observed chemical difference can be explained by assuming the different temperature structure and the age of cores.
As a result, by assuming that high temperature region ($\sim$200~K) is dominant in G10.47+0.03 than NGC6334F and G10.47+0.03 is evolved than NGC6334F, we succeeded to reproduce their observed abundances very well.

\item NH$_2$CHO, CH$_2$NH, and CH$_2$CHCN would be enhanced if the hot region inside the hot core is large since these species are susceptible to the hydrogenation processes if the temperature is low.
In addition, our modeling suggests that NH$_2$CHO, CH$_2$NH, and CH$_2$CHCN are produced after the completion of the warm-up phase, when CH$_3$OH, HCOOCH$_3$ and CH$_3$OCH$_3$ start to decrease.
Therefore, the different age of cores would also lead the correlations and anti-correlations between observed O- and N-bearing species.
We suggest that both of the different temperature structure inside the core and their evolutionary phase contribute to the observed molecular correlations.

\item Our calculations suggested the importance for both gas phase and grain surface chemistry of CH$_3$CH$_2$CN.
We found that current model tends to increase CH$_3$CH$_2$CN abundance under 120~K, which does not match with the observation. 
In addition, the possibility of gas phase formation process of ``C$_2$H$_6$ + CN $\rightarrow$ CH$_3$CH$_2$CN + H'' should be investigated.

\item CH$_3$CHO typically showed its excitation temperature of $\sim$20~K, and NH$_2$CHO in W51~e1/e2 and CH$_2$CHCN in G34.3+0.2 showed their excitation temperatures of $\sim$30~K.
The low excitation temperature of some COMs suggest that they exist in hot core envelopes.
While our chemical model for envelope suggested that CH$_2$CHCN can be non-thermally liberated from envelope, our model could not reproduce the observed NH$_2$CHO and CH$_3$CHO at all.
Further detailed studies would be required for the chemistry of hot core envelope.

\end{enumerate}

\acknowledgments
We are grateful to all the staff members of the Nobeyama Radio Observatory, the National Astronomical 
Observatory of Japan (NAOJ), and the Arizona Radio Observatory for their support throughout our observations.
We thank to Dr.Hideko Nomura, and staffs in Bordeaux University for fruitful discussions.
A part of the data analysis was made at the Astronomy Data Center, NAOJ.
This research has made use of NASA's Astrophysics Data System.
This study was supported by the Astrobiology Program of National Institutes of Natural Sciences (NINS) and by the JSPS Kakenhi Grant  Numbers 15H03646 and 14J03618.
Valentine Wakelam acknowledge the European Research Council (3DICE, grant agreement 336474) and the CNRS program ``Physique et Chimie du Milieu Interstellaire'' (PCMI) co-funded by the Centre National d'Etudes Spatiales (CNES).
Liton Majumdar acknowledges support from the NASA postdoctoral program. A portion of this research was carried out at the Jet Propulsion Laboratory, California Institute of Technology, under a contract with the National Aeronautics and Space Administration.






\appendix

\clearpage

\begin{figure}
 \begin{tabular}{l}
\includegraphics[scale=.6]{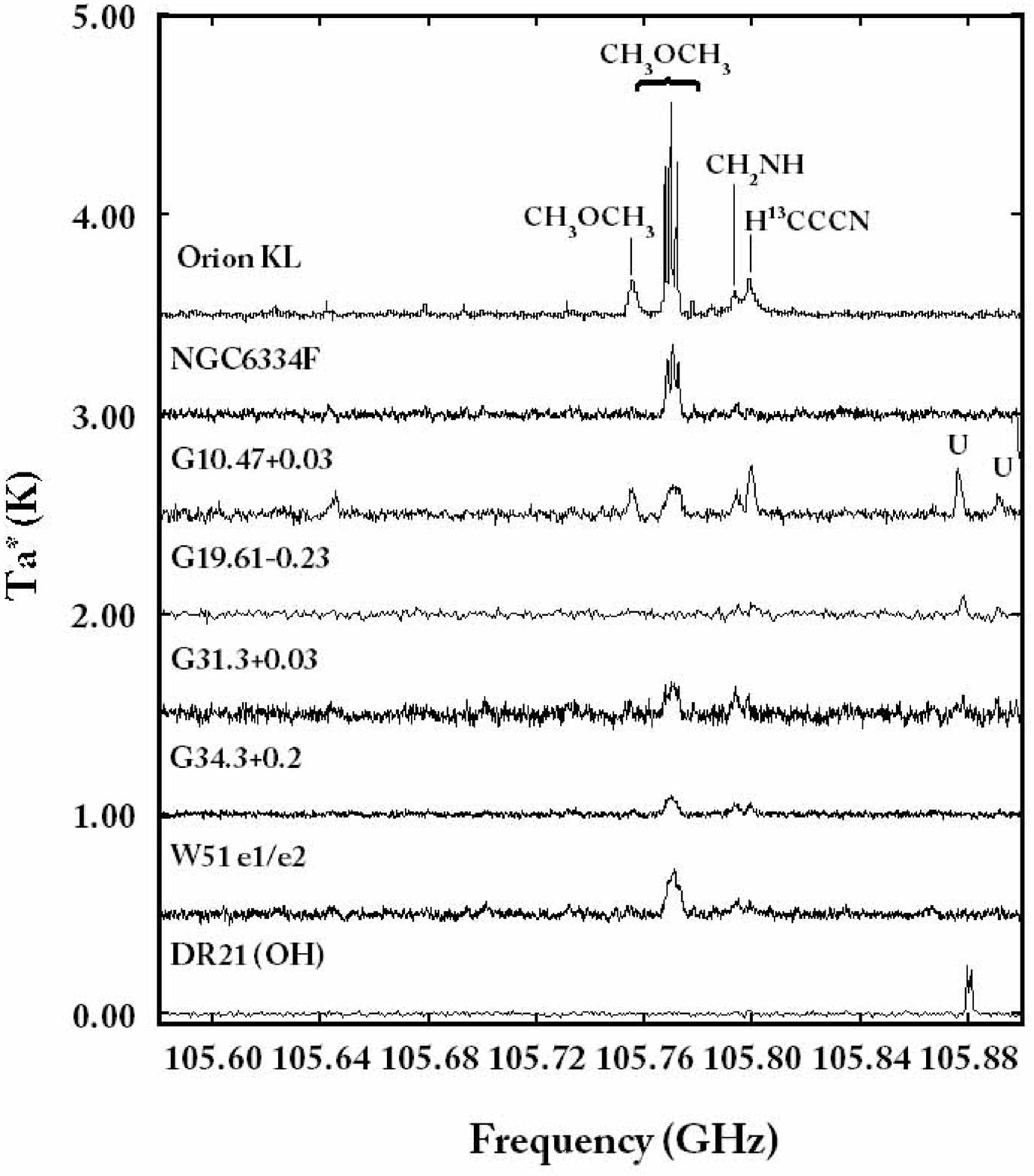}\\
  \end{tabular}
\caption{
Observed spectra for all sources are compared from 105.6 to 105.9~GHz.
Besides CH$_2$NH analyzed in \cite{Suzuki16}, CH$_3$OCH$_3$ transitions are detected. 
\label{fig:observed_spectra}
}
\end{figure}
\clearpage
\addtocounter{figure}{-1}

\begin{figure}
 \begin{tabular}{l}
\includegraphics[scale=.6]{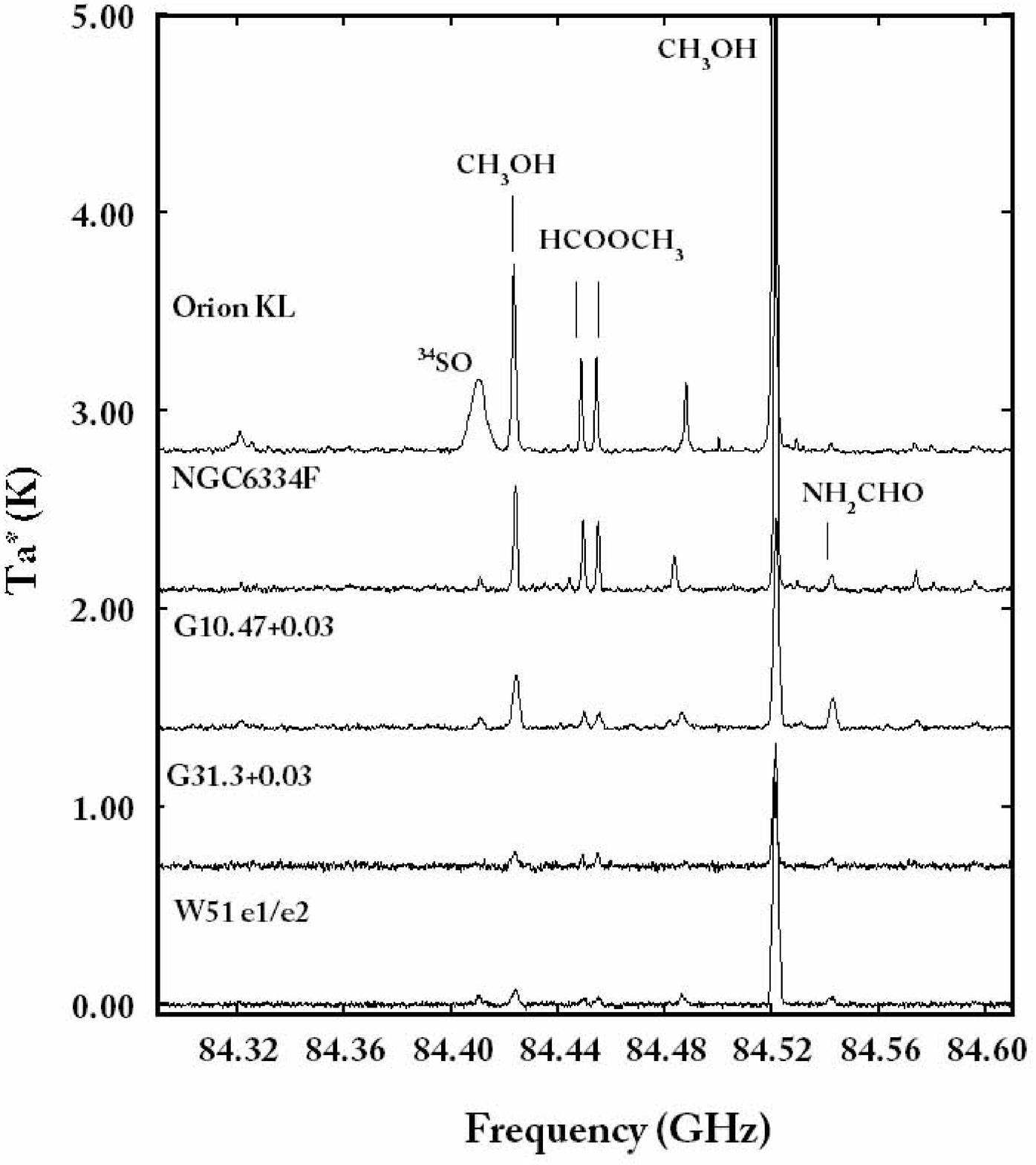}\\
  \end{tabular}
\caption{
(continued) Observed spectra for all sources are compared from 84.3 to 84.6~GHz.
CH$_3$OH, HCOOCH$_3$, and NH$_2$CHO are detected. 
This frequency was not observed for G19.61-0.23, G34.3+0.2, and DR21(OH).
\label{fig:observed_spectra}
}
\end{figure}
\clearpage
\addtocounter{figure}{-1}

\begin{figure}
 \begin{tabular}{l}
\includegraphics[scale=.6]{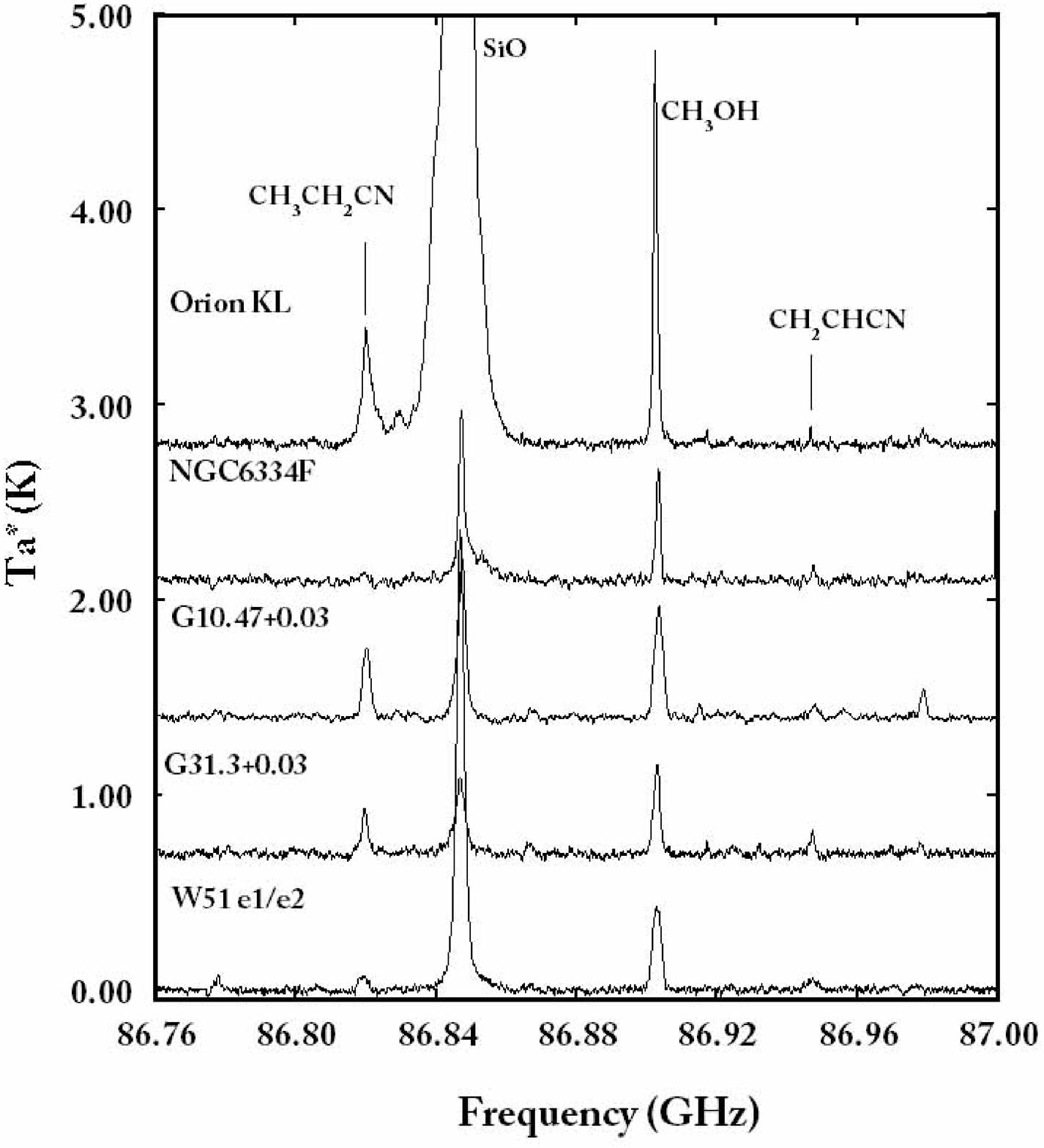}\\
  \end{tabular}
\caption{
(continued) Observed spectra for all sources are compared from 86.74 to 87~GHz.
CH$_3$OH, CH$_2$CHCN, and CH$_3$CH$_2$CN are detected. 
This frequency was not observed for G19.61-0.23, G34.3+0.2, and DR21(OH).
\label{fig:observed_spectra}
}
\end{figure}

\begin{figure}
 \begin{tabular}{ll}
\includegraphics[scale=.4]{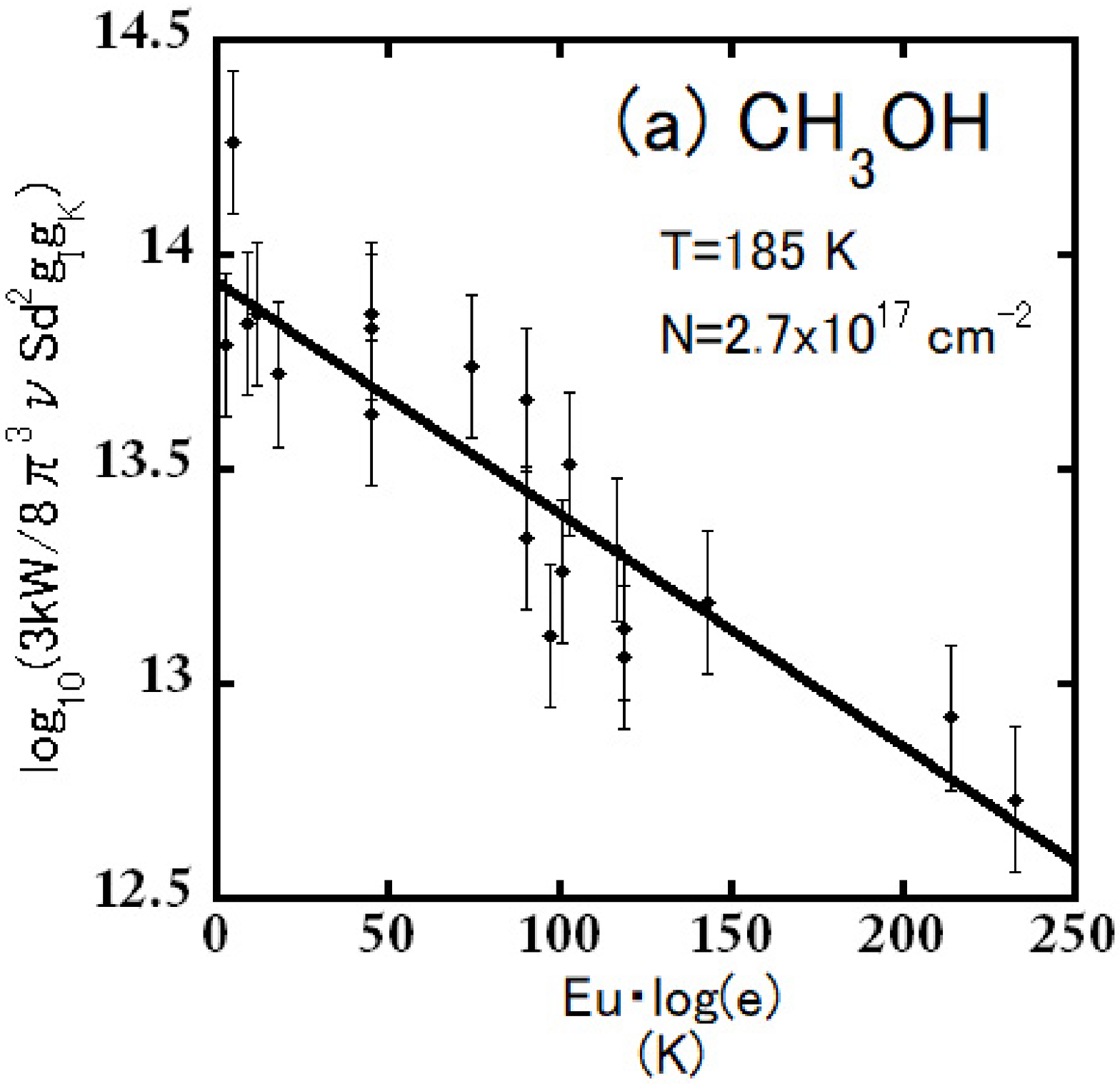}&
\includegraphics[scale=.4]{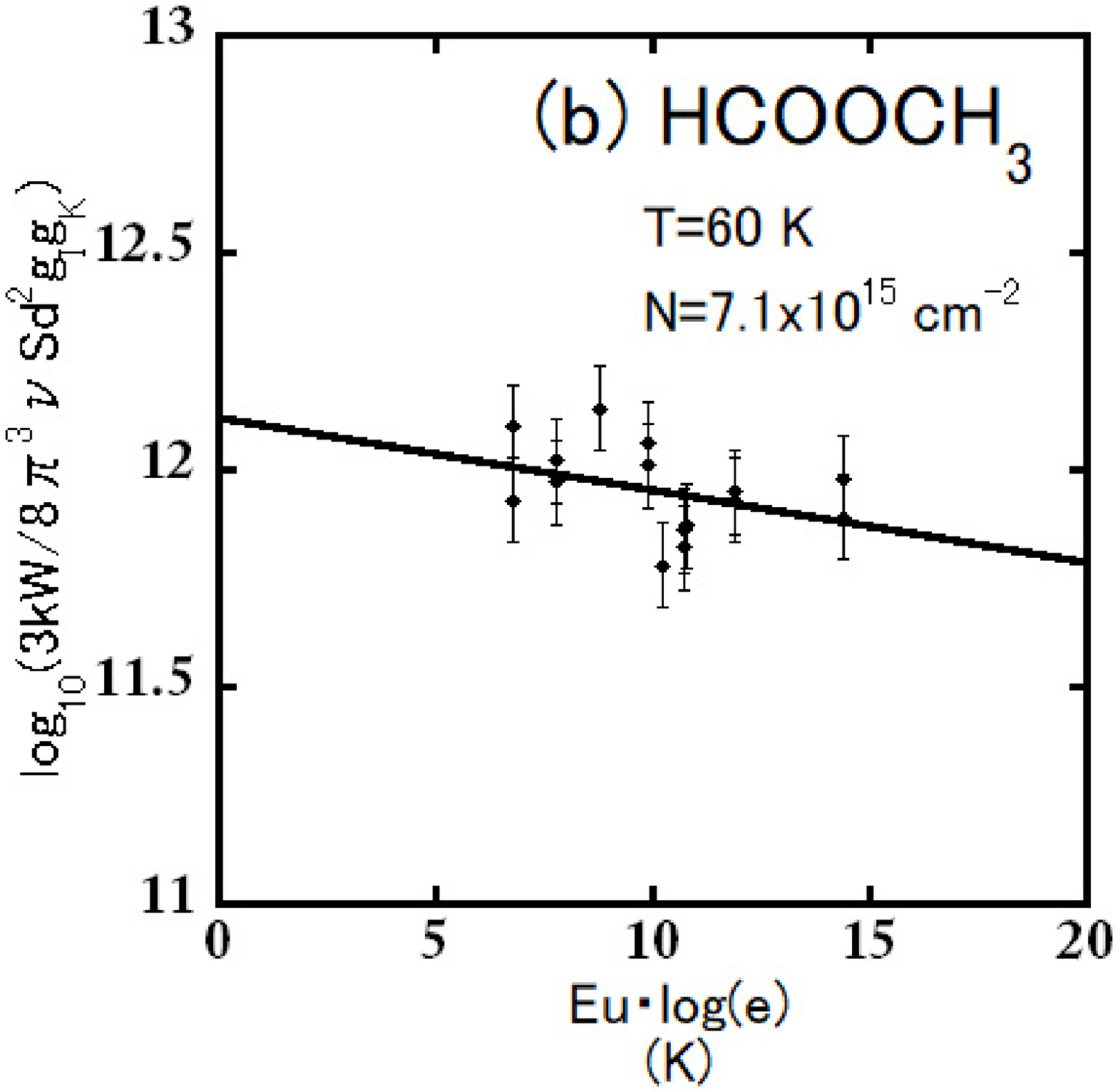}\\
\includegraphics[scale=.4]{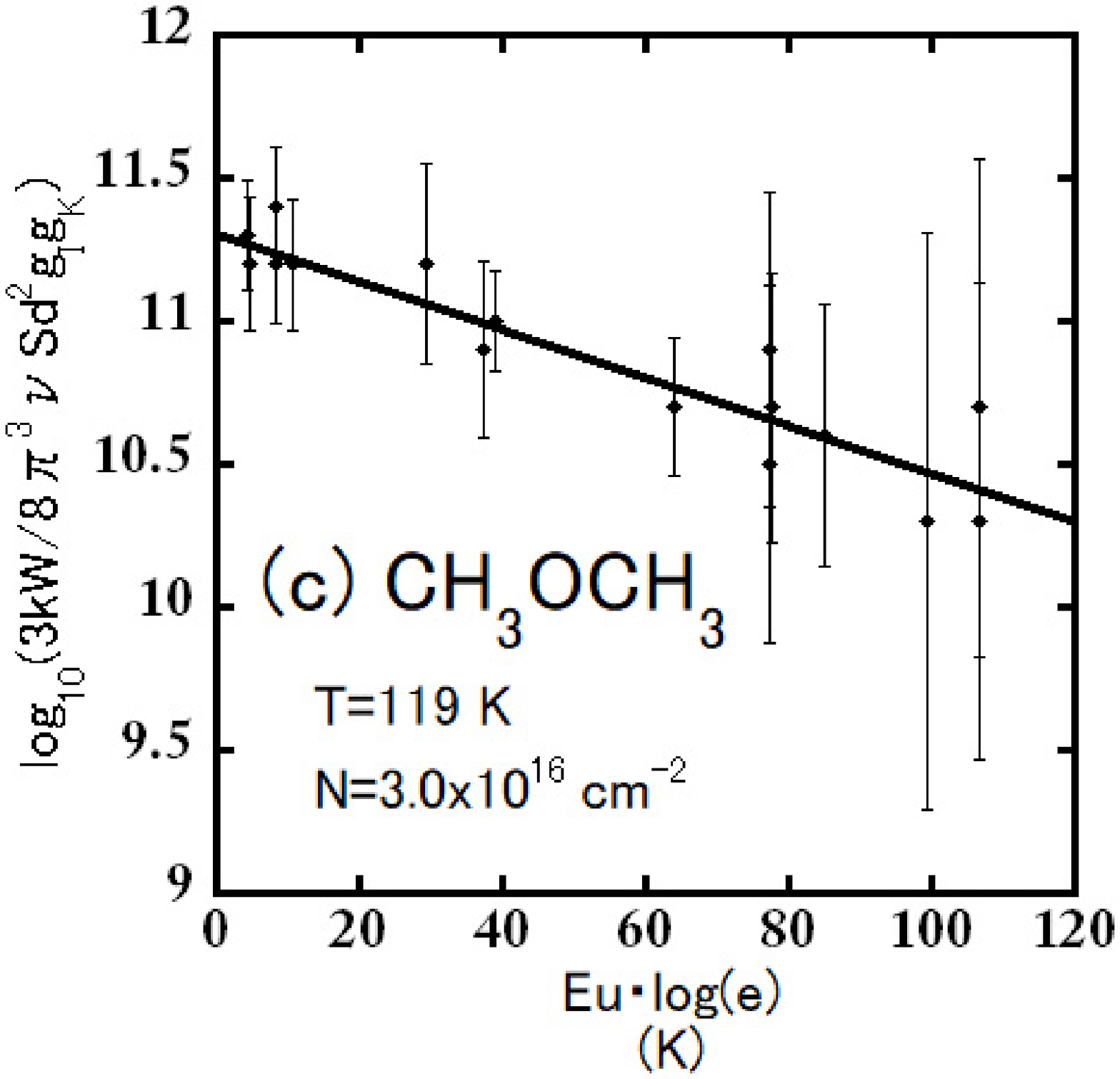}&
\includegraphics[scale=.4]{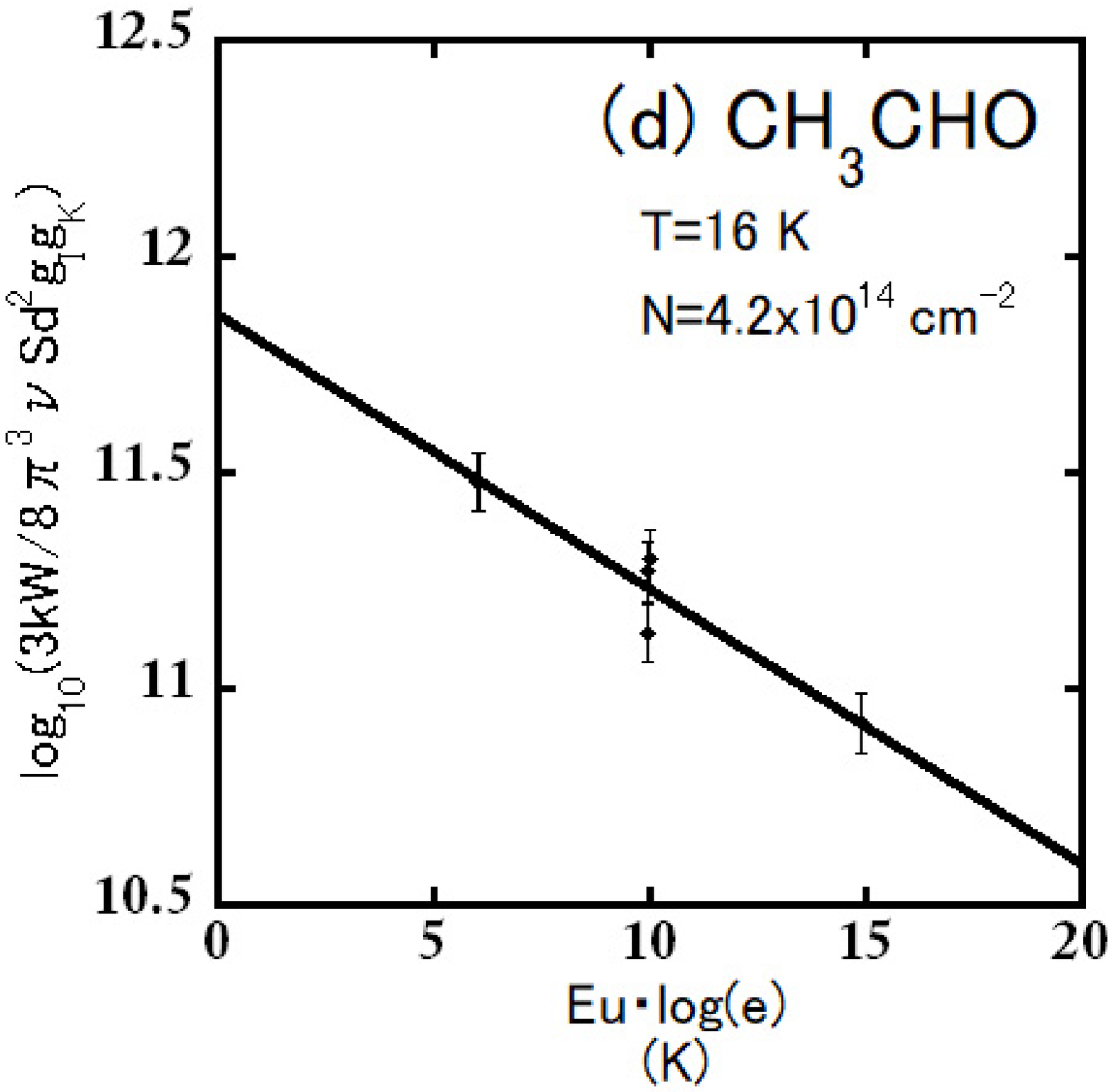}\\
\includegraphics[scale=.4]{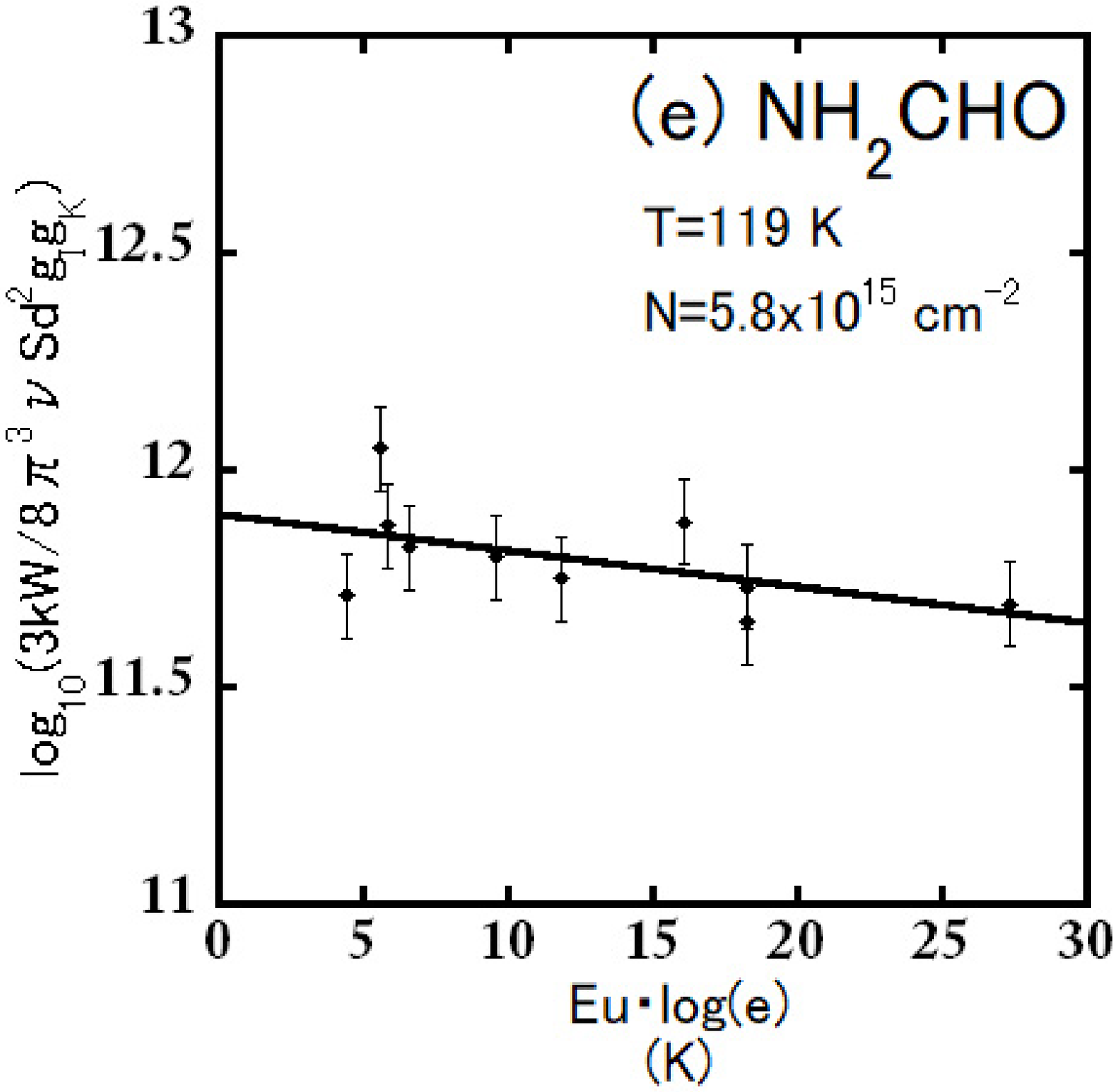}&
\includegraphics[scale=.4]{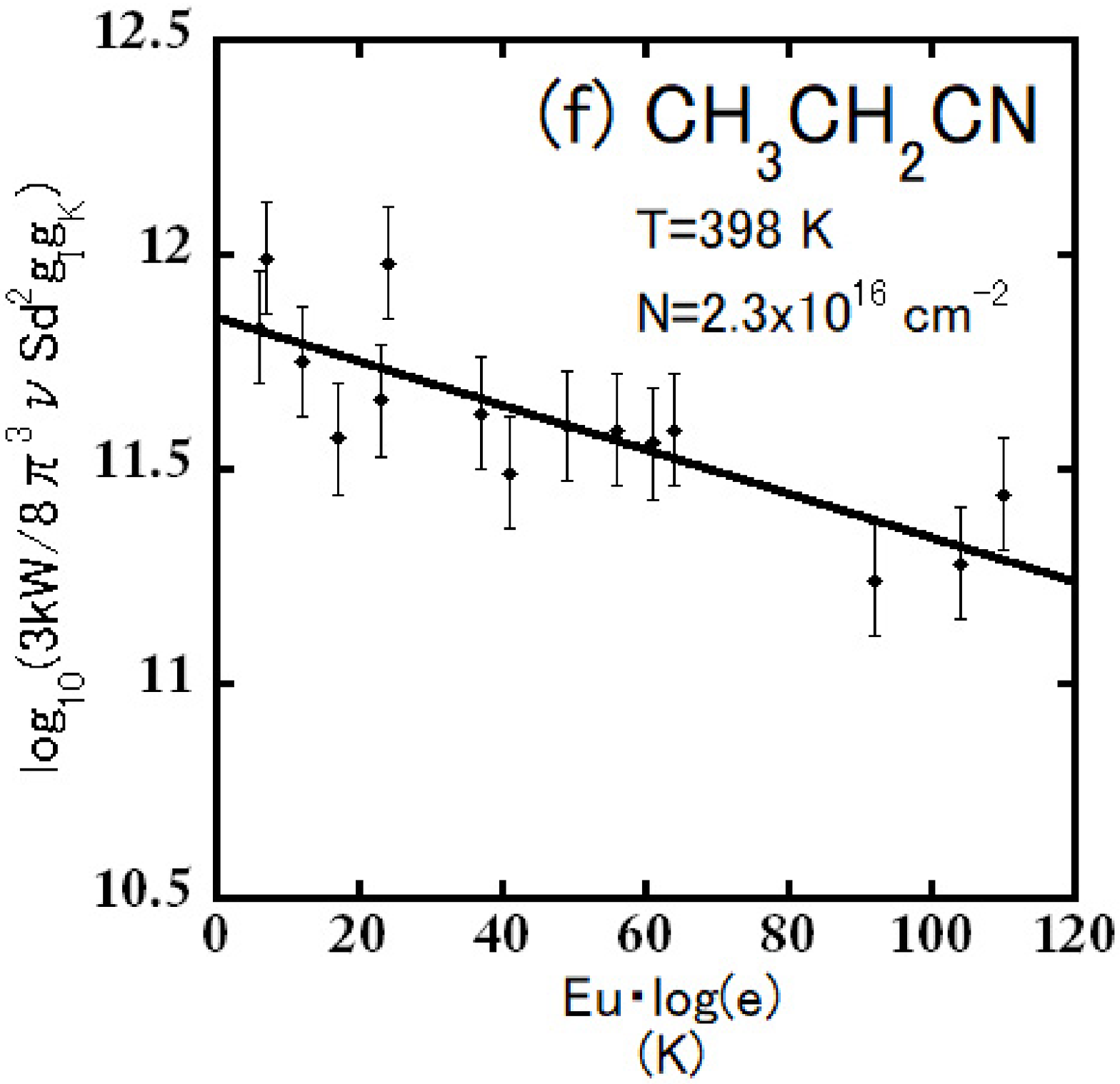}\\
  \end{tabular}
\caption{
The rotation diagrams for G10.47+0.03.
\label{fig:G10_rotation_diagrams}
}
\end{figure}
\clearpage
\addtocounter{figure}{-1}
\begin{figure}
 \begin{tabular}{ll}
\includegraphics[scale=.4]{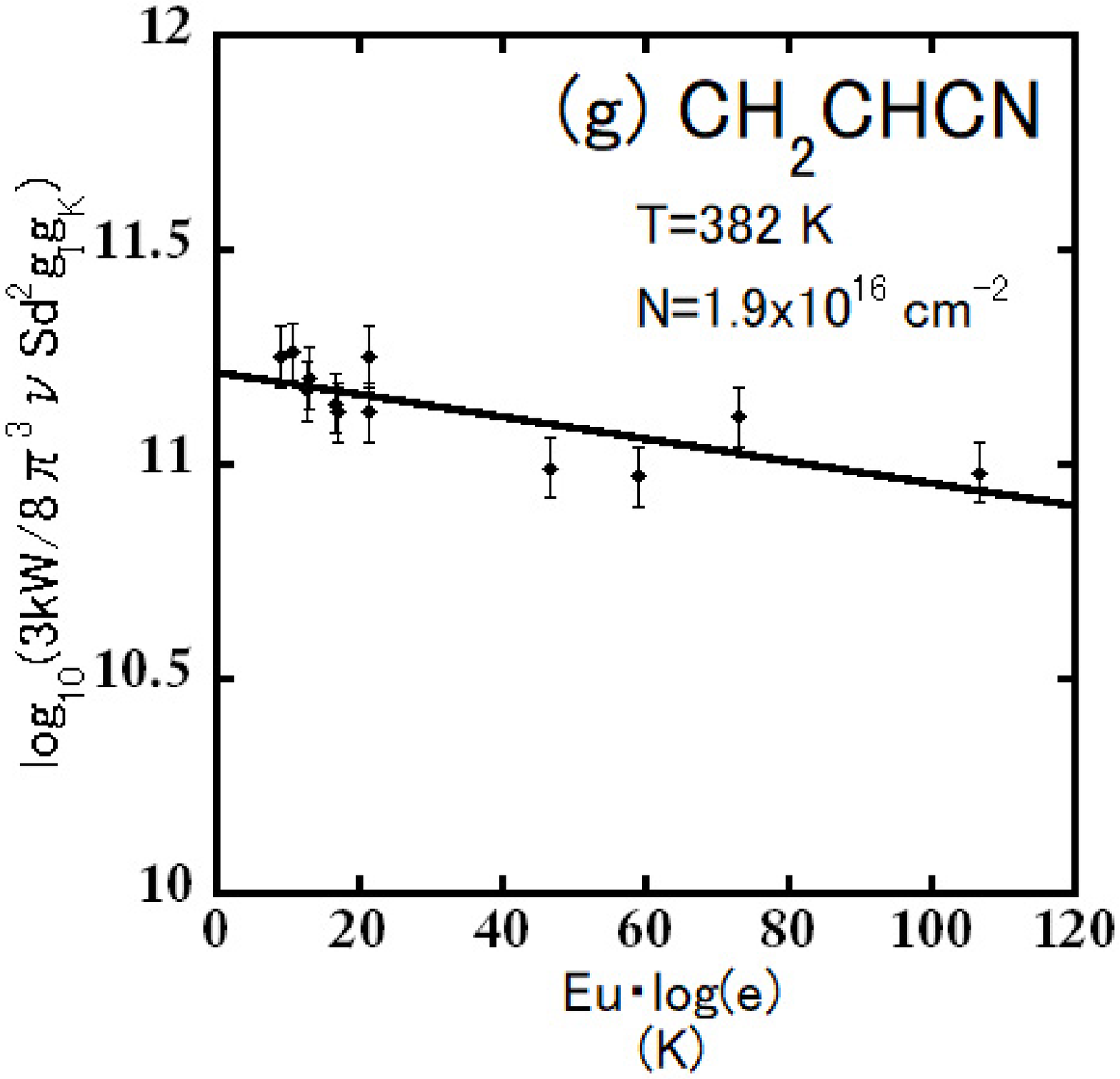}&\\
  \end{tabular}
\caption{
The rotation diagrams for G10.47+0.03.
\label{fig:G10_rotation_diagrams}
}
\end{figure}
\clearpage

\begin{figure}
 \begin{tabular}{ll}
\includegraphics[scale=.4]{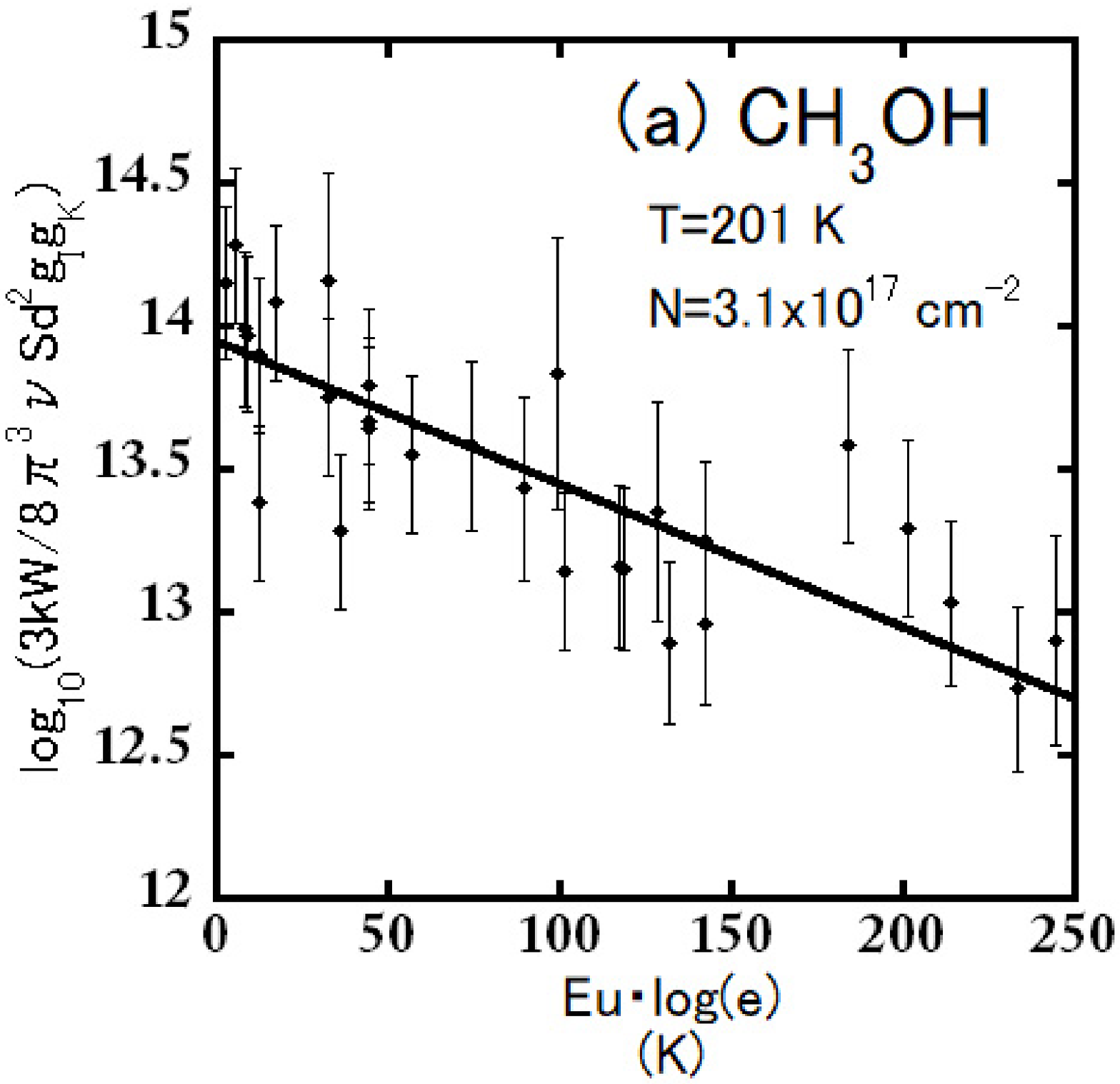}&
\includegraphics[scale=.4]{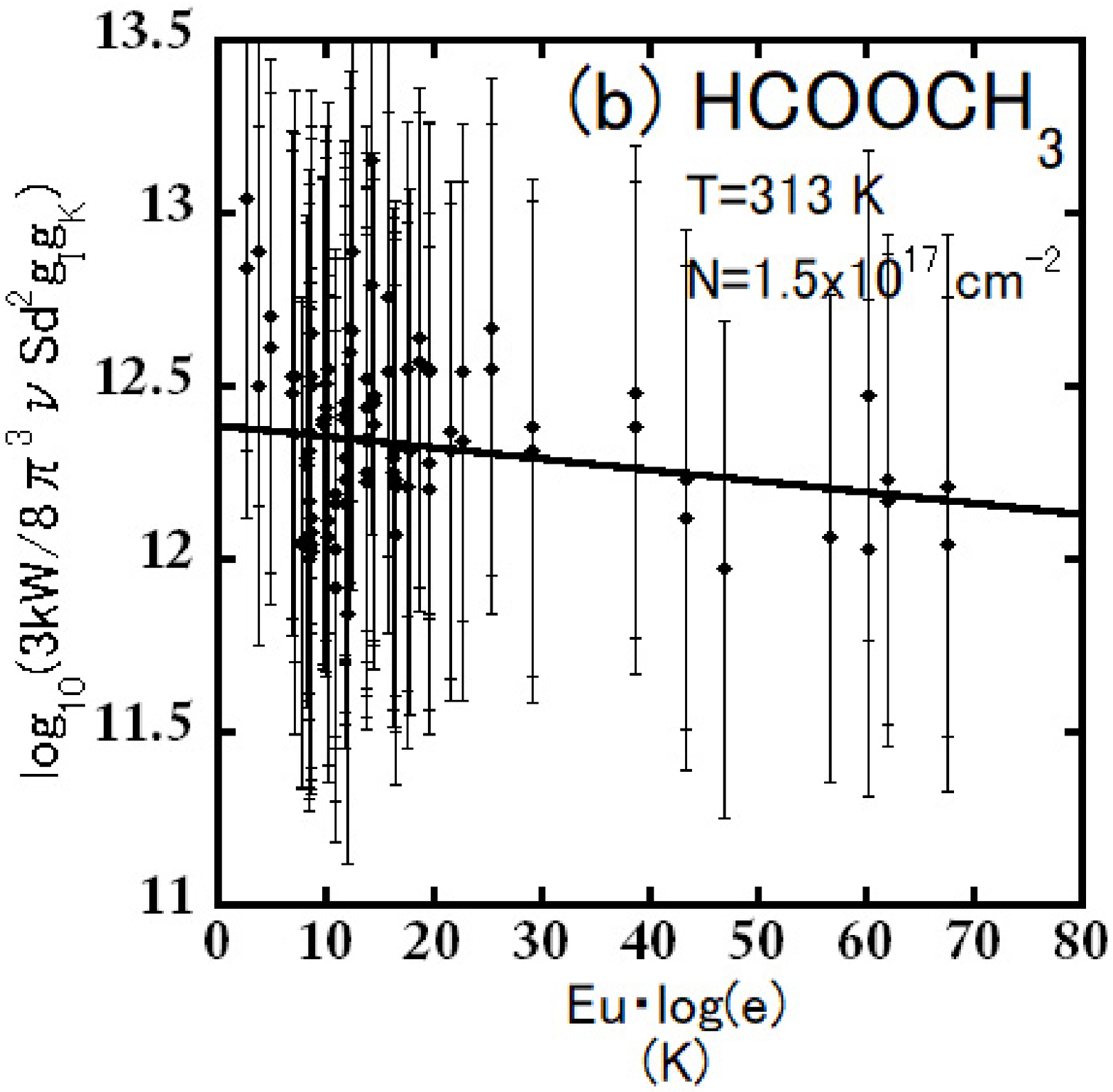}\\
\includegraphics[scale=.4]{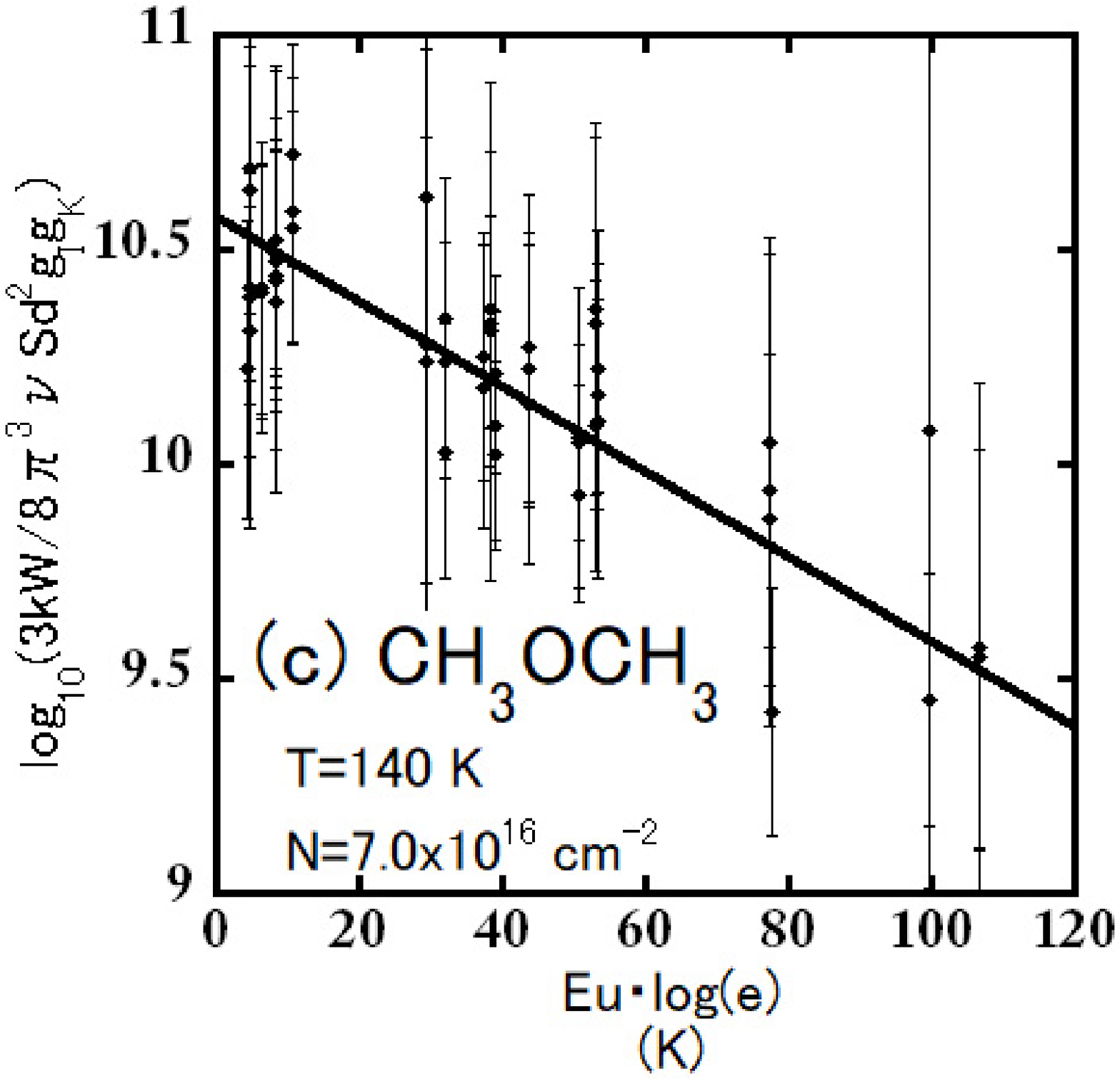}&
\includegraphics[scale=.4]{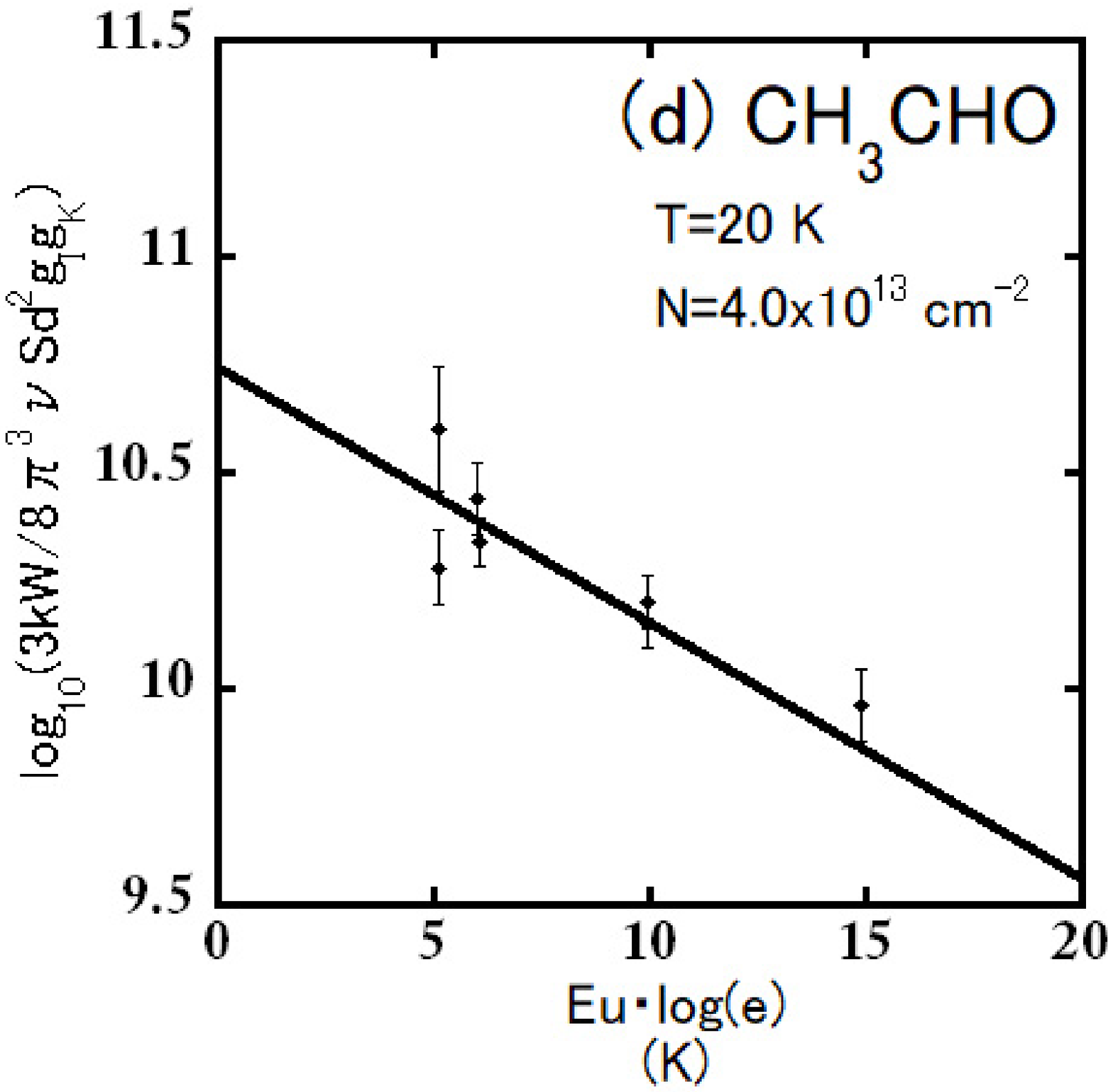}\\
\includegraphics[scale=.4]{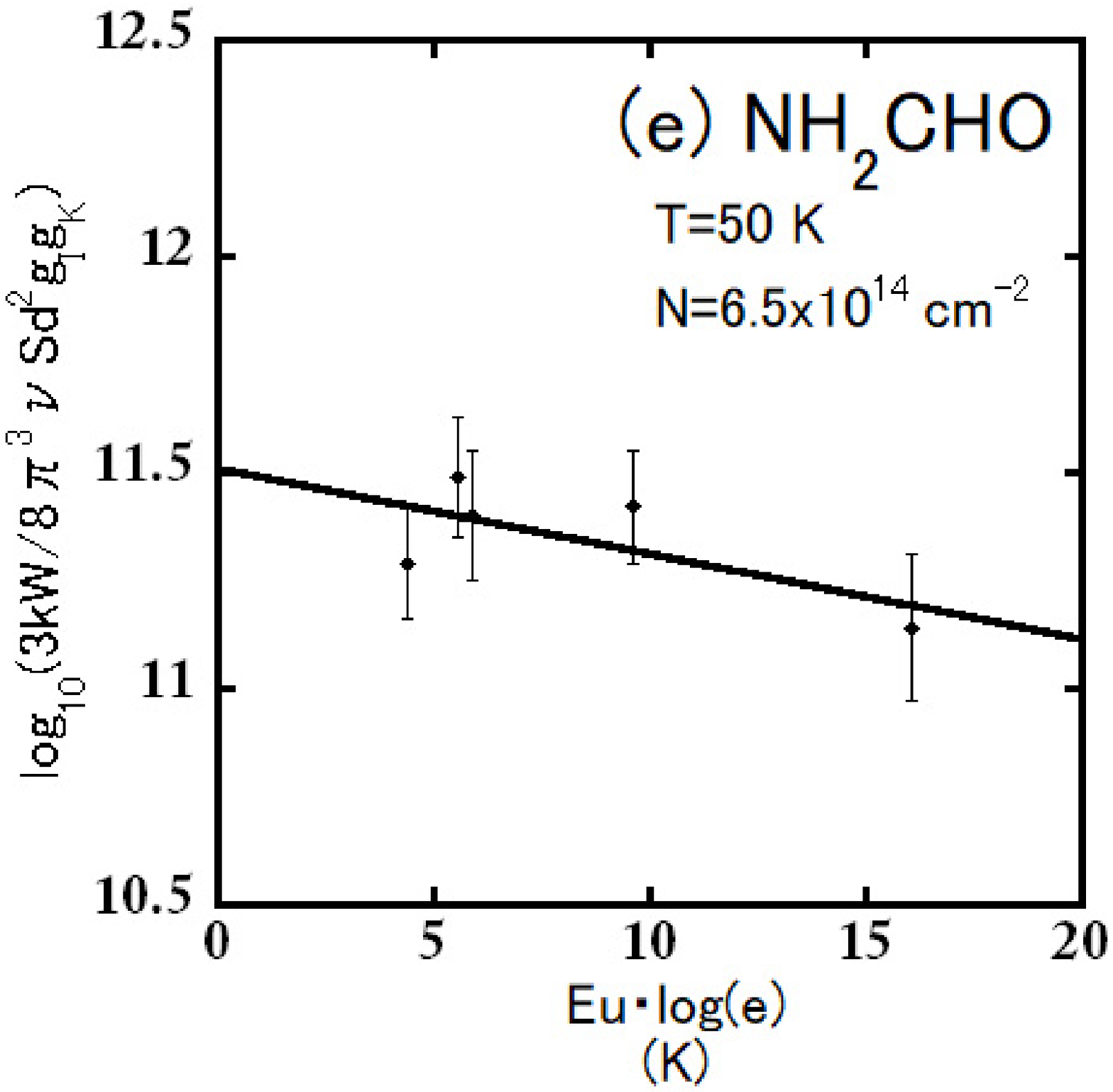}&
\includegraphics[scale=.4]{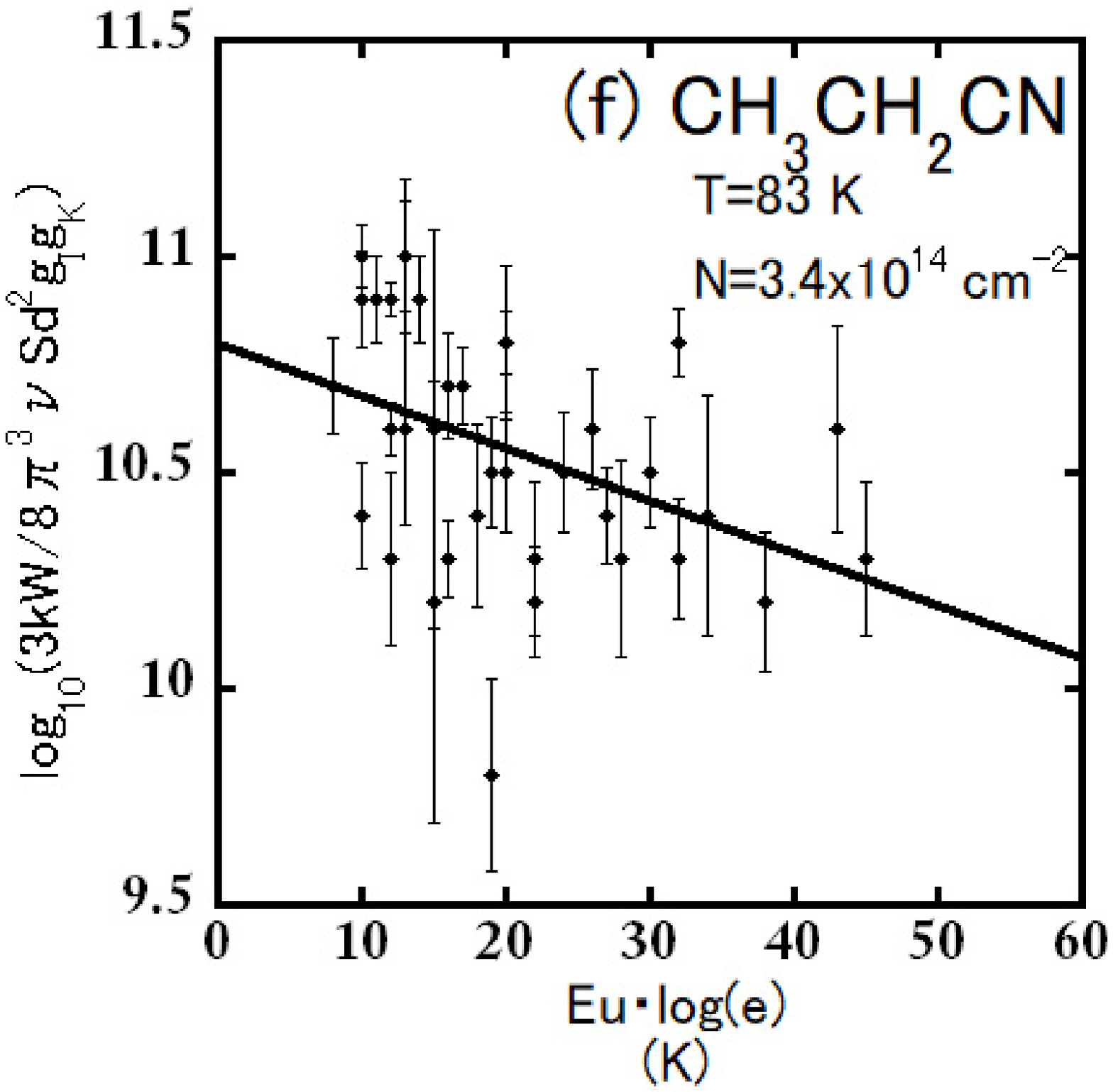}\\
  \end{tabular}
\caption{
The rotation diagrams for NGC6334F.
\label{fig:NGC6334F_rotation_diagrams}
}
\end{figure}
\clearpage

\begin{figure}
 \begin{tabular}{ll}
\includegraphics[scale=.4]{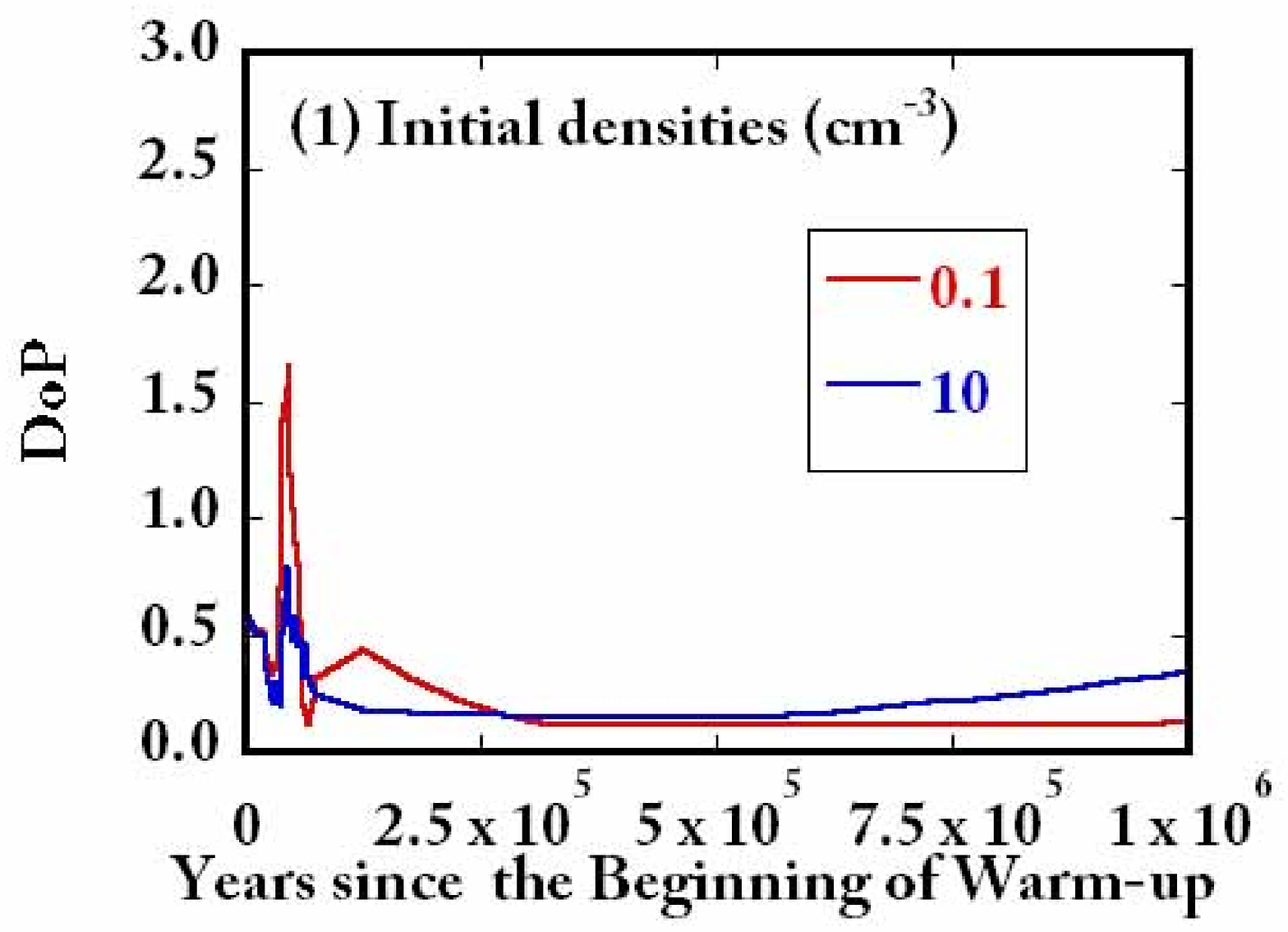}&
\includegraphics[scale=.4]{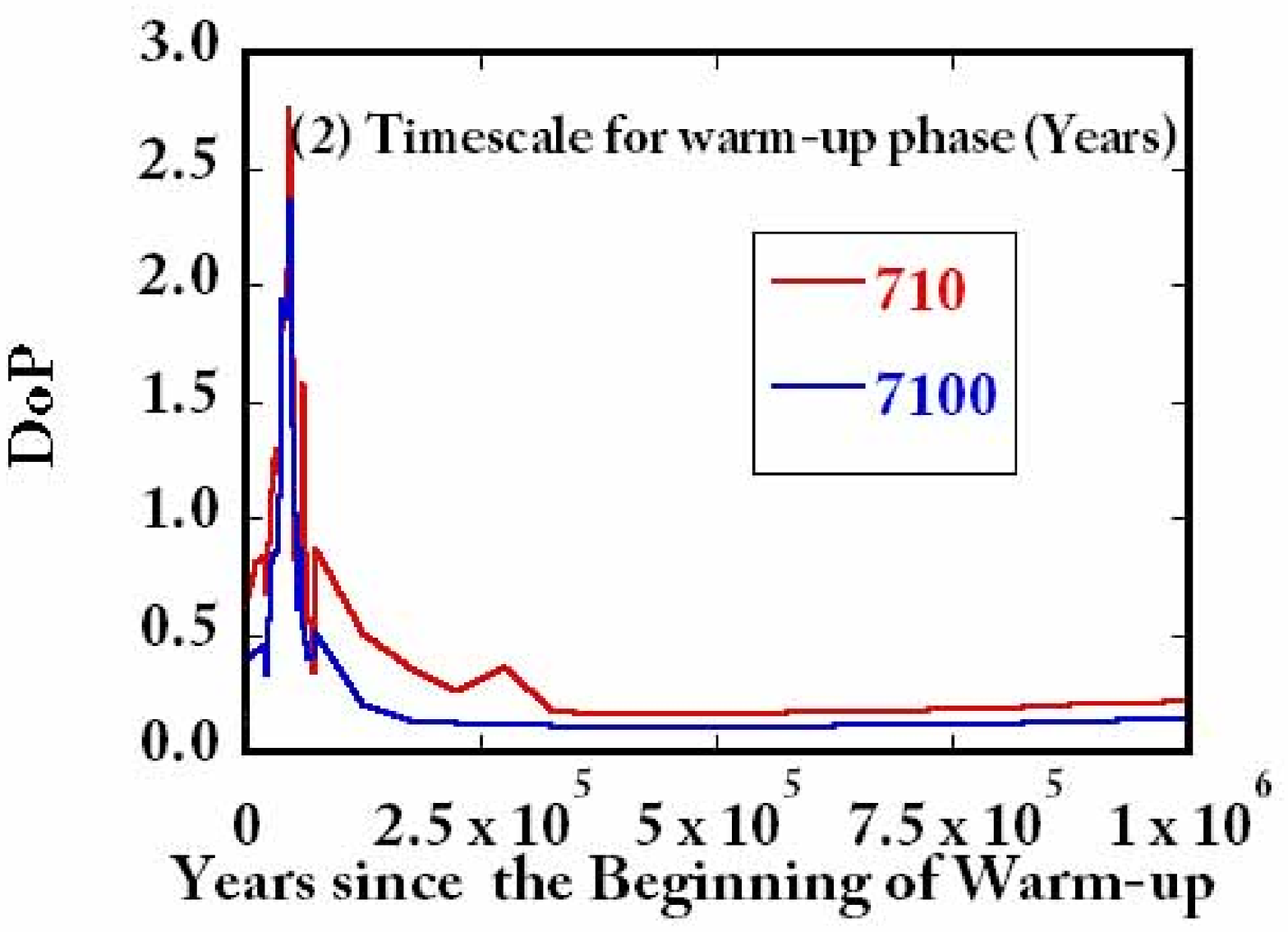}\\
\includegraphics[scale=.4]{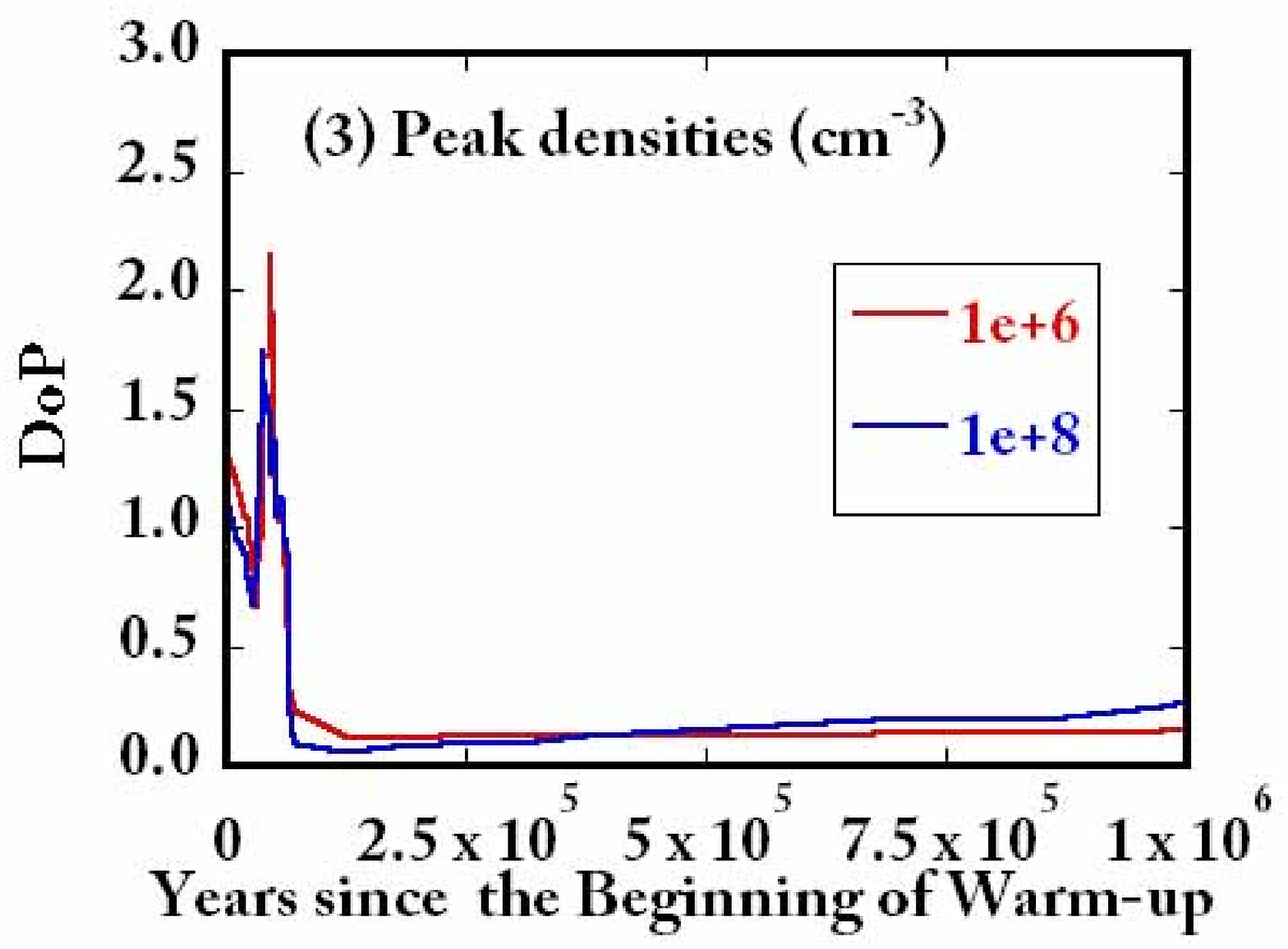}&
\includegraphics[scale=.4]{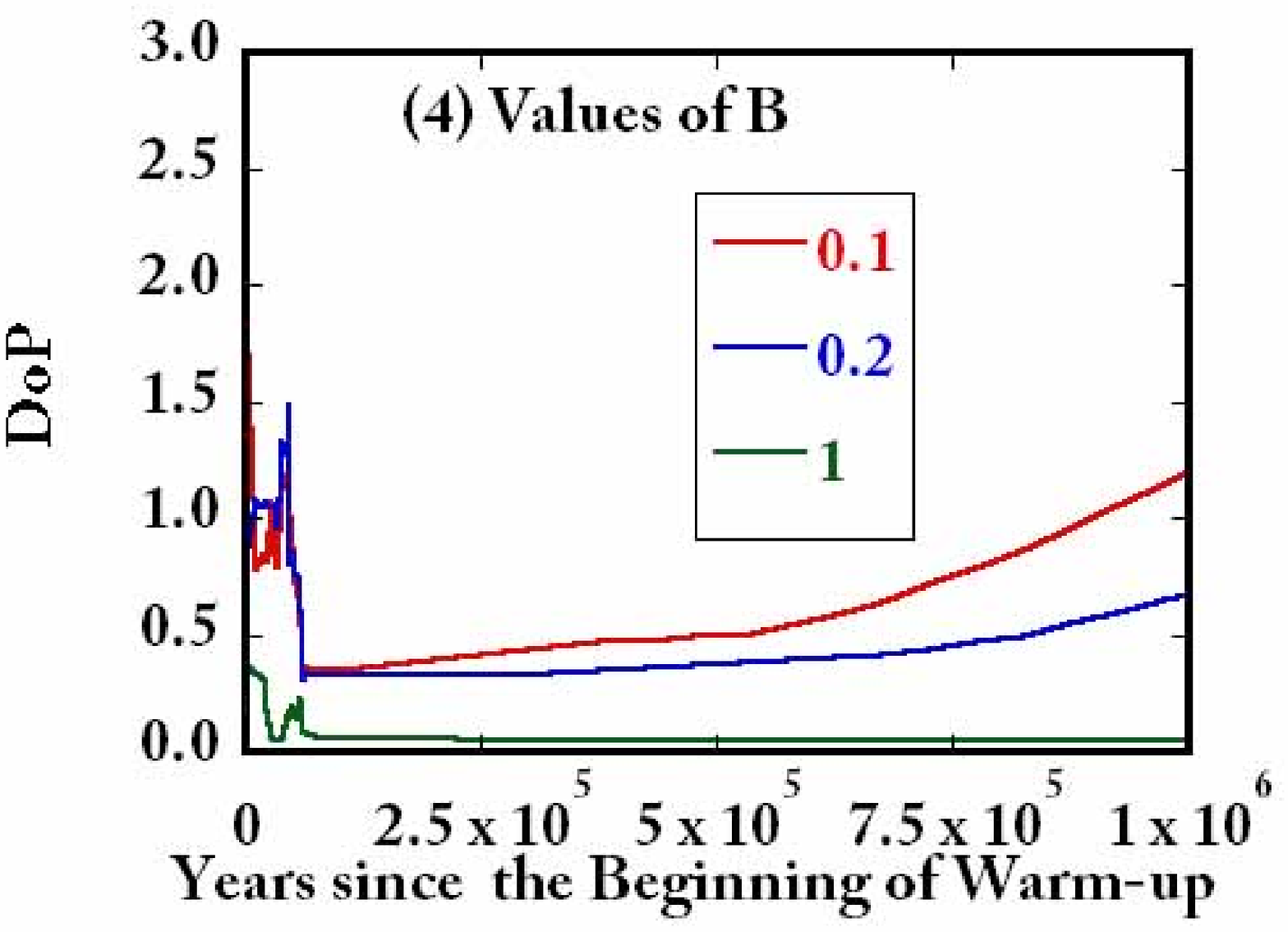}\\
\includegraphics[scale=.4]{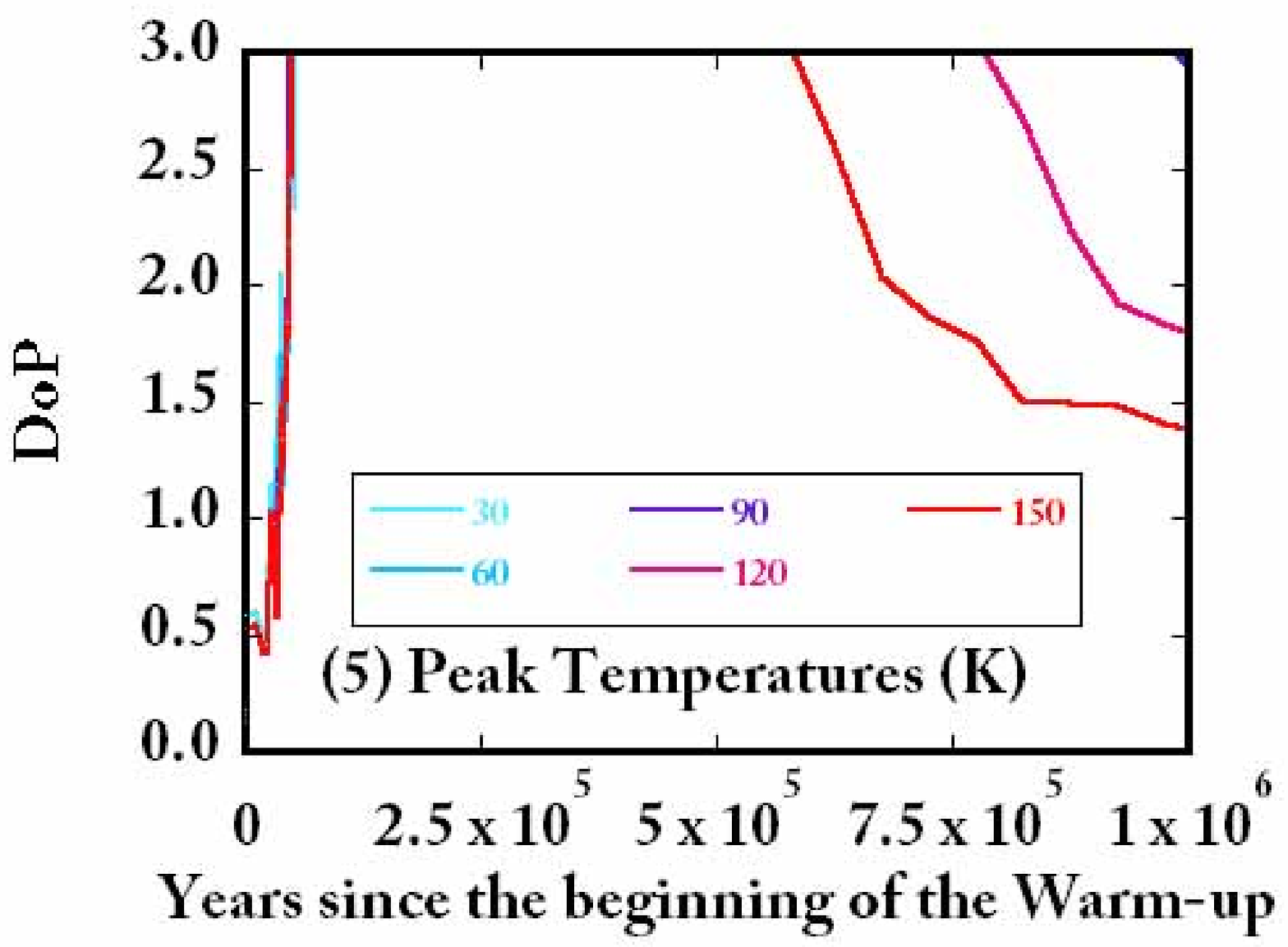}&\\
  \end{tabular}
\caption{
We calculated DoPs compared to a standard model, which employed the initial density of 1~cm$^{-3}$, B=0.7, the peak density of 1$\times$10$^7$~cm$^{-3}$, the warm-up timescale of 7.1$\times$10$^4$~years, and the peak temperature of 200~K.
The time of zero is corresponding to the beginning of the warm-up phase.
The calculated DoPs are shown for cases of (1) initial densities of 0.1 and 10~cm$^{-3}$, (2) warm-up timescales of 7.1$\times$10$^2$, and 7.1$\times$10$^3$~years (3) peak densities of 1$\times$10$^6$ and 1$\times$10$^8$~cm$^{-3}$, (4) the degree for resistance to gravitational collapsing, B, of 1, 0.2, and 0.1, and (5) different peak temperatures of 30, 60, 90, 120 and 150~K.
The DoPs for 30, 60, 90~K models in (3) were higher than 3.0.
\label{fig:DoPs}
}
\end{figure}
\clearpage

\begin{figure}
\begin{tabular}{ll}
\includegraphics[scale=.36]{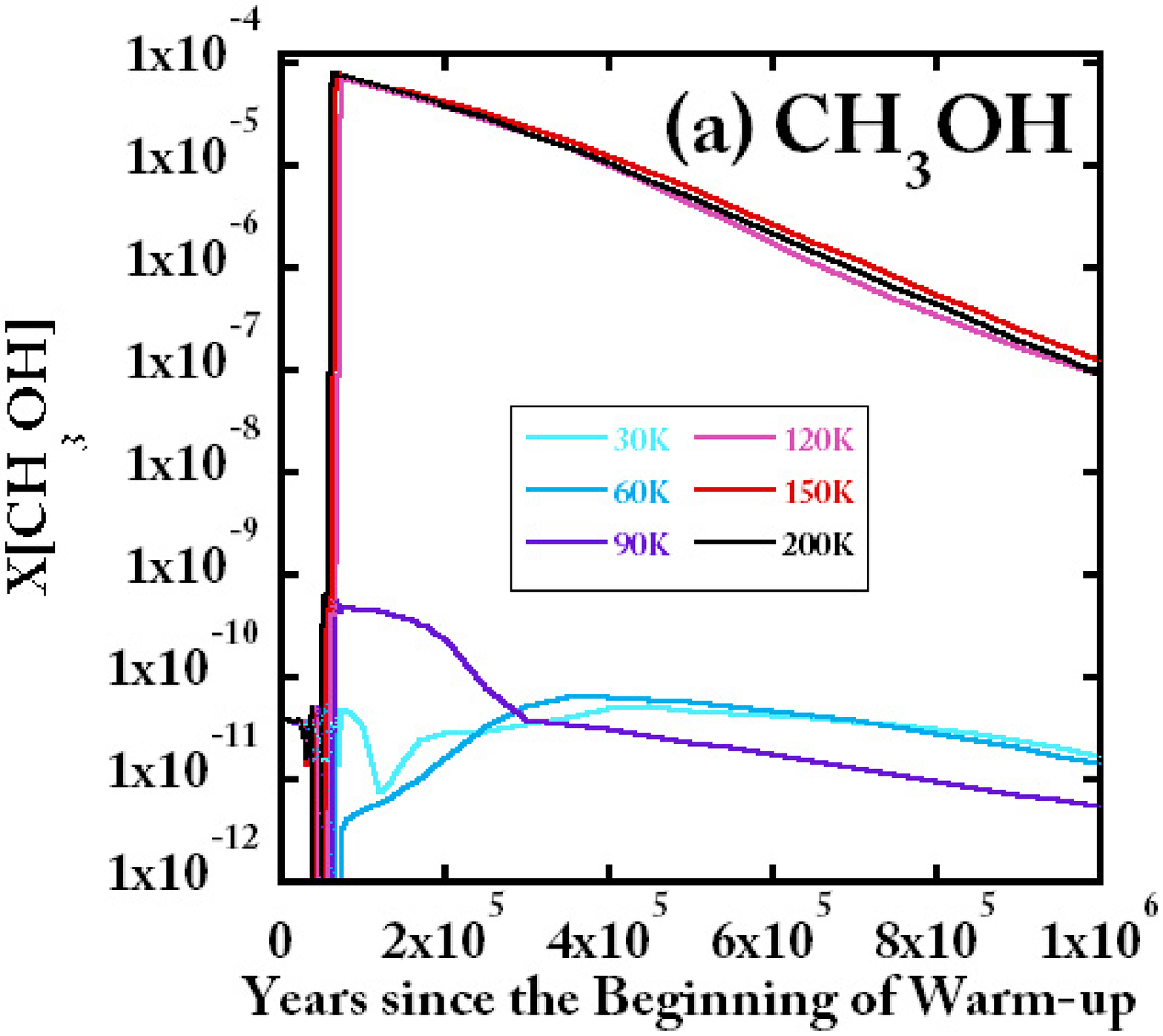}&
\includegraphics[scale=.36]{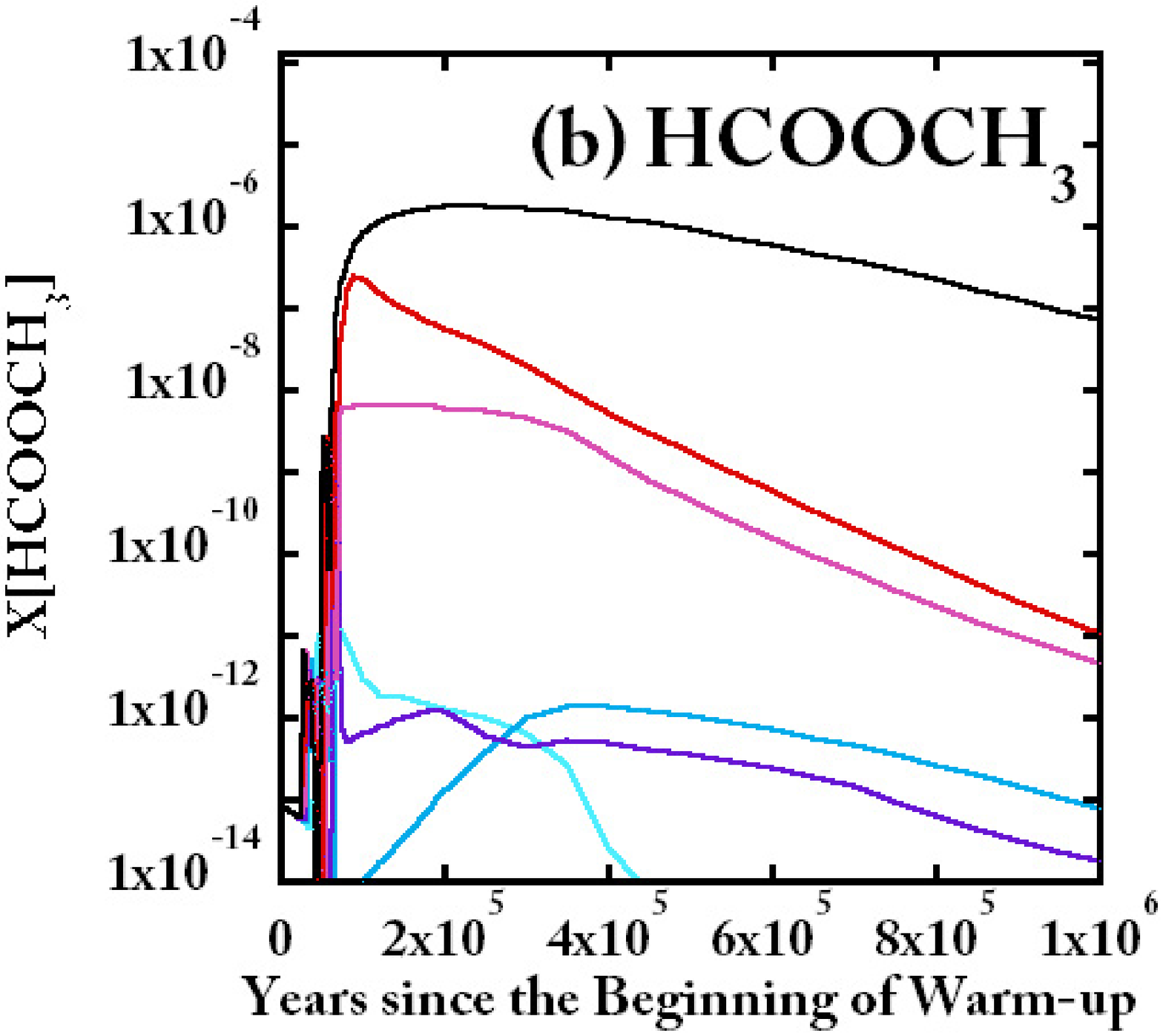}\\
\includegraphics[scale=.36]{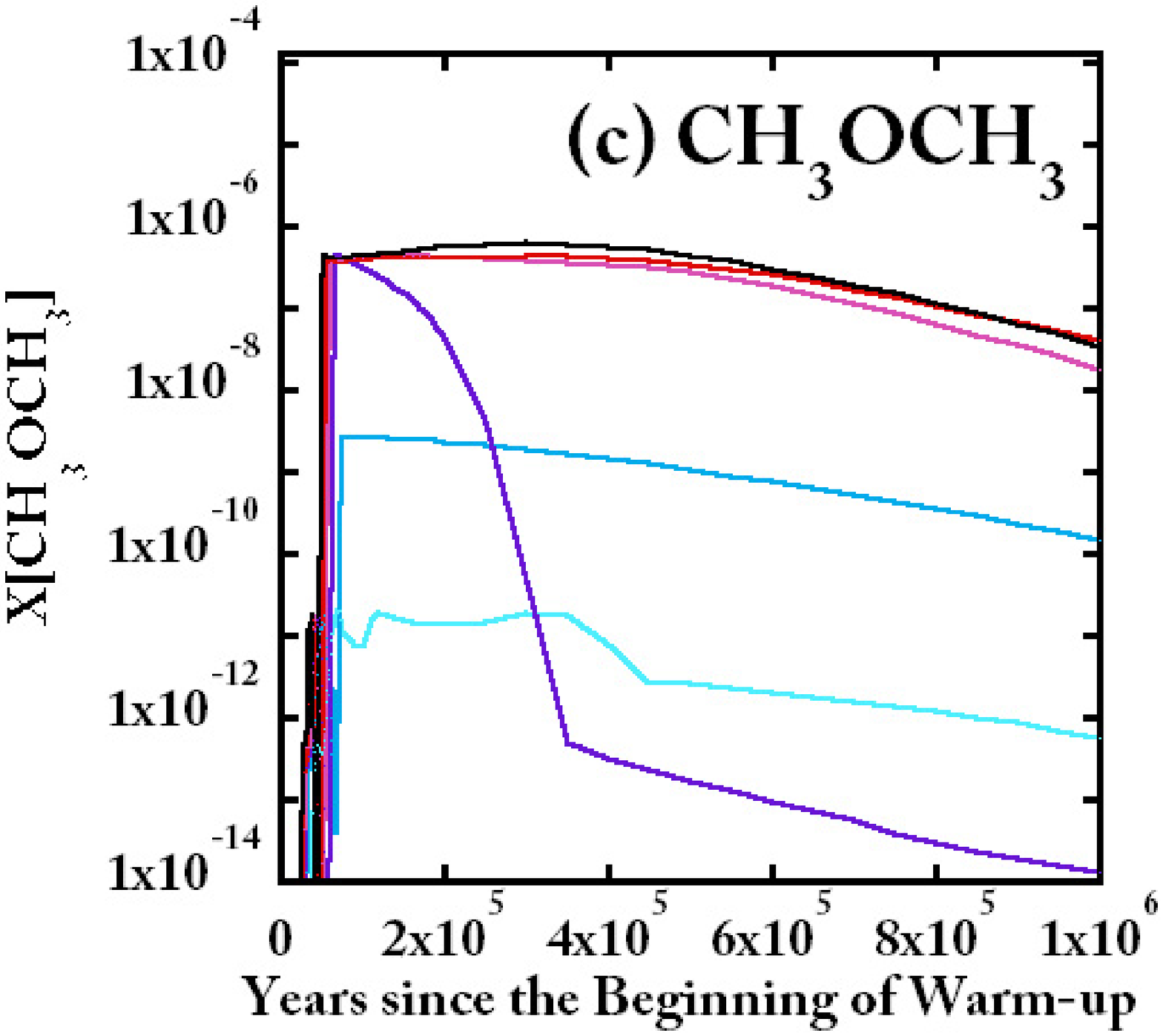}&
\includegraphics[scale=.36]{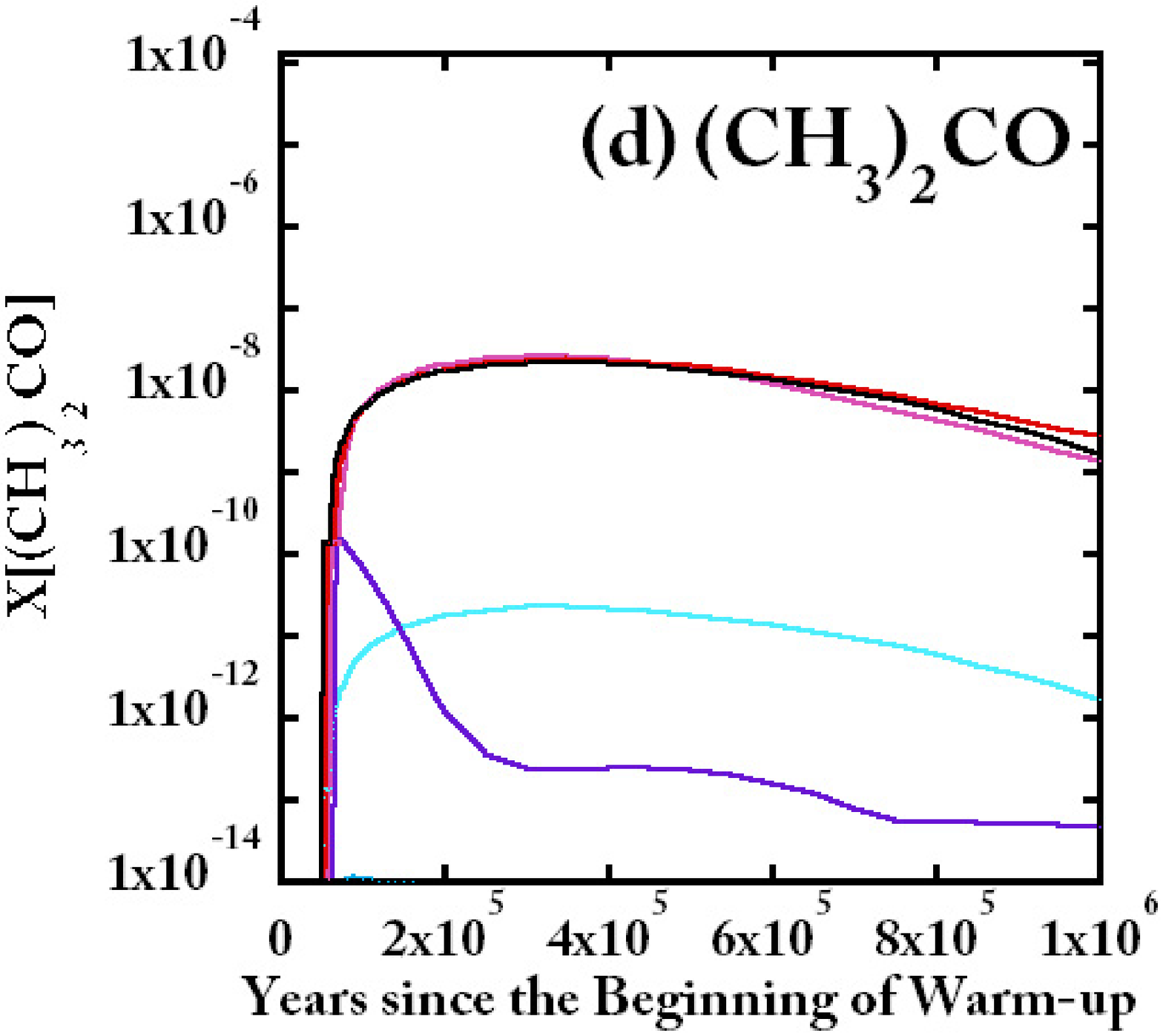}\\
\includegraphics[scale=.36]{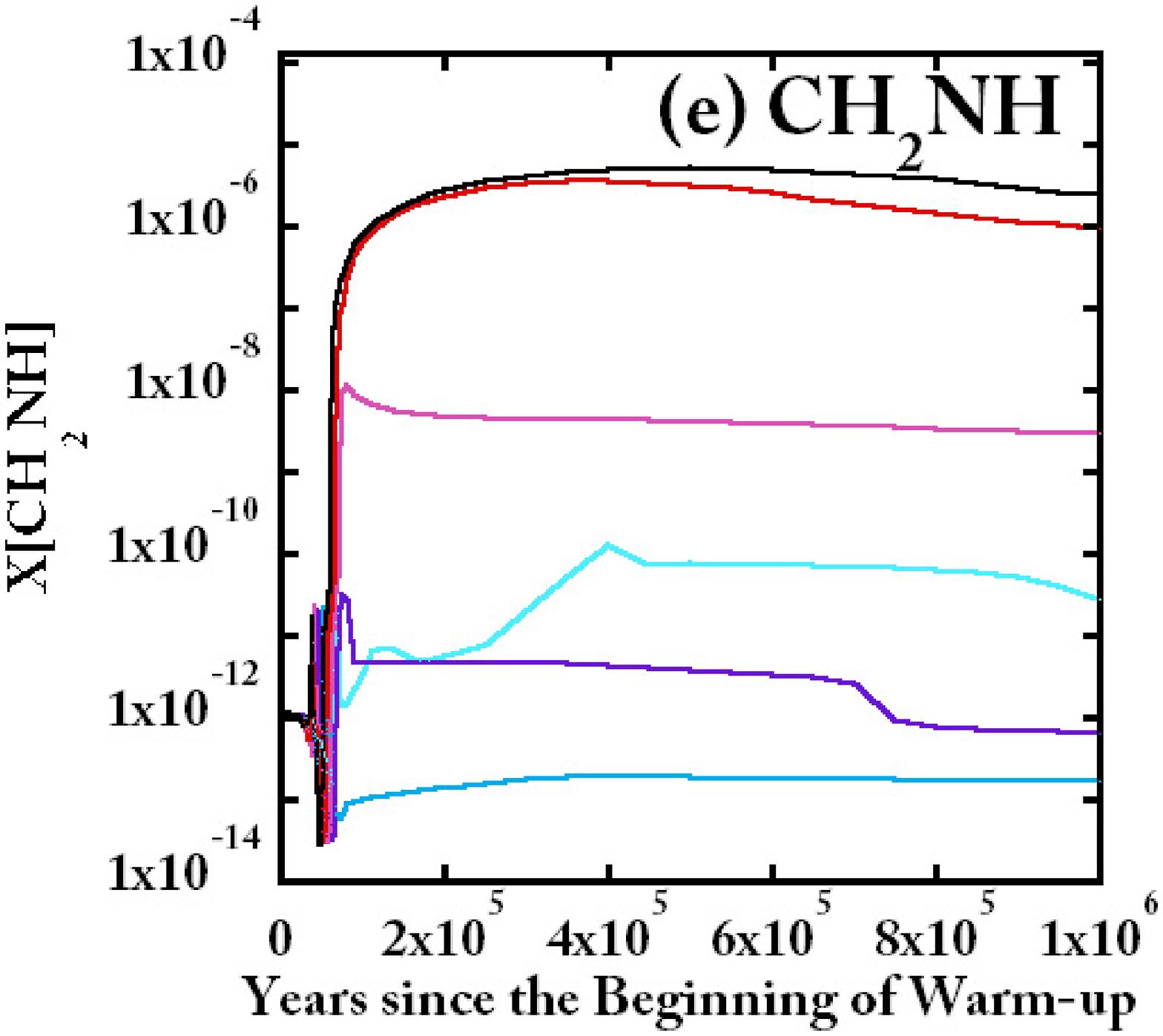}&
\includegraphics[scale=.36]{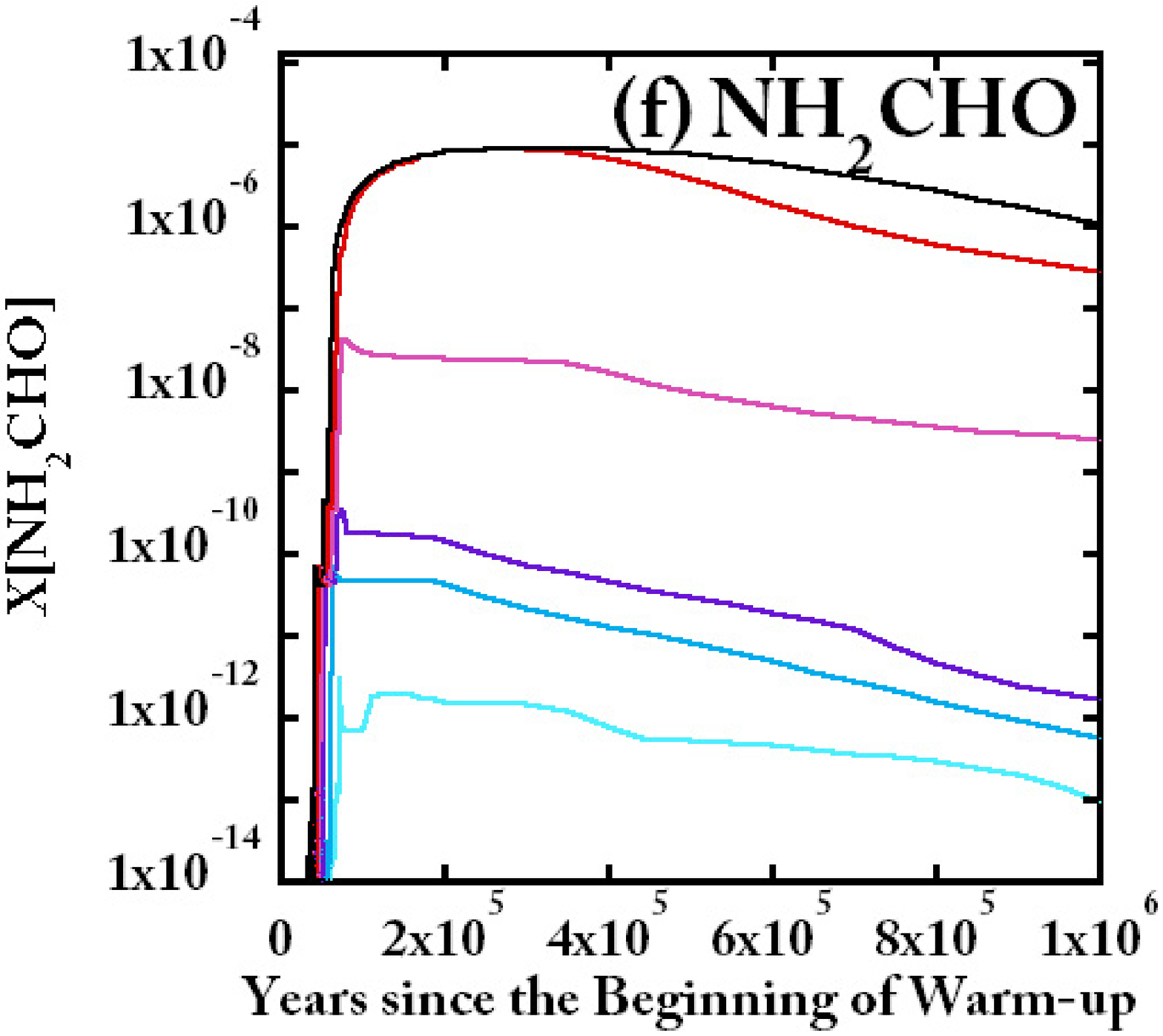}\\
  \end{tabular}
\caption{
Comparison of abundances of COMs simulated under the different peak temperatures, 30, 60, 90, 120, 150 and 200~K.
We added the time evolution of CH$_2$NH to compare with the observational results of \cite{Suzuki16}.
%
}
\end{figure}
\clearpage
\addtocounter{figure}{-1}
\begin{figure}
 \begin{tabular}{ll}
\includegraphics[scale=.36]{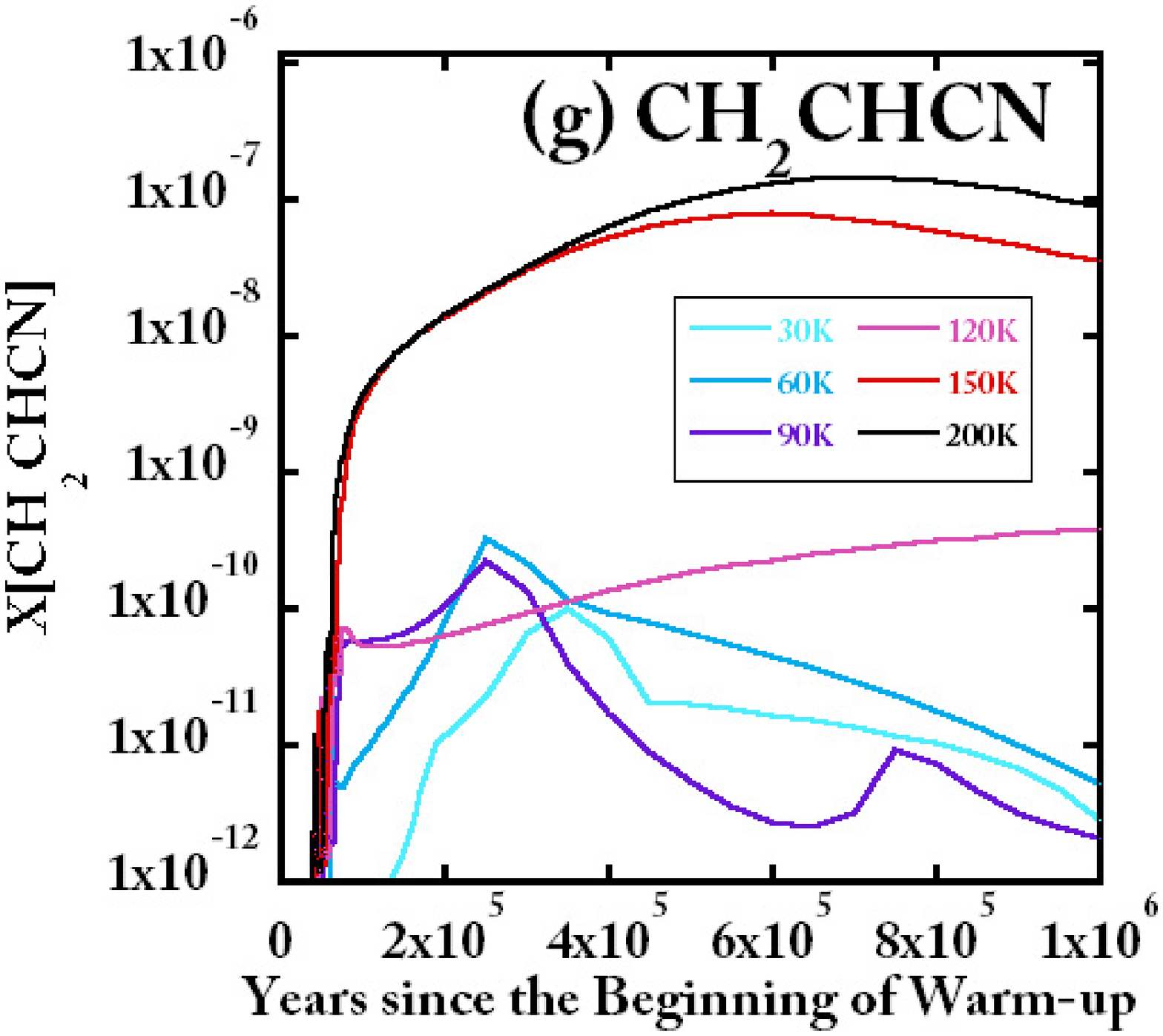}&
\includegraphics[scale=.36]{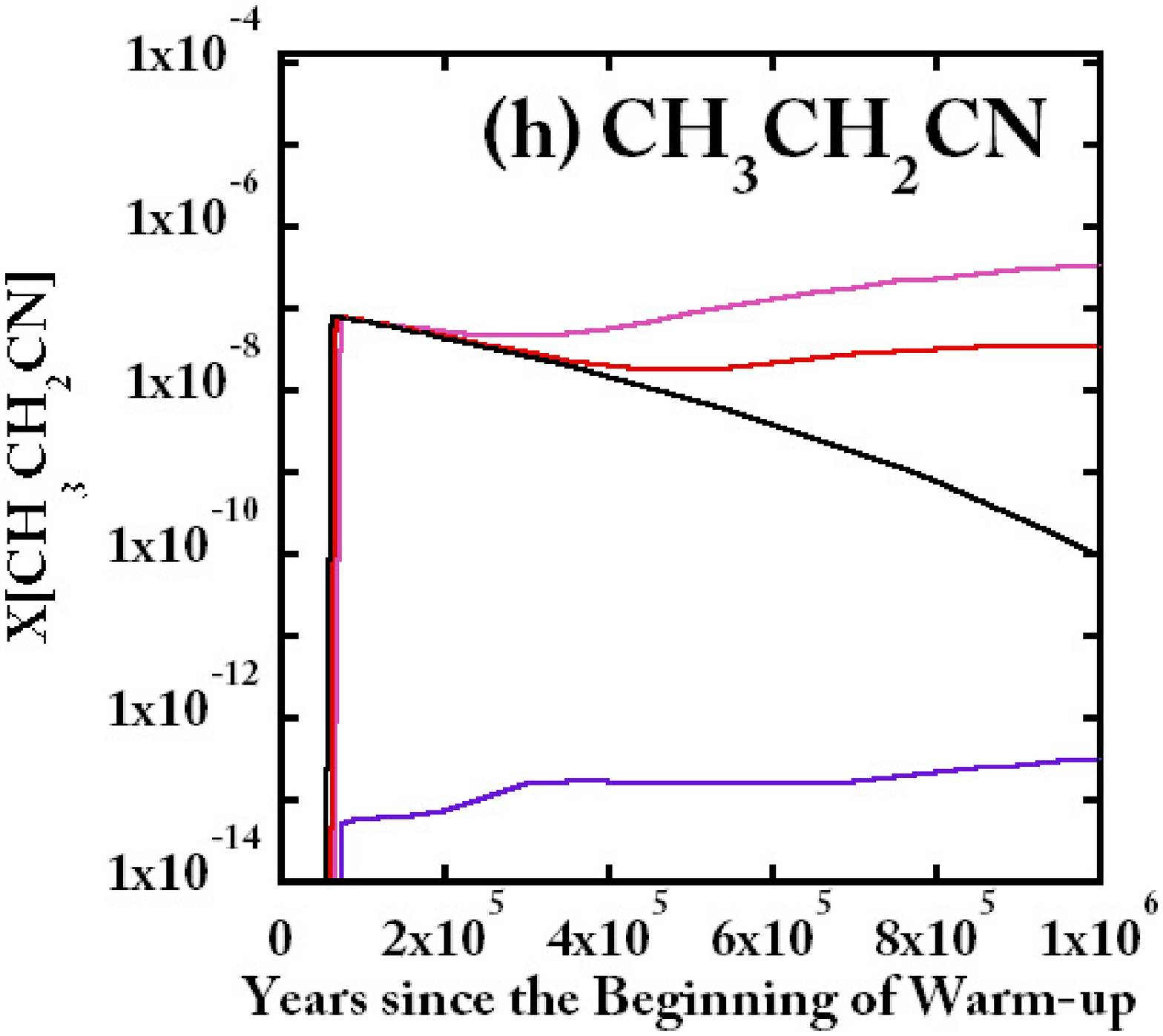}\\
\includegraphics[scale=.36]{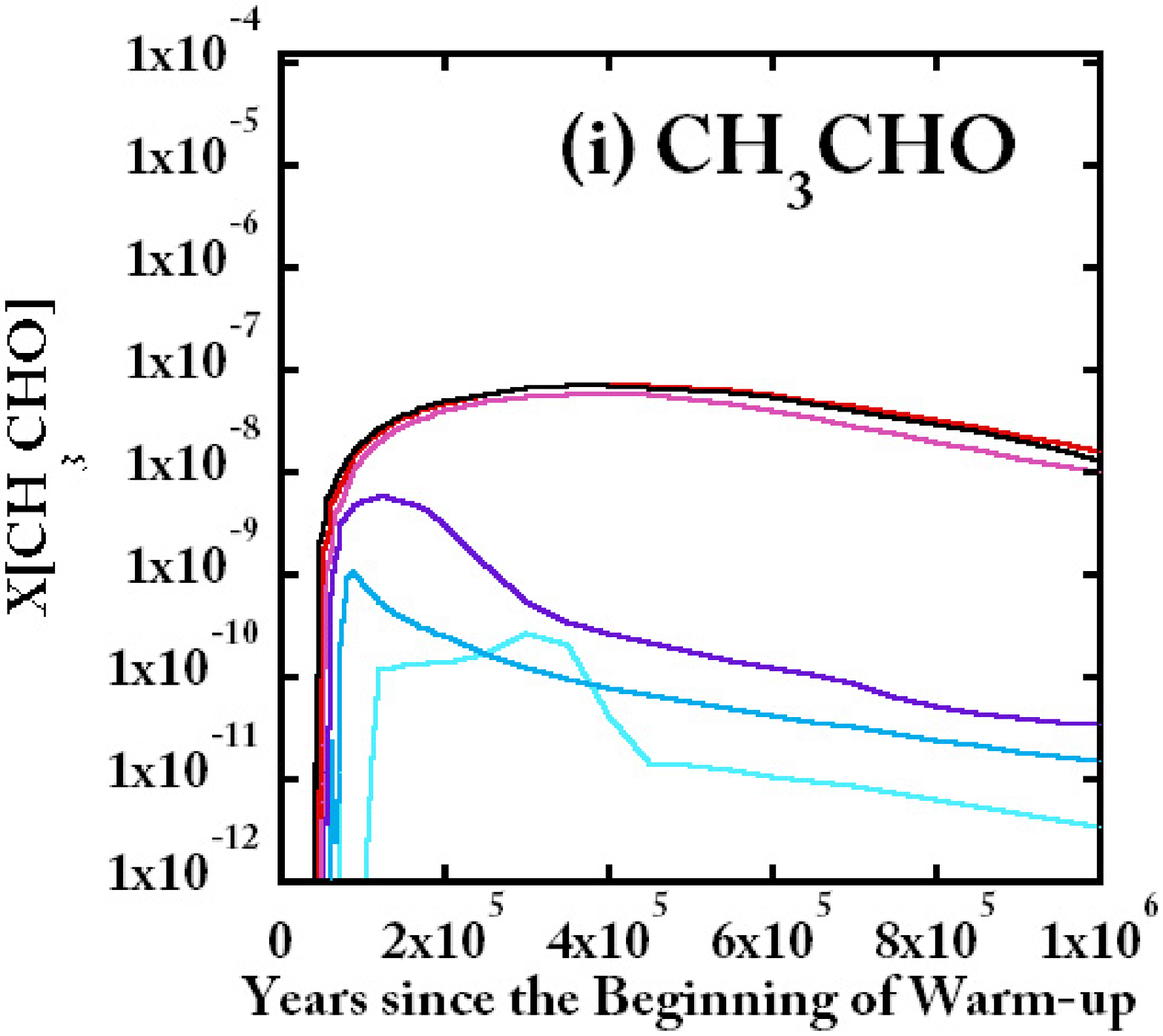}&\\
  \end{tabular}
\caption{
(continued) Comparison of abundances of COMs simulated under the different peak temperatures, 30, 60, 90, 120, 150 and 200~K.
\label{fig:different_peak_temperature}
}
\end{figure}
\clearpage

\begin{figure}
 \begin{tabular}{ll}
\includegraphics[scale=.4]{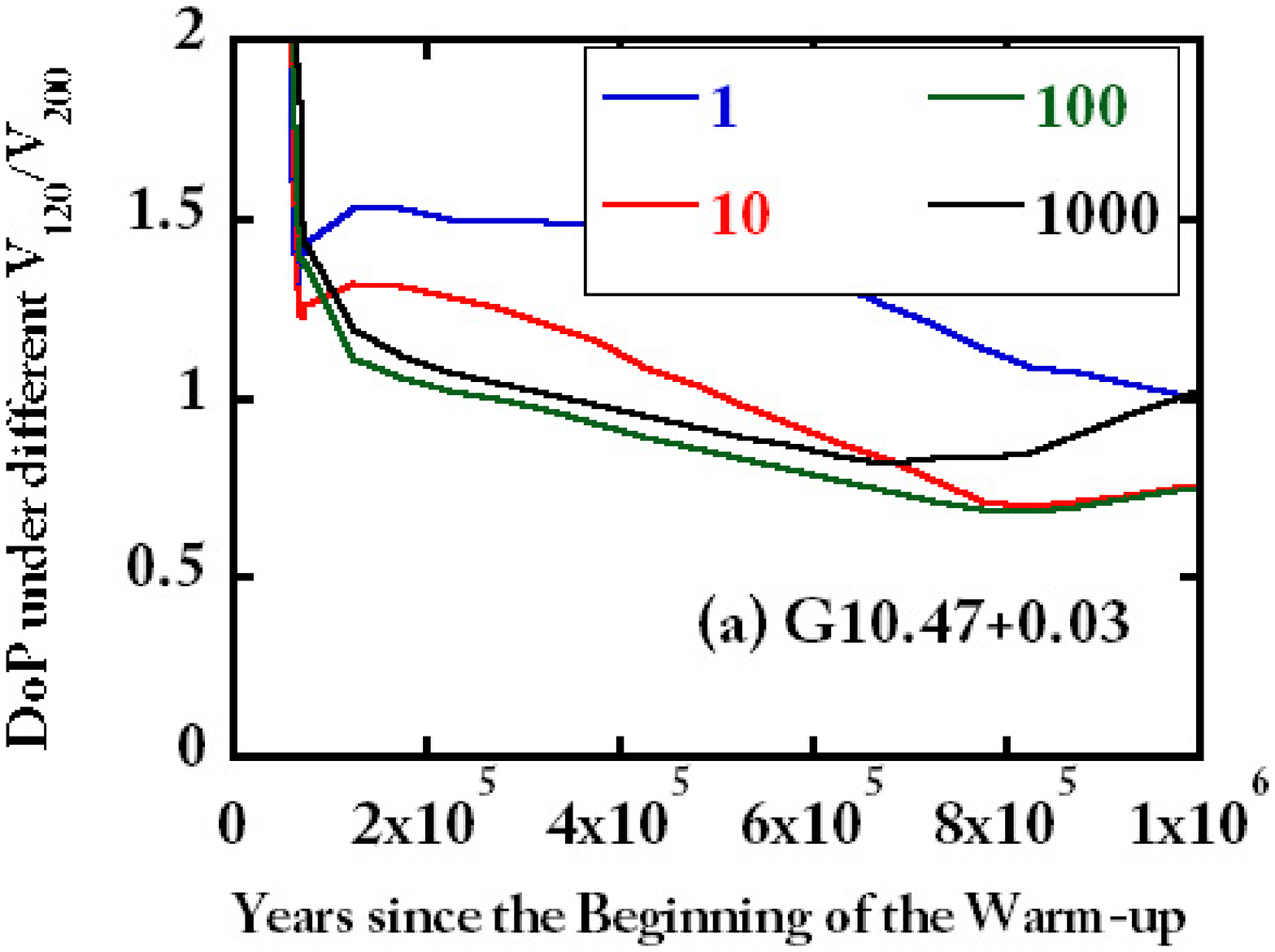}&
\includegraphics[scale=.4]{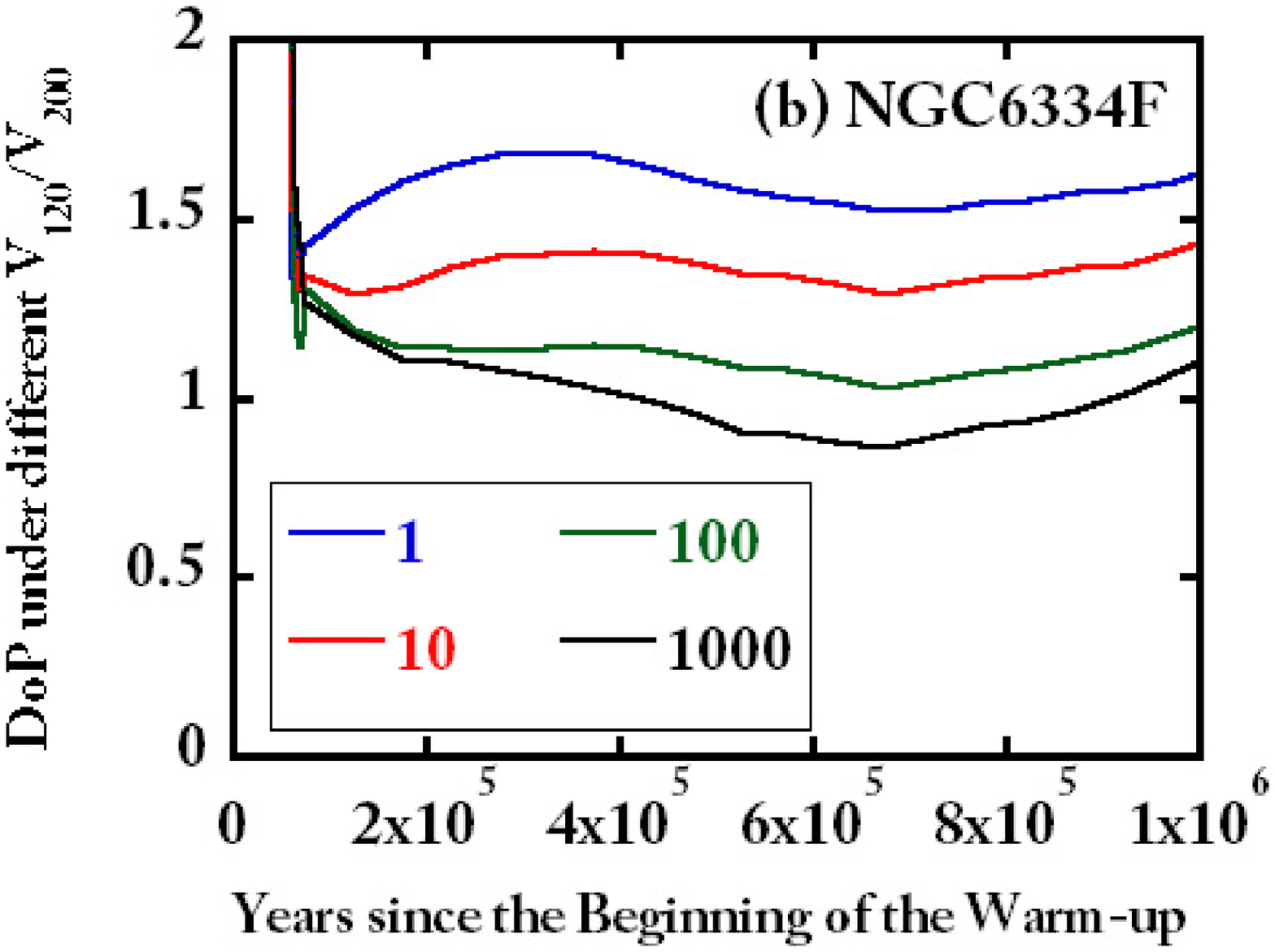}\\
  \end{tabular}
\caption{
DoPs under V$_{\rm 120K}$/V$_{\rm 200K}$=1, 10, 100, and 1000 were calculated for G10.47+0.03 and NGC6334F.
Smaller V$_{\rm 120K}$/V$_{\rm 200K}$ represent a core where high temperature region is dominant.
\label{fig:DoP_different_peak_temperature}
}
\end{figure}
\clearpage

\begin{figure}
 \begin{tabular}{ll}
\includegraphics[scale=.4]{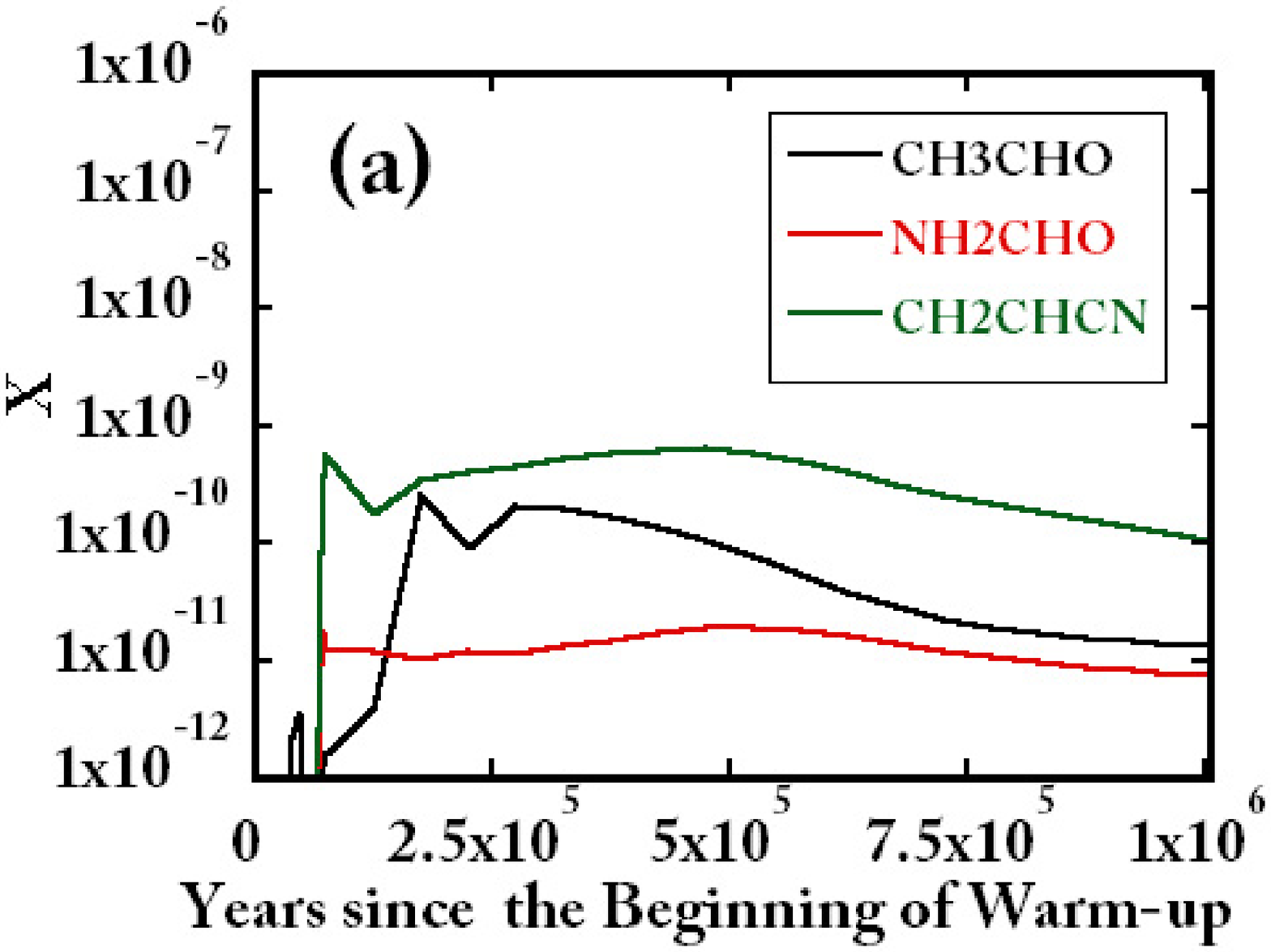}&
\includegraphics[scale=.4]{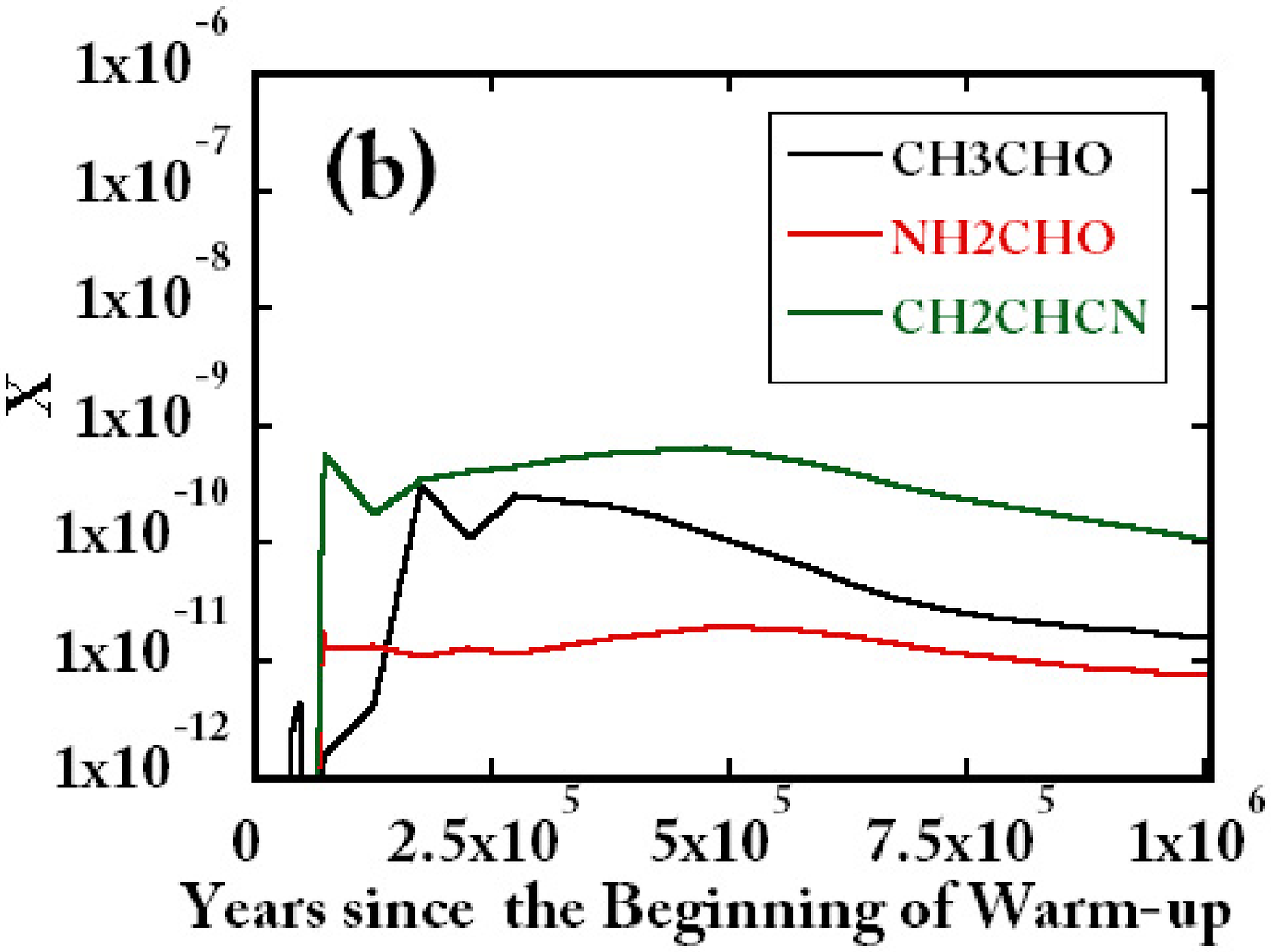}\\
\includegraphics[scale=.4]{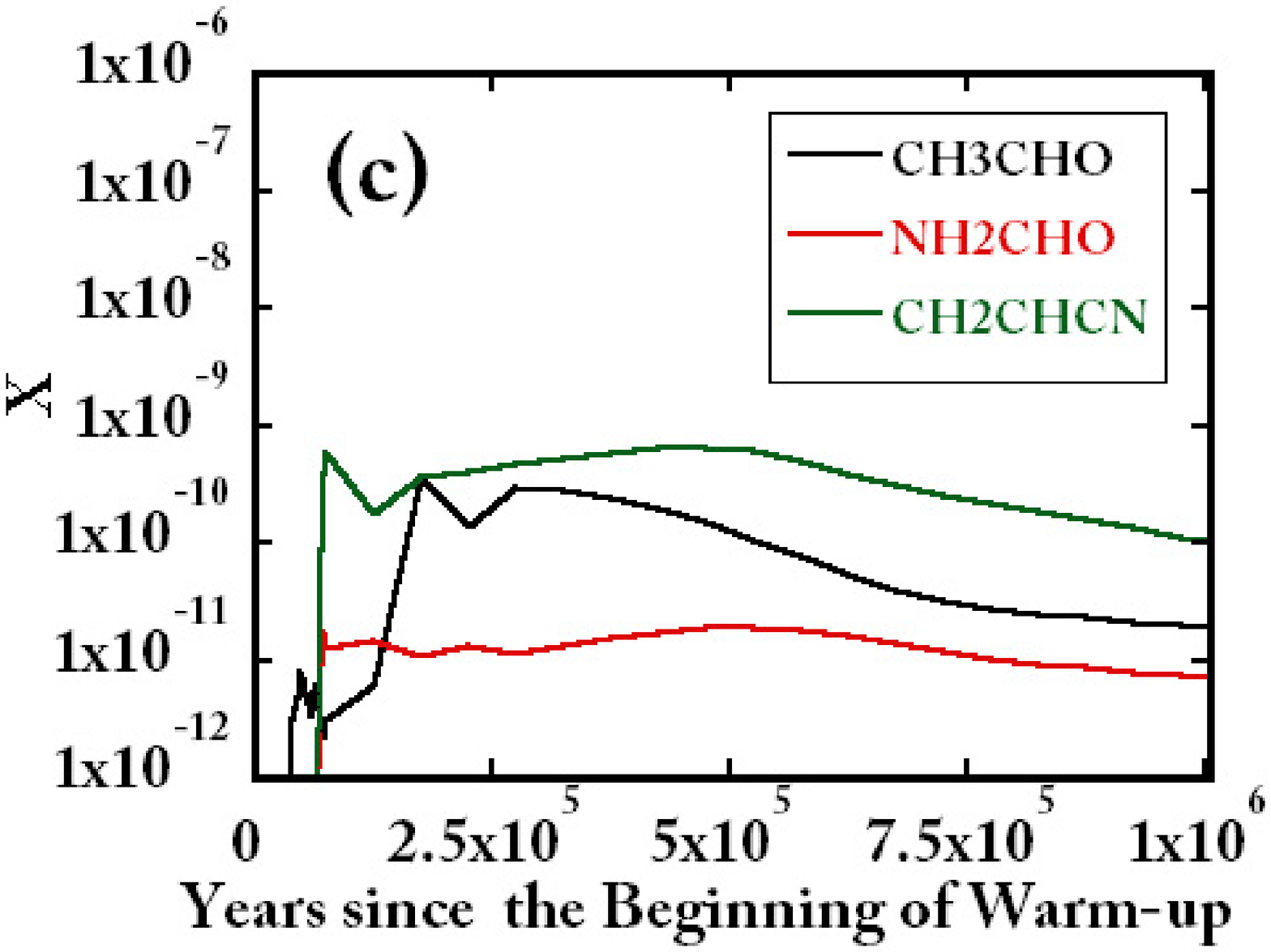}&\\
  \end{tabular}
\caption{
Chemical modeling of NH$_2$CHO, CH$_3$CHO, and CH$_2$CHCN for envelopes of hot cores. 
We assumed the peak temperature of 30~K and the peak density of 1$\times$10$^{6}$~cm$^{-2}$.
In (b) and (c), we lowered the binding energies of these species by 500~K and 1000~K, respectively.
\label{fig:envelope}
}
\end{figure}
\clearpage

\begin{deluxetable}{ccccccc}
\tabletypesize{\scriptsize}
\tablecaption{List of Observed Sources}
\tablewidth{0pt}
\tablehead{
\colhead{Source}
& \colhead{\shortstack {$\alpha$(J2000) \\  $^{\rm h}$ $^{\rm m}$ $^{\rm s}$}  }
& \colhead{\shortstack {$\delta$(J2000)  \\  $\degr$  '  "} } 
& \colhead{\shortstack {V$_{\rm LSR}$ \\ (km~s~$^{-1}$)}} 
& \colhead{\shortstack {distance \\ (kpc)}}
& \colhead{\shortstack {X[CH$_2$NH] \\ (x10$^{-8}$)}}
& \colhead{reference} 
}
\startdata
Orion KL&05 35 14.5&-05 22 30.6&6&0.4&3.3&1, 2\\
NGC6334F&17 20 53.4&-35 47 1.0&-7&1.3&0.24&1, 3\\
G10.47+0.03&18 08 38.13&-19 51 49.4&68&8.5&3.1&1, 3\\
G19.61-0.23&18 27 38.0&-11 56 42&41.9&4.0&0.14&1, 4\\
G31.41+0.3&18 47 34.6&-01 12 43&97&7.9&0.88&1, 5\\
G34.3+0.2&18 53 18.54&+01 14 57.0&58&1.6&0.24&1, 3\\
W51 e1/e2&19 23 43.77&+14 30 25.9&57&5.4&0.28&1, 3\\
DR21 (OH)&20 39 01.1&+42 22 50.2&-3&1.5&0.14&1, 3\\
\enddata
\tablecomments{
The coordinate, radial velocity, and the distance to sources are summarized. 
The fractional abundance of CH$_2$NH compared to hydrogen reported in \cite{Suzuki16} is also shown.
 References. (1) \cite{Ikeda01} (2)\cite{Menten07} (3)\cite{Reid14} (4)\cite{Hofner96} (5)\cite{Churchwell90}
}
\label{table:sources}
\end{deluxetable}
\clearpage

\begin{deluxetable}{ccc}
\tabletypesize{\scriptsize}
\tablecaption{The Observed Frequency Range}
\tablewidth{0pt}
\tablehead{
\colhead{Source}
& \colhead{\shortstack {From \\  (GHz)}}
& \colhead{\shortstack {To  \\  (GHz) }}
}
\startdata
Orion KL&78.80&79.80\\
&80.75&81.95\\
&83.70&85.00\\
&85.70&86.20\\
&86.50&87.00\\
&87.50&88.00\\
&88.20&88.70\\
&88.80&91.05\\
\enddata
\tablecomments{
The observed frequency range is shown.
}
\label{table:observed_frequency}
\end{deluxetable}
\clearpage

\startlongtable
\begin{deluxetable}{rrlrrr}
\tabletypesize{\scriptsize}
\tablecaption{Abundances Assuming 10" source size}
\tablewidth{0pt}
\tablehead{
\colhead{Source}
& \colhead{\shortstack {N[H$_2$] \\ (10$^{23}$~cm$^{-2}$)}}
& \colhead{Species}
& \colhead{\shortstack {N \\ (cm$^{-2}$)}}
& \colhead{\shortstack {Tex \\ (K)}}
& \colhead{\shortstack {X}}
}
\startdata
G10.47+0.03&1.3&CH$_3$OH&2.7$\pm$0.8 (17)&185$\pm$20&2.1$\pm$0.6 (-6)\\
&&HCOOCH$_3$&7.1$\pm$2.9 (15)&60$\pm$39&5.5$\pm$2.2 (-8)\\
&&CH$_3$OCH$_3$&3.0$\pm$0.9 (15)&119$\pm$14&2.3$\pm$0.7 (-8)\\
&&CH$_3$CHO&3.0$\pm$1.4 (14)&10$\pm$2&2.3$\pm$1.1 (-9)\\
&&NH$_2$CHO&1.9$\pm$29.0 (15)&119$\pm$58&1.5$\pm$22.4 (-8)\\
&&CH$_3$CH$_2$CN&2.3$\pm$0.6 (16)&398$\pm$209&1.8$\pm$0.5 (-7)\\
&&CH$_2$CHCN&6.2$\pm$1.4 (15)&382$\pm$104&4.8$\pm$1.1 (-8)\\
&&(CH$_3$)$_2$CO&1.3$\pm$0.4 (16)&[71]&1.0$\pm$0.3 (-7)\\
\enddata
\tablecomments{
a (b) means a$\times$ 10$^b$. 10" source size was assumed. Hydrogen column densities of these sources were summarized in \citep{Ikeda01}.
A reliable data for hydrogen column density under extended source was not available for G19.61-0.23.
\label{abundances_rotation_diagram}
}
\end{deluxetable}
\clearpage

\startlongtable
\begin{deluxetable}{rlrrrrrrr}
\tabletypesize{\scriptsize}
\tablecaption{Abundances in Compact Source Size}
\tablewidth{0pt}
\tablehead{
\colhead{Species}
&\colhead{Source}
& \colhead{\shortstack {$\Theta$\\initial\\ (sec)}} 
& \colhead{\shortstack {N\\initial \\ (cm$^{-2}$)}}
& \colhead{\shortstack {Tex\\initial\\ (K)}}
& \colhead{\shortstack {$\Theta$\\ (sec)}} 
& \colhead{\shortstack {N \\ (cm$^{-2}$)}}
& \colhead{\shortstack {Tex \\ (K)}}
& \colhead{X}
}
\startdata
CH$_3$OH&G10.47+0.03&&4.3 (18)&185&[2.5]&4.7$\pm$1.3 (18)&135$\pm$32&7.0$\pm$2.0 (-7)\\
&Orion~KL&&1.5 (18)&183&[5.2]&2.3$\pm$0.5 (18)&225$\pm$23&1.8$\pm$0.4 (-6)\\
&NGC6334F&3.5&2.5 (18)&201&3.8$\pm$0.5&2.6$\pm$1.3 (18)&116$\pm$37&1.8$\pm$0.9 (-6)\\
&W51 e1/e2&2&5.7 (18)&197&2.5$\pm$0.2&7.1$\pm$2.0 (18)&293$\pm$62&1.3$\pm$0.4 (-6)\\
&G31.41+0.3&1.7&3.7 (18)&217&1.8$\pm$0.1&3.8$\pm$1.1 (18)&185$\pm$51&1.1$\pm$0.3 (-6)\\
&G34.3+0.2&2.0&3.1 (18)&196&2.3$\pm$0.3&3.2$\pm$0.9 (18)&98$\pm$27&7.1$\pm$2.1 (-7)\\
&G19.61-0.23&&&&[2.5]&5.6$\pm$5.3 (17)&[131]&8.3$\pm$7.9 (-8)\\
HCOOCH$_3$&G10.47+0.03&&1.8 (17)&60&[2.0]&3.1$\pm$2.3 (17)&83$\pm$54&4.6$\pm$3.4 (-8)\\
\enddata
\tablecomments{
a (b) means a$\times$ 10$^b$. Fractional abundances were calculated from the H$_2$ column densities: 6.69$\times$10$^{24}$~cm$^{-2}$ for G10.47+0.03, 1.35$\times$10$^{24}$~cm$^{-2}$ for NGC6334F, 5.27$\times$10$^{24}$~cm$^{-2}$ for W51~e1/e2, 3.53$\times$10$^{24}$~cm$^{-2}$ for G31.41+0.03, \citep{Hernandez14}, 4.50$\times$10$^{24}$~cm$^{-2}$ for G34.3+0.2 \citep{Fontani07}, and 1.3$\times$10$^{24}$~cm$^{-2}$ \citep{Hirota15} for Orion~KL..
\label{abundances_compact}
}
\end{deluxetable}
\clearpage

\begin{deluxetable}{lrrrrrrr}
\tablecaption{The Calculated Correlation Coefficients}
\tablewidth{0pt}
\tablehead{
\colhead{} & \colhead{HCOOCH$_3$} & \colhead{CH$_3$OCH$_3$} & \colhead{(CH$_3$)$_2$CO} & \colhead{NH$_2$CHO} & \colhead{CH$_3$CH$_2$CN} & \colhead{CH$_2$CHCN} & \colhead{CH$_2$NH}
}
\startdata
CH$_3$OH&0.31&0.70&0.60&-0.12&0.51&-0.05&0.34\\
HCOOCH$_3$&&0.58&-0.34&-0.52&-0.62&-0.68&-0.74\\
CH$_3$OCH$_3$&&&0.22&-0.34&-0.14&-0.25&-0.21\\
(CH$_3$)$_2$CO&&&&0.09&0.54&0.55&0.78\\
NH$_2$CHO&&&&&0.01&0.88&0.66\\
CH$_3$CH$_2$CN&&&&&&0.26&0.92\\
CH$_2$CHCN&&&&&&&0.60\\
\enddata
\tablecomments{The correlation coefficients were calculated with our observed abundances for compact source sizes. The fractional abundances reported in \cite{Suzuki16} was also added in this discussion.}
\label{correlation}
\end{deluxetable}
\clearpage

\begin{deluxetable}{lrrrrrrr}
\tablecaption{The result of t-test}
\tablewidth{0pt}
\tablehead{
\colhead{} & \colhead{HCOOCH$_3$} & \colhead{CH$_3$OCH$_3$} & \colhead{(CH$_3$)$_2$CO} & \colhead{NH$_2$CHO} & \colhead{CH$_3$CH$_2$CN} & \colhead{CH$_2$CHCN} & \colhead{CH$_2$NH}
}
\startdata
CH$_3$OH&0.65&1.96&1.50&-0.27&1.33&-0.10&0.81\\
HCOOCH$_3$&&1.42&-0.72&-1.22&-1.58&-1.85&-2.20\\
CH$_3$OCH$_3$&&&0.45&-0.72&-0.28&-0.52&-0.43\\
(CH$_3$)$_2$CO&&&&0.18&1.28&1.32&{\bf2.49}\\
NH$_2$CHO&&&&&0.02&{\bf3.71}&1.96\\
CH$_3$CH$_2$CN&&&&&&0.54&{\bf5.25}\\
CH$_2$CHCN&&&&&&&1.50\\
\enddata
\tablecomments{The result of t-test to evaluate the reliability of correlation coefficients.
We showed the values with their significance level more than 95$\%$ in boldface. 
}
\label{t-test}
\end{deluxetable}
\clearpage

\startlongtable
\begin{deluxetable}{cc}
\tablecaption{Desorption Energies}
\tablewidth{0pt}
\tablehead{
\colhead{Species} & \colhead{E$_D$ (K)}
}
\startdata
CH$_3$OH&5534\\
HCOOCH$_3$&6295\\
CH$_3$OCH$_3$&3151\\
(CH$_3$)$_2$CO&3500\\
CH$_2$NH&5534\\
CH$_3$NH$_2$&6584\\
CH$_2$CHCN&5484\\
CH$_3$CH$_2$CN&5540\\
NH$_2$CHO&5556\\
CH$_3$CHO&2450\\
\enddata
\tablecomments{Desorption energies used our chemical reaction model and dipole moments cited from splatalogue. 
}
\label{dipole_moments_and_ED}
\end{deluxetable}
\clearpage

\begin{deluxetable}{llll}
\tablecaption{Important Gas Phase Reactions Related to Interested COMs}
\tablewidth{0pt}
\tablehead{
\colhead{Reaction}
& \colhead{$\alpha$}
& \colhead{$\beta$}
& \colhead{$\gamma$}
}
\startdata
CH$_3$OH$_2^+$ + e$^-$ $\longrightarrow$ H + H$_2$ + H$_2$CO&9.1 (-8)&-6.7 (-1)&0\\
CH$_3$OH$_2^+$ + e$^-$ $\longrightarrow$ H$_2$O + CH$_3$&8.19 (-8)&-6.7 (-1)&0\\
CH$_3$OH$_2^+$ + e$^-$ $\longrightarrow$ H + OH + CH$_3$&4.64 (-7)&-6.7 (-1)&0\\
CH$_3$OH$_2^+$ + e$^-$ $\longrightarrow$ H + CH$_2$ + H$_2$O&1.91 (-7)&-6.7 (-1)&0\\
CH$_3$OH$_2^+$ + e$^-$ $\longrightarrow$ H + CH$_3$OH&2.73 (-8)&-6.7 (-1)&0\\
HC(OH)OCH$_3^+$ + e$^-$ $\longrightarrow$ HCOOCH$_3$ + CH$_3$OH&1.5 (-7)&-5 (-1)&0\\
HC(OH)OCH$_3^+$ + e$^-$ $\longrightarrow$ HCOOCH$_3$ + H&1.5 (-7)&-5 (-1)&0\\
CH$_3$OCH$_4^+$ + e$^-$ $\longrightarrow$ O+CH$_3$ + CH$_4$&7.48 (-7)&-7 (-1)&0\\
\enddata
\tablecomments{
a (b) means a$\times$ 10$^b$.
$\alpha$, $\beta$, and $\gamma$ represent the coefficients for the Arrhenius equation.
}
\label{gas_reactions}
\end{deluxetable}
\clearpage

\begin{deluxetable}{ll}
\tablecaption{Dust Surface Reactions Related to Interested COMs}
\tablewidth{0pt}
\tablehead{
\colhead{Reaction}
& \colhead{\shortstack {E$_A$ \\ (K)}}
}
\startdata
CO + H $\longrightarrow$ HCO&2.5 (3)\\
HCO + H $\longrightarrow$ H$_2$CO&0\\
H$_2$CO + H $\longrightarrow$ CH$_3$O&2.2 (3)\\
H$_2$CO + H $\longrightarrow$ CH$_2$OH&5.4 (3)\\
CH$_3$O + H $\longrightarrow$ CH$_3$OH&0\\
CH$_2$OH + H $\longrightarrow$ CH$_3$OH&0\\
CH$_3$ + OH $\longrightarrow$ CH$_3$OH&0\\
CH$_3$O + CH$_3$ $\longrightarrow$ CH$_3$OCH$_3$&0\\
\enddata
\tablecomments{
The dust surface reactions related to CH$_2$NH are shown.
E$_A$ represents the value of the activation barrier.
Since radical species are so reactive, radical-radical reactions would have no activation barriers.
The activation barriers for HCN and CH$_2$NH were cited from the theoretical study by \cite{Woon02}.
}
\label{dust_reactions}
\end{deluxetable}
\clearpage

\begin{deluxetable}{lllll}
\tablecaption{DoPs for G10.47+0.03.}
\tablewidth{0pt}
\tablehead{
\multicolumn{1}{c}{ }  &
\multicolumn{4}{c}{V$_{\rm warm}$/V$_{\rm 200K}$}  \\
\cline{2-5}\\
\colhead{T$_{\rm warm}$} & \colhead{1} & \colhead{10} & \colhead{100} & \colhead{1000}
}
\startdata
30~K&1.05&0.83&0.76&1.27\\
60~K&1.05&0.83&0.73&1.10\\
90~K&1.05&0.83&0.75&1.09\\
120~K&0.99&\bf{0.68}&\bf{0.66}&0.81\\
150~K&1.04&0.95&1.05&1.18\\
\enddata
\tablecomments{
The temperature gradient inside of the hot core was represented by 200~K and warmer regions.
DoPs for G10.47+0.03 under the different volume ratio, ``V$_{\rm warm}$/V$_{\rm 200K}$" and the different temperature for the warm region, T$_{\rm warm}$
}
\label{DoP_G10}
\end{deluxetable}
\clearpage

\begin{deluxetable}{lllll}
\tablecaption{DoPs for NGC6334F}
\tablewidth{0pt}
\tablehead{
\multicolumn{1}{c}{ }  &
\multicolumn{4}{c}{V$_{\rm warm}$/V$_{\rm 200K}$}  \\
\cline{2-5}\\
\colhead{T$_{\rm warm}$} & \colhead{1} & \colhead{10} & \colhead{100} & \colhead{1000}
}
\startdata
30~K&1.25&1.10&1.20&1.56\\
60~K&1.24&1.10&1.16&1.39\\
90~K&1.24&1.21&1.06&1.24\\
120~K&1.26&1.24&1.03&\bf{0.86}\\
150~K&1.25&1.15&1.17&1.19\\
\enddata
\tablecomments{
The temperature gradient inside of the hot core was represented by 200~K and warmer regions.
DoPs for NGC6334F under the different volume ratio, ``V$_{\rm warm}$/V$_{\rm 200K}$" and the different temperature for the warm region, T$_{\rm warm}$.
}
\label{DoP_NGC6334F}
\end{deluxetable}
\clearpage

\begin{deluxetable}{ccccc}
\tablecaption{Comparison of Simulated Abundances with Observed Abundances}
\tablewidth{0pt}
\tablehead{
\colhead{Species} & \colhead{G10 (s)} & \colhead{G10 (o)} & \colhead{N63 (s)} & \colhead{N63 (o)}
}
\startdata
CH$_3$OH&3.0 (-7)&\bf{3.5 (-7)}&9.6 (-7)&\bf{9.0 (-7)}\\
HCOOCH$_3$&2.0 (-8)&\bf{2.3 (-8)}&5.0 (-10)&4.7 (-7)\\
CH$_3$OCH$_3$&7.3 (-8)&\bf{1.3 (-8)}&1.7 (-7)&\bf{4.3 (-8)}\\
(CH$_3$)$_2$CO&4.1 (-9)&\bf{2.5 (-8)}&9.2 (-9)&\bf{2.1 (-9)}\\
NH$_2$CHO&2.8 (-8)&\bf{1.0 (-8)}&9.6 (-9)&7.0 (-10)\\
CH$_2$CHCN&1.6 (-9)&\bf{1.1 (-8)}&3.9 (-10)&\bf{2.9 (-10)}\\
CH$_2$NH&3.9 (-8)&\bf{1.8 (-8)}&8.4 (-9)&\bf{1.2 (-9)}\\
\enddata
\tablecomments{
We searched for best time (DoP is minimum) to explain the observed abundances towards G10.47+0.03 (denoted as G10) and NGC6334F (denoted as N63) under the V$_{120K}$/V$_{200K}$ values of 1, 10, 100, and 1000.
The simulated abundances and the observed abundances are compared in under the best time and V$_{120K}$/V$_{200K}$ ratio, i.e., 8.7 $\times$10$^5$~years and V$_{120K}$/V$_{200K}$=100 for G10.47+0.03, and 8.2 $\times$10$^5$~years and V$_{120K}$/V$_{200K}$=1000 for G10.47+0.03.
s and o, respectively, represent simulated abundances and observed abundance.
A factor of 0.5 was multiplied to the observed fractional abundance to be compared with simulated values.
Observed abundances are presented in bold face if the difference from the simulated one is within a factor of 10.
}
\label{compare_model_with_obs}
\end{deluxetable}
\clearpage

\begin{deluxetable}{lrrrrr}
\tablecaption{The Predicted Correlation Coefficients between Observed Molecules}
\tablewidth{0pt}
\tablehead{
\colhead{} & \colhead{HCOOCH$_3$} & \colhead{CH$_3$OCH$_3$} & \colhead{NH$_2$CHO} & \colhead{CH$_2$CHCN} & \colhead{CH$_2$NH}
}
\startdata
CH$_3$OH&0.48&0.74&0.22&-0.89&-0.79\\
HCOOCH$_3$&&0.89&0.89&-0.78&-0.05\\
CH$_3$OCH$_3$&&&0.81&-0.90&-0.23\\
NH$_2$CHO&&&&-0.54&0.34\\
CH$_2$CHCN&&&&&0.60\\
\enddata
\tablecomments{
We calculated the correlation coefficients with our chemical model under the different ages, fixing V$_{120K}$/V$_{200K}$ to be 100 and T$_{\rm warm}$ to be 120~K.
In the calculation of the correlation coefficients, the simulated abundances in the gas phase were employed in the different age of the core, starting from 1$\times$10$^{5}$ up to 1$\times$10$^{6}$~years by the time interval of 1$\times$10$^{5}$~years.
}
\label{correlation-model}
\end{deluxetable}
\clearpage

\startlongtable
\begin{deluxetable}{lrrrrrrrrr}
\tabletypesize{\scriptsize}
\tablecaption{Observed Lines towards G10.47+0.03}
\tablewidth{0pt}
\tablehead{
\colhead{Species}
&\colhead{\shortstack {Obs. \\ freq \\ (GHz)}}  
& \colhead{\shortstack {Rest \\ freq \\ (GHz)}}
& \colhead{Transition}
&\colhead{\shortstack {Eu \\ (K)}}  
& \colhead{\shortstack {$\mu^2$S \\ (D$^2$)}}
& \colhead{\shortstack {T$_{\rm A}$* \\ (mK)} }
& \colhead{\shortstack {$\Delta$v \\ (km~s$^{-1}$)} }
& \colhead{\shortstack {V$_{\rm LSR}$ \\ (km~s$^{-1}$) }}
& \colhead{\shortstack {rms \\ noise \\ (mK)} }
}
\startdata
CH$_3$OH&80.99371&80.99324&7(2) - 8(1) A$^-$&103&2.5&488&9.9&66.3&9\\
&81.65344&81.65293&18(4) -19(3) E&493&5.9&228&9.9&66.1&8\\
&84.42421&84.42378&13(-3) - 14(-2) E&274&4.3&262&9.7&66.5&6\\
&84.52160&84.52117&5(1) - 4(0) E&40&3.1&1022&8.0&66.5&16\\
&84.74446&84.74390&19(4) - 18(5) E&537&5.2&111&13.0&66.0&11\\
&86.61607&86.61560&7(2) - 6(3) A$^-$&103&1.4&515&9.8&66.4&15\\
&86.90346&86.90295&7(2) - 6(3) A$^+$&103&1.4&560&9.8&66.2&14\\
&88.94051&88.93999&15(3,12) - 14(4) A$-$&328&4.2&404&9.5&66.3&13\\
\enddata
\label{G10_lines}
\end{deluxetable}

\newpage
\startlongtable
\begin{deluxetable}{lrrrrrrrrr}
\tabletypesize{\scriptsize}
\tablecaption{Observed Lines towards Orion KL}
\tablewidth{0pt}
\tablehead{
\colhead{Species}
&\colhead{\shortstack {Obs. \\ freq \\ (GHz)}}  
& \colhead{\shortstack {Rest \\ freq \\ (GHz)}}
& \colhead{Transition}
&\colhead{\shortstack {Eu \\ (K)}}  
& \colhead{\shortstack {$\mu^2$S \\ (D$^2$)}}
& \colhead{\shortstack {T$_{\rm A}$* \\ (mK)} }
& \colhead{\shortstack {$\Delta$v \\ (km~s$^{-1}$)} }
& \colhead{\shortstack {V$_{\rm LSR}$ \\ (km~s$^{-1}$) }}
& \colhead{\shortstack {rms \\ noise \\ (mK)} }
}
\startdata
CH$_3$OH&80.99277&80.99326&7(2) - 8(1) A$^-$&103&2.5&2450&4.3&7.8&15\\
&81.65268&81.65293&18(4) - 19(3) E&493&5.9&439&4.2&6.9&20\\
&84.42337&84.42371&13(-3) - 14(-2) E&274&4.3&888&5.8&7.2&24\\
&84.52058&84.52121&5(-1) - 4(0) E&40&3.1&6136&2.5&8.2&14\\
&84.57350&84.57402&19(2) - 18(-3) E&463&0.4&48&3.0&7.8&12\\
&84.74384&84.74390&19(4) - 18(5) E&537&5.2&409&9.8&6.2&24\\
&86.61514&86.61560&7(2) - 6(3) A$^-$&103&1.4&1606&5.0&7.6&21\\
&86.90250&86.90295&7(2) - 6(3) A$^+$&103&1.4&1787&4.9&7.6&18\\
\enddata
\label{Orion_lines}
\end{deluxetable}

\clearpage

\startlongtable
\begin{deluxetable}{lrrrrrrrrr}
\tabletypesize{\scriptsize}
\tablecaption{Observed Lines towards NGC6334F}
\tablewidth{0pt}
\tablehead{
\colhead{Species}
&\colhead{\shortstack {Obs. \\ freq \\ (GHz)}}  
& \colhead{\shortstack {Rest \\ freq \\ (GHz)}}
& \colhead{Transition}
&\colhead{\shortstack {Eu \\ (K)}}  
& \colhead{\shortstack {$\mu^2$S \\ (D$^2$)}}
& \colhead{\shortstack {T$_{\rm A}$* \\ (mK)} }
& \colhead{\shortstack {$\Delta$v \\ (km~s$^{-1}$)} }
& \colhead{\shortstack {V$_{\rm LSR}$ \\ (km~s$^{-1}$) }}
& \colhead{\shortstack {rms \\ noise \\ (mK)} }
}
\startdata
CH$_3$OH&80.99346&80.99326&7(2) - 8(1) A$^-$&103&2.5&1086&5.9&-7.7&7\\
&81.65317&81.65293&18(4) - 19(3) E&493&5.9&506&5.2&-7.9&21\\
&83.79289&83.79265&18(-2) - 18(2) E&423&0.2&54&4.9&-7.9&7\\
&84.42402&84.42371&13(-3) - 14(-2) E&274&4.3&522&5.4&-8.1&15\\
&84.52163&84.52121&5(-1) - 4(0) E&40&3.1&4014&4.3&-8.5&8\\
&84.57398&84.57402&19(2) - 18(-3) E&463&0.4&80&4.7&-6.9&6\\
&84.74419&84.74390&19(4) - 18(5) E&537&5.2&204&6.4&-8.0&9\\
&85.56844&85.56807&6(-2) - 7(-1) E&75&2.0&974&5.5&-8.3&18\\
\enddata
\label{NGC6334F_lines}
\end{deluxetable}
\clearpage

\startlongtable
\begin{deluxetable}{lrrrrrrrrr}
\tabletypesize{\scriptsize}
\tablecaption{Observed Lines towards W51~e1/e2}
\tablewidth{0pt}
\tablehead{
\colhead{Species}
&\colhead{\shortstack {Obs. \\ freq \\ (GHz)}}  
& \colhead{\shortstack {Rest \\ freq \\ (GHz)}}
& \colhead{Transition}
&\colhead{\shortstack {Eu \\ (K)}}  
& \colhead{\shortstack {$\mu^2$S \\ (D$^2$)}}
& \colhead{\shortstack {T$_{\rm A}$* \\ (mK)} }
& \colhead{\shortstack {$\Delta$v \\ (km~s$^{-1}$)} }
& \colhead{\shortstack {V$_{\rm LSR}$ \\ (km~s$^{-1}$) }}
& \colhead{\shortstack {rms \\ noise \\ (mK)} }
}
\startdata
CH$_3$OH&80.99335&80.99326&7(2) - 8(1) A$^-$&103&2.5&326&9.6&56.7&11\\
&81.65328&81.65293&18(4) - 19(3) E&493&5.9&349&8.1&55.7&11\\
&84.52135&84.52121&5(-1) - 4(0) E&40&3.1&1262&8.2&56.5&24\\
&85.56834&85.56807&6(-2) - 7(-1) E&75&2.0&690&9.7&56.0&20\\
&86.61561&86.61560&7(2) - 6(3) A$^-$&103&1.4&417&9.2&57.0&13\\
&86.90293&86.90295&7(2) - 6(3) A$^+$&103&1.4&438&9.7&57.1&16\\
&88.59507&88.59481&15(3) - 14(4) A$^+$&328&4.2&575&8.8&56.1&15\\
&88.94004&88.93999&15(3) - 14(4) A$^-$&328&4.2&267&9.6&56.8&8\\
\enddata
\label{W51_lines}
\end{deluxetable}
\clearpage

\startlongtable
\begin{deluxetable}{lrrrrrrrrr}
\tabletypesize{\scriptsize}
\tablecaption{Observed Lines towards G31.41+0.3}
\tablewidth{0pt}
\tablehead{
\colhead{Species}
&\colhead{\shortstack {Obs. \\ freq \\ (GHz)}}  
& \colhead{\shortstack {Rest \\ freq \\ (GHz)}}
& \colhead{Transition}
&\colhead{\shortstack {Eu \\ (K)}}  
& \colhead{\shortstack {$\mu^2$S \\ (D$^2$)}}
& \colhead{\shortstack {T$_{\rm A}$* \\ (mK)} }
& \colhead{\shortstack {$\Delta$v \\ (km~s$^{-1}$)} }
& \colhead{\shortstack {V$_{\rm LSR}$ \\ (km~s$^{-1}$) }}
& \colhead{\shortstack {rms \\ noise \\ (mK)} }
}
\startdata
CH$_3$OH&80.99327&80.99324&7(2) - 8(1) A$^-$&103&2.5&241&6.1&96.9&9\\
&81.65291&81.65293&18(4) - 19(3) E&493&5.9&114&7.3&97.1&11\\
&84.42373&84.42378&13(-3) - 14(-2) E&274&4.3&70&7.9&97.2&12\\
&84.52089&84.52117&5(-1) - 4(0) E&40&3.1&481&5.7&98.0&15\\
&84.74390&84.74390&19(4) - 18(5) E&537&5.2&51&8.7&97.0&11\\
&86.61558&86.61560&7(2) - 6(3) A$^-$&103&1.4&387&6.4&97.1&16\\
&86.90297&86.90295&7(2) - 6(3) A$^+$&103&1.4&449&6.9&96.9&22\\
&88.93988&88.93999&15(3) - 14(4) A$^-$&328&4.2&276&7.2&97.4&15\\
\enddata
\label{G31_lines}
\end{deluxetable}
\clearpage

\startlongtable
\begin{deluxetable}{lrrrrrrrrr}
\tabletypesize{\scriptsize}
\tablecaption{Observed Lines towards G34.3+0.2}
\tablewidth{0pt}
\tablehead{
\colhead{Species}
&\colhead{\shortstack {Obs. \\ freq \\ (GHz)}}  
& \colhead{\shortstack {Rest \\ freq \\ (GHz)}}
& \colhead{Transition}
& \colhead{\shortstack {Eu \\ (K)}}  
& \colhead{\shortstack {$\mu^2$S \\ (D$^2$)}}
& \colhead{\shortstack {T$_{\rm A}$* \\ (mK)} }
& \colhead{\shortstack {$\Delta$v \\ (km~s$^{-1}$)} }
& \colhead{\shortstack {V$_{\rm LSR}$ \\ (km~s$^{-1}$) }}
& \colhead{\shortstack {rms \\ noise \\ (mK)} }
}
\startdata
CH$_3$OH&80.99306&80.99326&7(2) - 8(1) A$^-$&103&2.5&539&7.1&59.4&12\\
&81.65298&81.65293&18(4) - 19(3) E&493&5.9&210&6.3&58.4&9\\
&85.56790&85.56807&6(-2) - 7(-1) E&75&2.0&495&7.0&59.2&10\\
&88.59464&88.59481&15(3) - 14(4) A$^+$&328&4.2&369&7.0&59.2&11\\
&89.50548&89.50578&8(-4) - 9(-3) E&171&1.6&206&6.9&59.6&6\\
&94.54160&94.54181&8(3) - 9(2) E&131&2.2&511&6.8&59.3&16\\
&94.81494&94.81507&19(7) - 20(6) A$^+$&685&4.6&71&5.2&59.0&6\\
&blend&94.81508&19(7) - 20(6) A$^-$&685&4.6&&&&\\
\enddata
\label{G34_lines}
\end{deluxetable}
\clearpage

\startlongtable
\begin{deluxetable}{lrrrrrrrrr}
\tabletypesize{\scriptsize}
\tablecaption{Observed Lines towards G19.61-0.23}
\tablewidth{0pt}
\tablehead{
\colhead{Species}
&\colhead{\shortstack {Obs. \\ freq \\ (GHz)}}  
& \colhead{\shortstack {Rest \\ freq \\ (GHz)}}
& \colhead{Transition}
&\colhead{\shortstack {Eu \\ (K)}}  
& \colhead{\shortstack {$\mu^2$S \\ (D$^2$)}}
& \colhead{\shortstack {T$_{\rm A}$* \\ (mK)} }
& \colhead{\shortstack {$\Delta$v \\ (km~s$^{-1}$)} }
& \colhead{\shortstack {V$_{\rm LSR}$ \\ (km~s$^{-1}$) }}
& \colhead{\shortstack {rms \\ noise \\ (mK)} }
}
\startdata
CH$_3$OH&103.38181&103.38115&12(-2) - 12(1) E&207&0.8&69&5.3&40.0&15\\
&104.06136&104.06066&13(-3) - 12(-4) E&274&3.3&91&6.9&39.9&17\\
&104.33730&104.33656&13(-2) - 13(1) E&237&1.2&67&4.7&39.8&15\\
&104.35532&104.35482&10(4) - 11(3) A$^-$&208&2.5&96&7.2&40.5&18\\
&105.06443&105.06369&13(1) - 12(2) A$^+$&224&4.3&146&6.6&39.8&17\\
&105.57686&105.57629&14(-2) - 14(1) E&270&1.8&95&4.6&40.3&18\\
NH$_2$CHO&105.46481&105.46422&5(0,5)-4(0,4)&15&65.3&101&9.1&40.2&14\\
&105.97357&105.97260&5(2,4)-4(2,3)&27&54.9&84&6.2&39.1&18\\
\enddata
\label{G19_lines}
\end{deluxetable}
\clearpage

\startlongtable
\begin{deluxetable}{lrrrrrrrrr}
\tabletypesize{\scriptsize}
\tablecaption{Observed Lines towards DR21(OH)}
\tablewidth{0pt}
\tablehead{
\colhead{Species}
&\colhead{\shortstack {Obs. \\ freq \\ (GHz)}}  
& \colhead{\shortstack {Rest \\ freq \\ (GHz)}}
& \colhead{Transition}
&\colhead{\shortstack {Eu \\ (K)}}  
& \colhead{\shortstack {$\mu^2$S \\ (D$^2$)}}
& \colhead{\shortstack {T$_{\rm A}$* \\ (mK)} }
& \colhead{\shortstack {$\Delta$v \\ (km~s$^{-1}$)} }
& \colhead{\shortstack {V$_{\rm LSR}$ \\ (km~s$^{-1}$) }}
& \colhead{\shortstack {rms \\ noise \\ (mK)} }
}
\startdata
CH$_3$OH&105.06372&105.06369&13(1) -12(2) A$^+$&224&4.3&41&4.4&-3.1&8\\
HCOOCH$_3$&103.46663&103.46729&8(2,6)-7(2,5)E&25&20.2&22&3.9&-1.1&9\\
&103.47894&103.47927&8(2,6)-7(2,5)A&25&20.2&27&2.9&-2.0&8\\
\enddata
\label{DR21_lines}
\end{deluxetable}




\end{document}